\documentclass{jfm}

\usepackage{graphicx}
\usepackage{newtxtext}
\usepackage{newtxmath}
\usepackage{natbib}
\usepackage{hyperref}
\hypersetup{
    colorlinks = true,
    urlcolor   = blue,
    citecolor  = black,
}

\newcommand{\RomanNumeralCaps}[1]

\title{On the wavenumber--frequency spectra of the wall pressure fluctuations in turbulent channel flow}

\author{Bowen Yang\aff{1},
  Zixuan Yang\aff{1,2}
  \corresp{\email{yangzx@imech.ac.cn}}
}

\affiliation{\aff{1}The State Key Laboratory of Nonlinear Mechanics,
Institute of Mechanics, Chinese Academy of Sciences, Beijing 100190, China
\aff{2}School of Engineering Sciences, University of
Chinese Academy of Sciences, Beijing 100049, China}

\begin{document}
\maketitle

\begin{abstract}
Direct numerical simulations (DNS) of turbulent channel flows up to $\Rey_\tau\approx 1000$ are conducted to investigate the three-dimensional (consisting of streamwise wavenumber, spanwise wavenumber and frequency) spectrum of wall pressure fluctuations.   To develop a predictive model of the wavenumber--frequency spectrum from the wavenumber spectrum, the time decorrelation mechanisms of wall pressure fluctuations are investigated.  It is discovered that the energy-containing part of the wavenumber--frequency spectrum of wall pressure fluctuations can be well predicted using a similar random sweeping model for streamwise velocity fluctuations. To refine the investigation, we further decompose the spectrum of the total wall pressure fluctuations into the auto spectra of rapid and slow pressure fluctuations, and the cross spectrum between them. We focus on evaluating the assumption applied in many predictive models, that is, the magnitude of the cross spectrum is negligibly small.  The present DNS shows that neglecting the cross spectrum causes a maximum error up to 4.7dB in the sub-convective region for all Reynolds numbers under test.  Our analyses indicate that the assumption of neglecting the cross spectrum needs to be applied carefully in the investigations of acoustics at low Mach numbers, in which the sub-convective components of wall pressure fluctuations make important contributions to the radiated acoustic power.
\end{abstract}

\begin{keywords}
Authors should not enter keywords on the manuscript, as these must be chosen by the author during the online submission process and will then be added during the typesetting process (see \href{https://www.cambridge.org/core/journals/journal-of-fluid-mechanics/information/list-of-keywords}{Keyword PDF} for the full list).  Other classifications will be added at the same time.
\end{keywords}

{\bf MSC Codes }  {\it(Optional)} Please enter your MSC Codes here

\section{Introduction}\label{sec:introduction}
Wall pressure fluctuations are dominant sources of flow-generated noise in wall-bounded turbulence~\citep{Wang06,Graham97}. The sound radiated by a solid wall depends on the intensity and spatial--temporal variations of the pressure fluctuations. Understanding the characteristics of the wavenumber--frequency spectrum of the wall pressure fluctuations is crucial for developing predictive models of flow-generated noise.

Experimental measurement~\citep{Willmarth75,Corcos64,Blake70,Farabee91} is an important approach for acquiring data of wall pressure fluctuations. \citet{Arguillat05} and \citet{Salze14} used an array of sensors to measure the spatial correlations of wall pressure fluctuations at different frequencies, and applied Fourier transform in space to obtain the wavenumber--frequency spectrum. The energy-containing parts of the wavenumber--frequency spectrum, including the convective peak and acoustic part, were identified in their experiments, while other parts with lower energy level were unresolved. 

Another approach for investigating the wall pressure fluctuations is numerical simulation. \citet{Kim89} pioneered the direct numerical simulation (DNS) study of pressure fluctuations in a turbulent channel flow at $\Rey_\tau=180$, with focuses on the root-mean-square (RMS) profiles, one-dimensional (either wavenumber or frequency) spectra, and probability density functions. The database established by \citet{Kim89} was further analyzed by \citet{Choi90} to investigate the wavenumber--frequency spectrum of wall pressure fluctuations. They identified the convective characteristic of wall pressure fluctuations, and showed that the similarity form of the spectrum proposed in an early study of \citet{Corcos64} was only accurate at small length and time scales. \citet{Chang99} also conducted DNS of turbulent channel flow at $\Rey_\tau=180$ to study the wall pressure fluctuations. They concluded that the velocity gradients in the buffer layer were dominant sources of wall pressure fluctuations for most wavenumbers. \citet{Hu02} used DNS data of turbulent channel flow at $\Rey_\tau=180$ to study the wavenumber--frequency spectrum of wall pressure fluctuations at low wavenumbers corresponding to large-scale turbulent motions. They argued that the wavenumber--frequency spectrum of wall pressure fluctuations showed no $k^2$-scaling at low wavenumbers as predicted theoretically by \citet{Kraichnan56} and \citet{Phillips56}, where $k$ refers to the norm of the wavenumber vector consisting of streamwise and spanwise components. While the above DNS studies of pressure fluctuations are confined to $\Rey_\tau=180$, \citet{Abe05} conducted DNS of turbulent channel flows up to $\Rey_\tau\approx1000$ to investigate the effect of the Reynolds number on the pressure fluctuations. \citet{Hu06} analyzed DNS data of turbulent channel flows from $\Rey_\tau=90$ to 1440 to investigate the scaling of the frequency spectrum of wall pressure fluctuations at different characteristic frequencies. \citet{Xu20} derived a theoretical model of the streamwise-wavenumber spectrum of pressure fluctuations in the logarithmic layer based on Kolmogorov's theory~\citep{Kolmogorov41} and Townsend's attached eddy hypothesis~\citep{Townsend76}.  This model shows a good agreement with the DNS results in turbulent channel flows up to $\Rey_\tau=5200$~\citep{Lee2015}. The above investigations of the pressure fluctuations at higher Reynolds numbers focused on their RMS profiles and one-dimensional spectra, while owing to the requirements of long simulation time and large storage capacity, DNS investigations of three-dimensional wavenumber--frequency spectrum of wall pressure fluctuations are limited to $\Rey_\tau=180$~\citep{Choi90,Hu02}. Besides the above DNS studies, large-eddy simulations (LES) were also conducted to investigate the wall pressure fluctuations in recent years. \citet{Gloerfelt13} performed LES of compressible turbulent boundary layer flow, which resolved the acoustic part of the wavenumber--frequency spectrum of wall pressure fluctuations. \citet{Viazzo01} and \citet{Park16} used both wall-resolved and wall-modelled LES to investigate the wavenumber--frequency spectrum of wall pressure fluctuations in turbulent channel flows.

Besides the experimental and numerical studies, numerous models were proposed to facilitate a fast prediction of the wavenumber--frequency spectrum of wall pressure fluctuations.  To develop a model of the wavenumber--frequency spectrum (or equivalently, the space--time correlation in physical space), it is useful to understand the physical mechanisms of the time decorrelation process~\citep{He17}. In previous studies of velocity fluctuations, it was discovered that in turbulent shear flows, the time decorrelation was mainly caused by the convection effect of mean flow~\citep{Taylor38} and the sweeping effect of large-scale turbulent eddies~\citep{Kraichnan64,Tennekes75}. Models for space--time correlations~\citep{He06,Zhao09} and wavenumber--frequency spectra~\citep{Wilczek12,Wilczek15} of velocity fluctuations, which took both convection and sweeping into consideration, were proposed and tested. It was verified that the wavenumber--frequency spectra of velocity fluctuations obtained from these models were consistent with the DNS results in turbulent channel flows. By contrast, the time decorrelation mechanisms of  wall pressure fluctuations were rarely investigated in literature. In isotropic turbulence, \citet{Yao08} showed analytically that the pressure fluctuations followed the same random advection equations of velocity fluctuations. 
However, in wall-bounded shear turbulence, time decorrelation mechanisms of wall pressure fluctuations have not been discussed in literature, which are investigated using the present DNS data.

The DNS database is also useful for validating models.  The existing models of wall pressure fluctuations can be categorized into two classes, namely the semi-empirical models and those based on the Poisson equation of pressure fluctuations~\citep{Slama18}. In semi-empirical models~\citep[e.g.][]{Corcos64,Chase80,Chase87,Smolyakov06,Frendi20}, data fitting is usually used to directly construct the wavenumber--frequency spectrum of wall pressure fluctuations. Various semi-empirical models predict similar spectral magnitudes around the convective peak, but show discrepancies at wavenumber--frequency combinations located in the sub-convective region (which refers to the region in the wavenumber--frequency space with lower streamwise wavenumber than the convective peak). Detailed comparisons of different semi-empirical models can be found in the review articles of \citet{Graham97} and \citet{Hwang09}.

Different from the semi-empirical models, the models based on the Poisson equation of the pressure fluctuations~\citep[e.g.][]{Panton74,Peltier07,Slama18,Grasso19} do not predict the wavenumber--frequency spectrum of wall pressure fluctuations directly. Instead, the wall pressure fluctuations are expressed as the solution of the following Poisson equation
\begin{equation}
\frac{1}{\rho }{\nabla ^2}p =  - 2\frac{{\partial {u_i}}}{{\partial {x_j}}}\frac{{\partial {U_j}}}{{\partial {x_i}}} - \frac{{{\partial ^2}}}{{\partial {x_i}\partial {x_j}}}({u_i}{u_j} - \langle  {{u_i}{u_j}} \rangle ),
\label{eq:eq1.1}
\end{equation}
where $p$ denotes the pressure fluctuations, $\rho$ is the fluid density, $u_i$ and $U_i$ $(i=1, 2, 3)$ represent the velocity fluctuations and mean velocity, respectively, and a pair of angular brackets denotes averaging over time and streamwise--spanwise plane.  In equation~(\ref{eq:eq1.1}), the right-hand side consists of a linear term and a quadratic term with respect to the velocity fluctuations $u_i$, which are called the rapid and slow source terms, respectively~\citep{Kim89}. Because equation~(\ref{eq:eq1.1}) is linear about the pressure fluctuations, the total pressure fluctuations can be decomposed into rapid and slow components, corresponding to the rapid and slow source terms, respectively. Thus, the wavenumber--frequency spectrum of the total wall pressure fluctuations can be decomposed into three components as the auto spectra of rapid (AS-Rapid) and slow (AS-Slow) components, respectively, and the cross spectrum between them (CS-RS).  It is commonly assumed in the models based on the Poisson equation of  pressure fluctuations that  velocity fluctuations are Gaussian variables satisfying joint normal distributions~\citep{Peltier07,Slama18,Grasso19}. Because CS-RS is a cubic function of velocity fluctuations, it is neglected in these models, since odd-order moments of Gaussian variables equal to zero theoretically. In some semi-empirical models, such as the Chase model~\citep{Chase80,Chase87}, this assumption is also applied. However, this assumption has not been confirmed by any numerical or experimental data.

In the present study, we conduct DNS of turbulent channel flows at four Reynolds numbers ranging from $Re_\tau =179$ to $998$ to investigate the characteristics of the wavenumber--frequency spectrum of wall pressure fluctuations. The  objectives of the present study include: (1) to establish DNS database of the wavenumber--frequency spectrum of wall pressure fluctuations up to $\Rey_\tau\approx1000$; (2) to investigate the time decorrelation mechanisms of wall pressure fluctuations; and (3) to validate the assumption that CS-RS is negligibly small in comparison with AR-Rapid and AR-Slow. The remainder of this paper is organized as follows. Numerical details and the method to compute the wavenumber--frequency spectrum are described in \S\,\ref{sec:numerical}. The characteristics of the wavenumber--frequency spectrum of the total wall pressure fluctuations are analyzed in \S\,\ref{sec:total}. The decorrelation mechanisms of  wall pressure fluctuations are discussed in \S\,\ref{sec:decorrelation}. The decomposition of the wavenumber--frequency spectrum of  wall pressure fluctuations is further investigated in \S\,\ref{sec:decomposition}, followed by the conclusions in \S\,\ref{sec:conclusion}.

\section{Simulation parameters and numerical methods}\label{sec:numerical}

\begin{table}
  \begin{center}
\def~{\hphantom{0}}
  \begin{tabular}{lcccccccc}
      Cases  & $Re_b$ &  $Re_\tau$ & $L_x\times L_y\times L_z$ & $N_x\times N_y\times N_z$  & $\Delta x^+$ & $\Delta y^+$ & $\Delta_z^+$ & $T_s^+$\\[3pt]
      CH180   & 2800  & 179 & $4\pi h\times 2h\times 2\pi h$ & $192\times129\times192$   & 11.7 & $0.054\sim4.39$ & 5.9 & 0.57\\
      CH330   & 5700  & 333 & $4\pi h\times 2h\times 2\pi h$ & $384\times193\times384$   & 11.1 & $0.046\sim5.56$ & 5.6 & 0.39\\
      CH550   & 10150 & 551 & $4\pi h\times 2h\times 2\pi h$ & $576\times257\times576$   & 12.0 & $0.041\sim6.75$ & 6.0 & 0.30\\
      CH1000  & 20000 & 998 & $4\pi h\times 2h\times 2\pi h$ & $1152\times385\times1152$ & 10.9 & $0.034\sim8.18$ & 5.5 & 0.60\\
  \end{tabular}
  \caption{Key parameters of DNS, including the Reynolds numbers $Re_b$ and $Re_\tau$, computational domain size $L_x\times L_y\times Lz$, number of grid points $N_x\times N_y\times N_z$, grid resolution $\Delta x^+\times\Delta y^+\times\Delta z^+$ and the time separation $T_s^+$ of data storage.}
  \label{tab:DNS}
  \end{center}
\end{table}

\begin{figure}
	\centering{\includegraphics[width=0.48\textwidth]{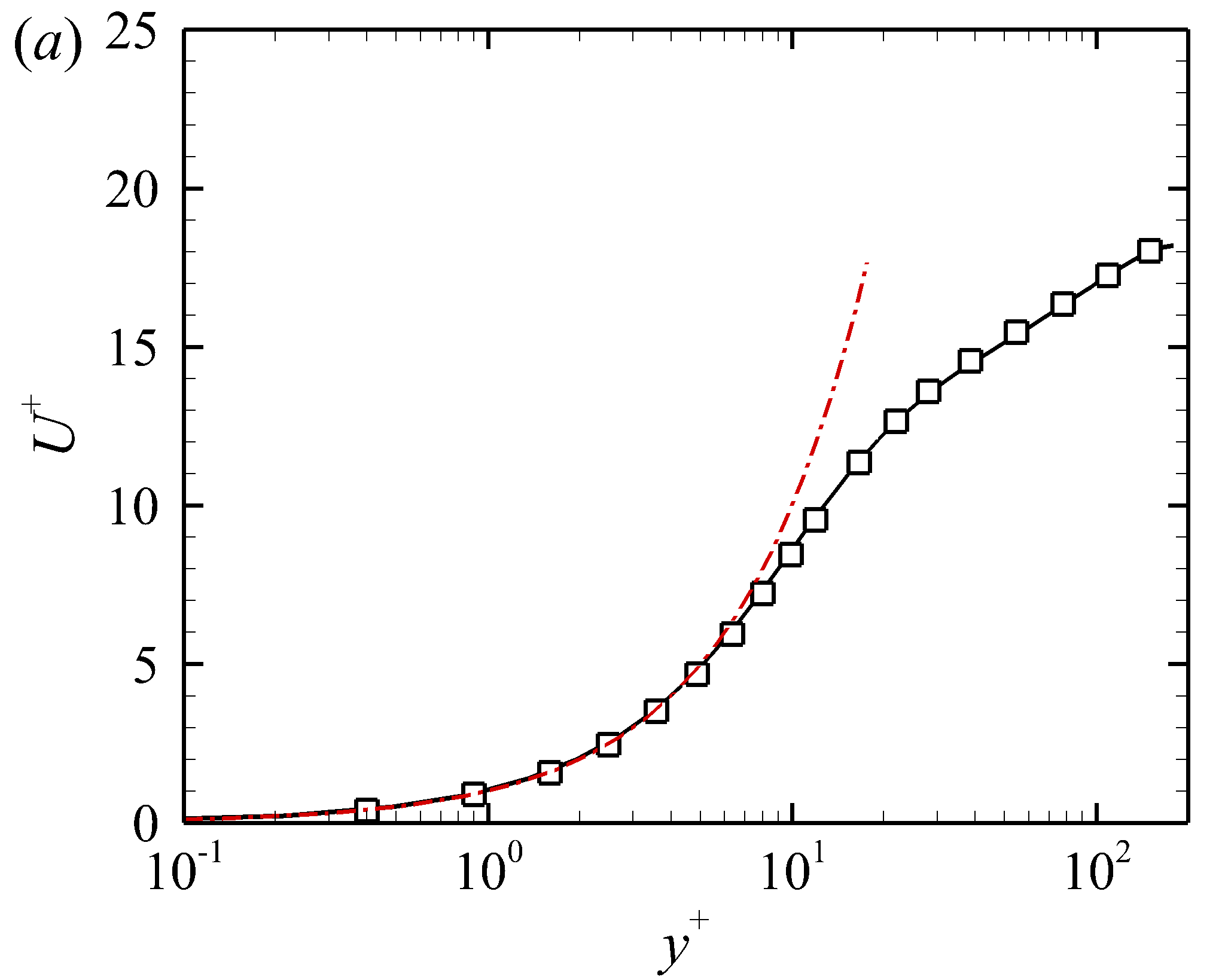}$\quad$
          	  {\includegraphics[width=0.48\textwidth]{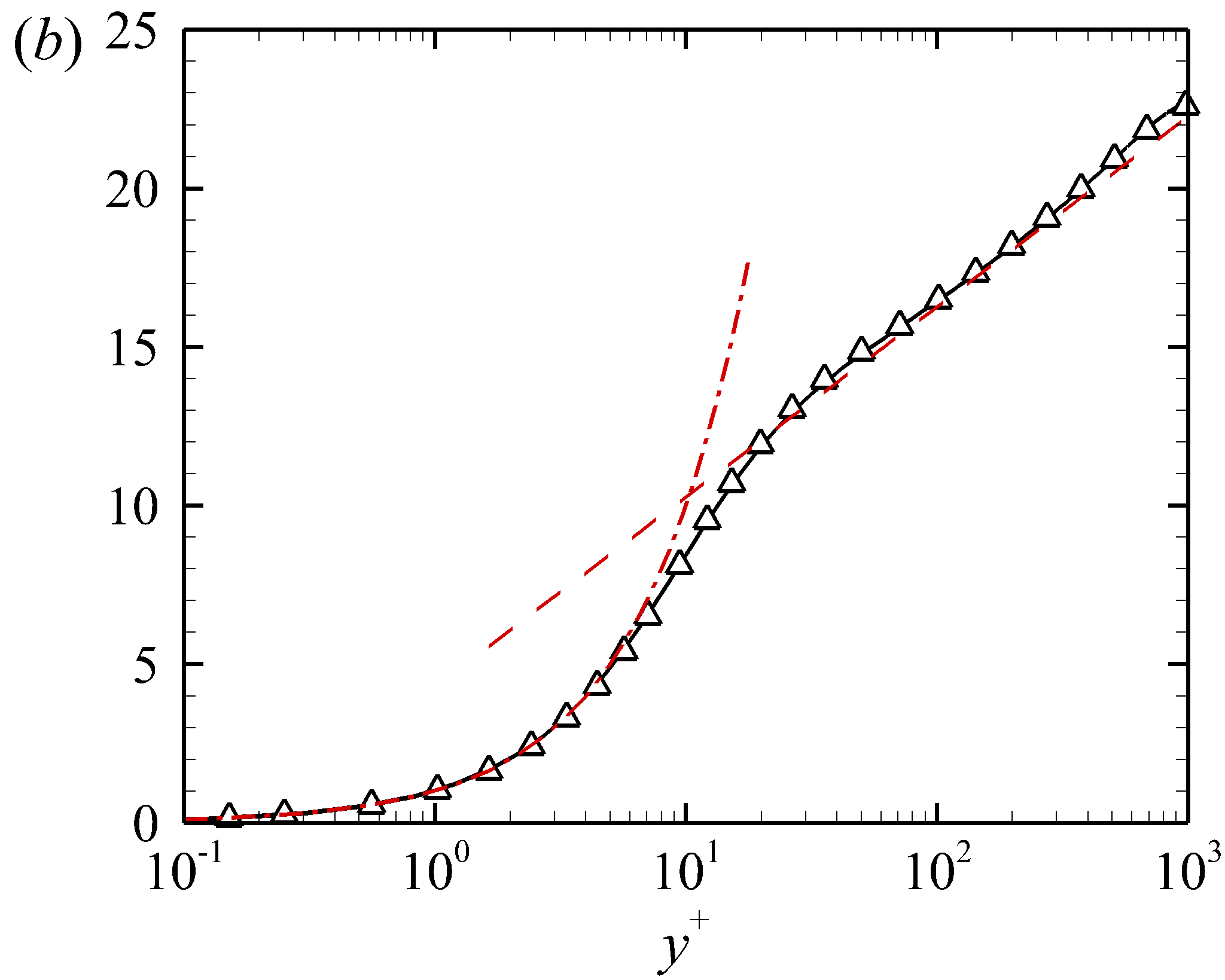}}
              {\includegraphics[width=0.48\textwidth]{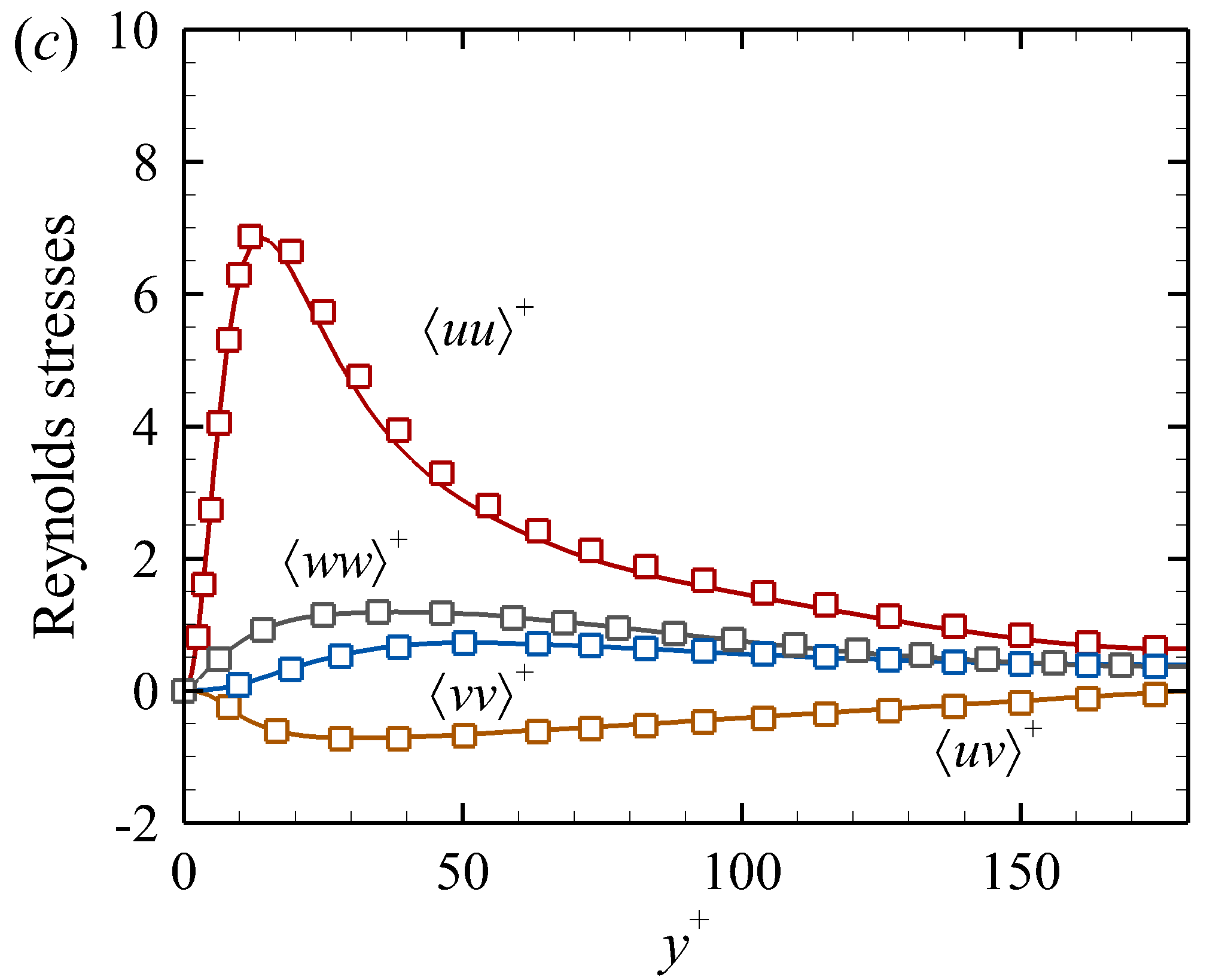}}$\quad$
              {\includegraphics[width=0.48\textwidth]{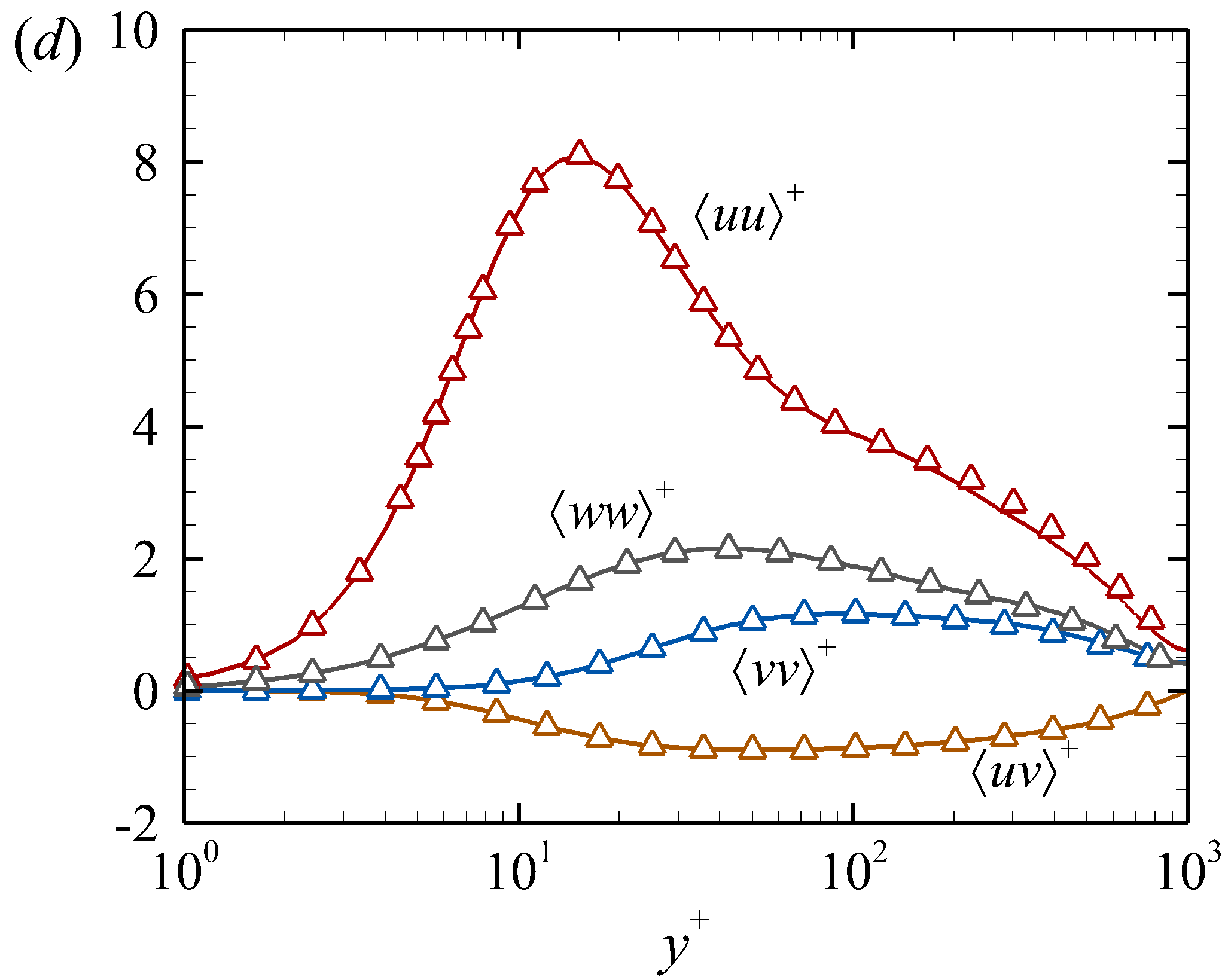}}}
    \caption{Profiles of (a) mean velocity at $\Rey_{\tau}=179$, (b) mean velocity at $\Rey_{\tau}=998$, (c) Reynolds stresses at $\Rey_{\tau}=179$, and (d) Reynolds stresses at $\Rey_{\tau}=998$.  The solid lines represent the results of the present DNS.  The symbols in panels (a) and (c) denote the DNS results of \citet{Hoyas06} at $Re_\tau = 180$, while those in panels (b) and (d) denote the DNS results of \citet{Lee2015} at $Re_\tau = 1000$. The dash-dotted lines in panels (a) and (b) represent the linear law $U^+=y^+$ and the dashed line in panel (b) represents the logarithmic law $U^+=\log y^+/\kappa+B$, where $\kappa = 0.384$ and $B=4.27$ are constants~\citep{Lee2015}.} 
    \label{fig:fig1}
\end{figure}

Table~\ref{tab:DNS} summarizes the key parameters of DNS. We have conducted DNS at four Reynolds numbers $Re_\tau = u_\tau h / \nu = 179$, 333, 551 and 998, where $u_\tau=\sqrt{\tau_w/\rho}$ represents the wall-friction velocity, $h$ denotes one-half the channel height, and $\nu$ is the kinematic viscosity. Here, $\tau_w$ is the mean wall shear stress, computed as

\begin{equation}
	\tau_w=\frac{\tau_{w, low}+\tau_{w, up}}{2}=\frac{1}{2}\biggl(\mu\left.\dfrac{dU}{dy}\right|_{y=-h}-\mu\left.\dfrac{dU}{dy}\right|_{y=h}\biggl),
	\label{eq:utau}
\end{equation}
where $\tau_{w,low}$ and $\tau_{w,up}$ are the mean wall stresses at the lower wall $y=-h$ and upper wall $y=h$, respectively, $\mu=\rho\nu$ is the dynamic viscosity. The corresponding Reynolds number based on the bulk mean velocity $U_b$ is $\Rey_b=U_bh/\nu=2800$, 5700, 10150 and 20000, respectively. As mentioned in \S\,\ref{sec:introduction}, existing DNS results of the wavenumber--frequency spectrum of the wall pressure fluctuation are limited to a low Reynolds number at $\Rey_{\tau}=180$ in the literature~\citep{Choi90,Hu02}. In the present study, we conduct DNS at higher Reynolds numbers to further investigate the Reynolds number effects. The computational domain size is set to $L_x\times L_y\times L_z=4\pi h\times 2h\times 2\pi h$ in all cases. Hereinafter, the streamwise, wall-normal and spanwise directions are denoted using $x$, $y$ and $z$, respectively. The number of grid points $N_x\times N_y\times N_z$ is chosen to match the grid resolution of~\citet{Hoyas06}.  Specifically, in the streamwise and spanwise directions, the grid resolution is set to $\Delta x^+\approx 11$ and $\Delta z^+\approx 5.5$, respectively. In this paper, the superscript `$+$' denotes variables non-dimensionalized using the wall units $\nu/u_\tau$ and $u_\tau$ as characteristic length and velocity scales, respectively. The wall-normal grids locate on the Chebyshev collocation points, and the number of wall-normal grid points is similar to~\citet{Hoyas06}. The smallest spatial scales are comparable to the local Kolmogorov scales, and owing to the high accuracy of the pseudo-spectral method used for conducting the DNS, the numerical dissipation is negligible~\citep{Hoyas06}. The flow is driven by a streamwise pressure gradient, which is adjusted to sustain the bulk mean velocity to a constant. Periodic boundary conditions are applied in the streamwise and spanwise directions, while no-slip and no-penetration conditions are prescribed at the solid walls.  
As pointed out by \citet{Choi90},  the use of the periodic boundary condition in the wall-parallel directions can induce a numerical artifact called the artificial acoustic.  However, as is discussed in both \citet{Choi90} and the present study (\S\,\ref{subsec:total_two}), the influence of the artificial acoustic is confined at a few smallest wavenumbers, corresponding to large length scales. Increasing the computational domain can reduce the influence of the artificial acoustic.  The wavenumber--frequency spectrum of wall pressure fluctuations around the convection line considered in \S\,\ref{sec:decorrelation} are merely influenced by the artificial acoustic, while in \S\,\ref{sec:decomposition}, we mainly focus on the maximum error that occurs out of the wavenumbers and frequencies influenced by the artificial acoustic. In this regard, the present computational domain size of $L_x = 4\pi h$ is sufficient for the problems under investigation. The continuity and momentum equations for incompressible flows are solved using an in-house pseudo-spectral method code. Velocity and pressure are expanded into Fourier series in the streamwise and spanwise directions, and into Chebyshev polynomials in the wall-normal direction. The nonlinear terms are calculated in physical space, and the 3/2 rule is used to remove the aliasing errors~\citep{Kim87,Patterson71}. The third-order time splitting method of~\citet{Karniadakis91} is used for time advancement. The computational time step is chosen to satisfy the CFL condition, which is $\Delta t^+ = 0.057$, 0.039, 0.030, 0.040 for $Re_\tau = 179$, 333, 551, and 998, respectively. The DNS code has been tested systematically in previous studies~\citep{Deng12,Deng16}. Figure~\ref{fig:fig1} compares the profiles of the mean velocity and Reynolds stresses $\langle uu\rangle^+$, $\langle vv\rangle^+$, $\langle ww\rangle^+$, $\langle uv\rangle^+$ of cases CH180 and CH1000 with the DNS results of \citet{Hoyas06} and \citet{Lee2015}. It is seen that the present results are in good agreement with the results in the literature.

Once the turbulence is fully developed to a statistically stationary state, the instantaneous flow fields are stored with a time separation of $T_s^+$, ranging from 0.35 to 0.60 at different Reynolds numbers (see table~\ref{tab:DNS}). Since $T_s$ determines the largest resolved frequency, it should satisfy $T_s\le \Delta x/U_{bc}$ in order to resolve the energy-containing convection line, where $U_{bc}$ is the bulk convection velocity (defined later in \S\,\ref{subsec:two}). The above criterion is satisfied in all test cases. The wavenumber--frequency spectrum of  wall pressure fluctuations are then calculated using the method described by~\citet{Choi90}. The time series of wall pressure fluctuations $p(x,z,t)$ are divided into $M$ intervals with 50\% overlapping between two neighboring intervals. Each interval contains $N_t=512$ successive snapshots. Fourier transform is then performed over the wall pressure fluctuation in each time interval in the streamwise and spanwise directions and in time to obtain the Fourier modes as
\begin{equation}
\begin{aligned}
	\hat p({k_x},{k_z};\omega ) &= \frac{1}{{{L_x}{L_z}\sqrt {T\int_0^T {w{{(t)}^2}{\rm{d}}t} } }}\\
	&\times\int_0^T {{\rm{d}}t\int_0^{{L_x}} {{\rm{d}}x\int_0^{{L_z}} {{\rm{d}}z \cdot w(t)p(x,z;t){e^{ - {\rm{i}}(x{k_x} + z{k_z} - \omega t)}}} } },
	\label{eq:eq2.1}
\end{aligned}
\end{equation}
where $T=N_t\cdot T_s$ is the time duration of each interval and $w(t)$ is a standard Hanning window. Since all the DNS data are calculated on discrete grids, practically the discrete Fourier transform (DFT) is used to compute $\hat p(k_x,k_z;\omega)$ as
\begin{equation}
\begin{aligned}
	\hat p({k_x},{k_z};\omega ) &= \frac{\sqrt {8/3}}{{{N_x}{N_z}{N_t}}}\\
	&\times\sum_{l=0}^{N_x-1}\sum_{m=0}^{N_z-1}\sum_{n=0}^{N_t-1}w\biggl(\frac{nT}{N_t}\biggl)p\biggl(\frac{lL_x}{N_x},\frac{mL_z}{N_z},\frac{nT}{N_t}\biggl)e^{ - {\rm{i}}(lL_x{k_x}/N_x + mL_z{k_z}/N_z - nT\omega/N_t)},
	\label{eq:DFT}
\end{aligned}
\end{equation}
where a coefficient $\sqrt{8/3}$ is included to keep the RMS value of $p$ unchanged after applying the window function $w(t)$. The wavenumber--frequency spectrum of  wall pressure fluctuations are then calculated as
\begin{equation}
	{\phi _{pp}}({k_x},{k_z},\omega ) = \frac{{\overline {\hat p({k_x},{k_z},\omega ){{\hat p}^*}({k_x},{k_z},\omega )} }}{{\Delta {k_x}\Delta {k_z}\Delta \omega }},
	\label{eq:eq2.2}
\end{equation}
where $\Delta k_x=2\pi/L_x$ and $\Delta k_z=2\pi/L_z$ represent the streamwise- and spanwise-wavenumber resolution, respectively, and $\Delta\omega=2\pi/T$ is the frequency resolution. The overbar denotes time averaging over all of the $M$ intervals.  In this paper, the results for $M=38$ are presented. To be specific, 5120 fields are stored for each case. These fields are then divided into 19 windows of 512 fields with 50\% overlap, and averaging over two walls doubles the number of samples $M$ to 38. The corresponding total time duration is 16.3, 6.0, 2.8 and 3.1 eddy turnover time ($h/u_{\tau}$) for $\Rey_{\tau}=179$, 333, 551 and 998, respectively. Because the windows are not randomly chosen, it is desired that the time duration of the data set for time averaging is sufficiently long.  To verify this, we have conducted a data convergence analysis by changing the number of averaging samples $M$.  The results are provided in appendix~\ref{appA}. It is seen that reducing the value of $M$ from 38 to 19 does not cause any qualitative change in the main conclusions of this paper.

The mean-square wall pressure fluctuation can be calculated as the summation of the discrete three-dimensional wavenumber--frequency spectrum over all spatial and temporal scales as
\begin{equation}
	\overline{p^2} = \sum_{l=-N_x/2}^{N_x/2-1}\sum_{m=-N_z/2}^{N_z/2-1}\sum_{n=-N_t/2}^{N_t/2-1}\phi_{pp}(l\Delta k_x,m\Delta k_z,n\Delta\omega)\Delta k_x\Delta k_z\Delta \omega.
	\label{eq:meansquare}
\end{equation}
Similarly, the reduced-dimension spectra are calculated by integrating the three-dimensional wavenumber--frequency spectrum given by   equation~(\ref{eq:eq2.2}). For example, the one-dimensional streamwise-wavenumber spectrum $\phi_{pp}(k_x)$ is the integration of $\phi_{pp}(k_x,k_z,\omega)$ over the spanwise wavenumber $k_z$ and frequency $\omega$.

\section{The wavenumber-frequency spectrum of the total wall pressure fluctuations}\label{sec:total}
\noindent
The wavenumber--frequency spectrum of  wall pressure fluctuations is a key input of predictive models of turbulent-generated noise. In literature, DNS investigations of the wavenumber--frequency spectrum of wall pressure fluctuations are limited to $\Rey_\tau=180$. In this section, we analyze the DNS results of the wavenumber--frequency spectrum of the total wall pressure fluctuations up to $\Rey_\tau=998$.

\subsection{One-dimensional spectra}\label{subsec:total_one}
We start our results analyses with the one-dimensional spectra. We note that the one-dimensional spectra of the wall pressure fluctuation have been previously studied~\citep{Abe05,Anantharamu20}. The purpose of presenting the results of the one-dimensional spectra is to provide a further validation of our DNS data.

\begin{figure}
	\centering{\includegraphics[width=0.3\textwidth]{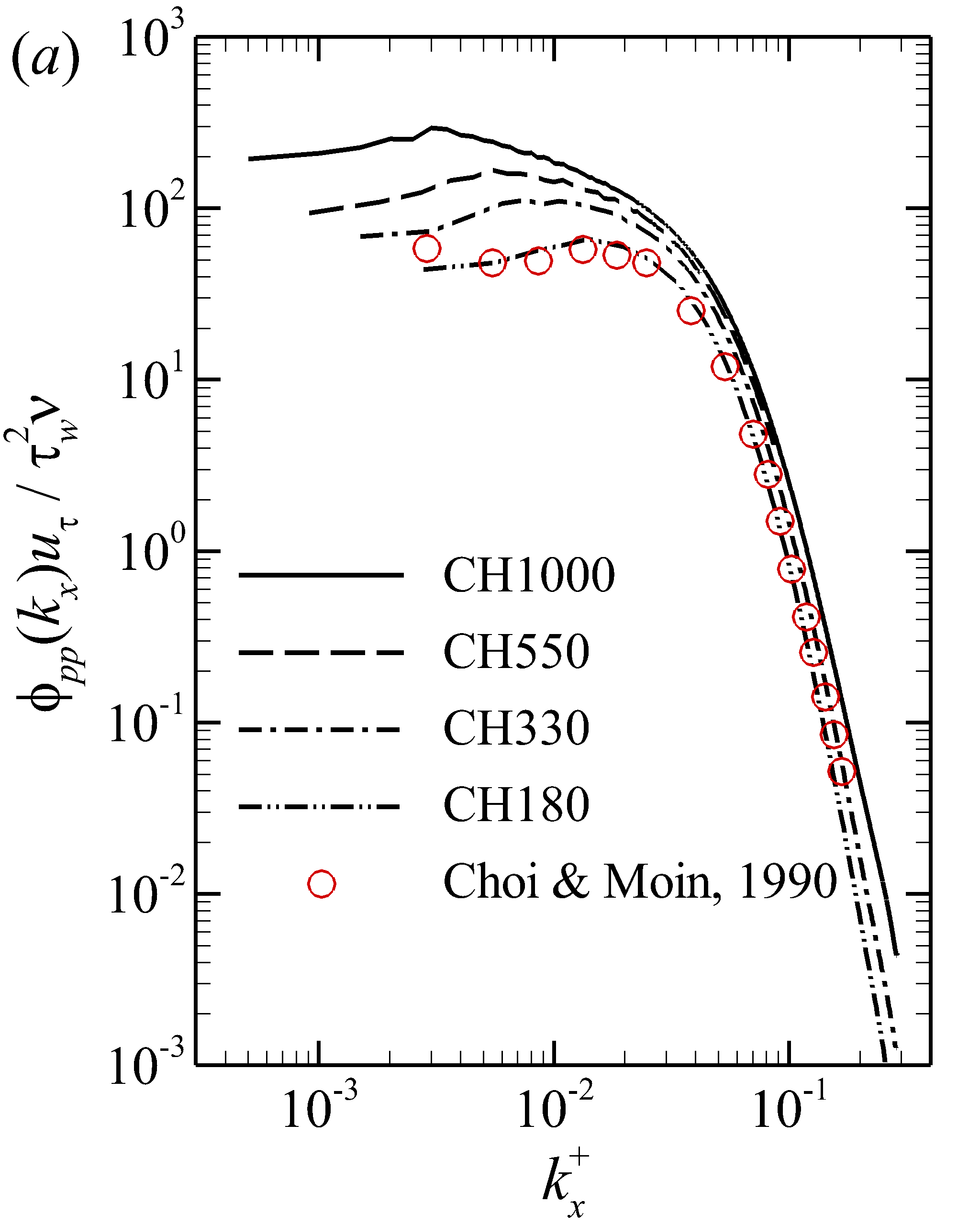}$\quad$
	          {\includegraphics[width=0.3\textwidth]{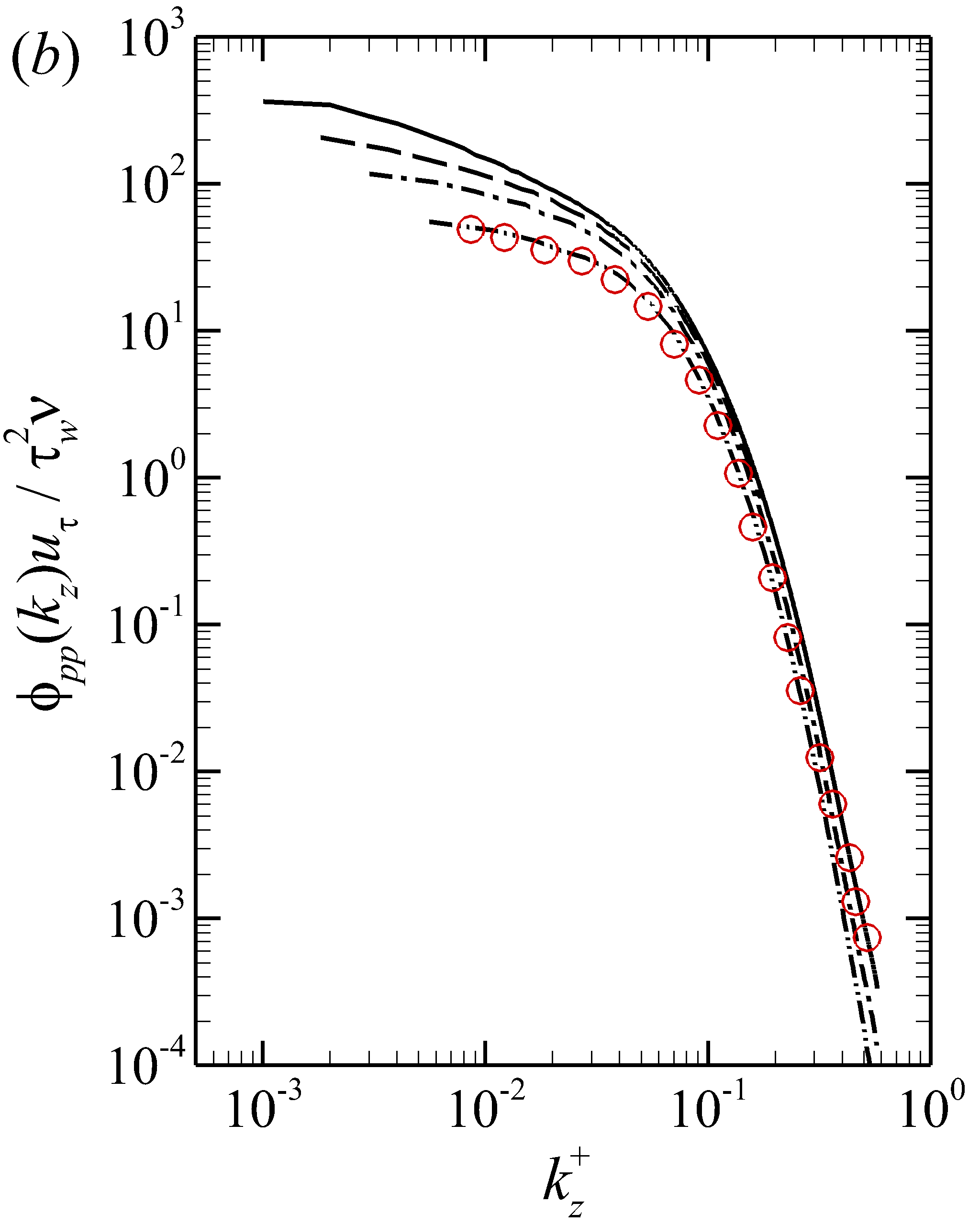}}$\quad$
	          {\includegraphics[width=0.3\textwidth]{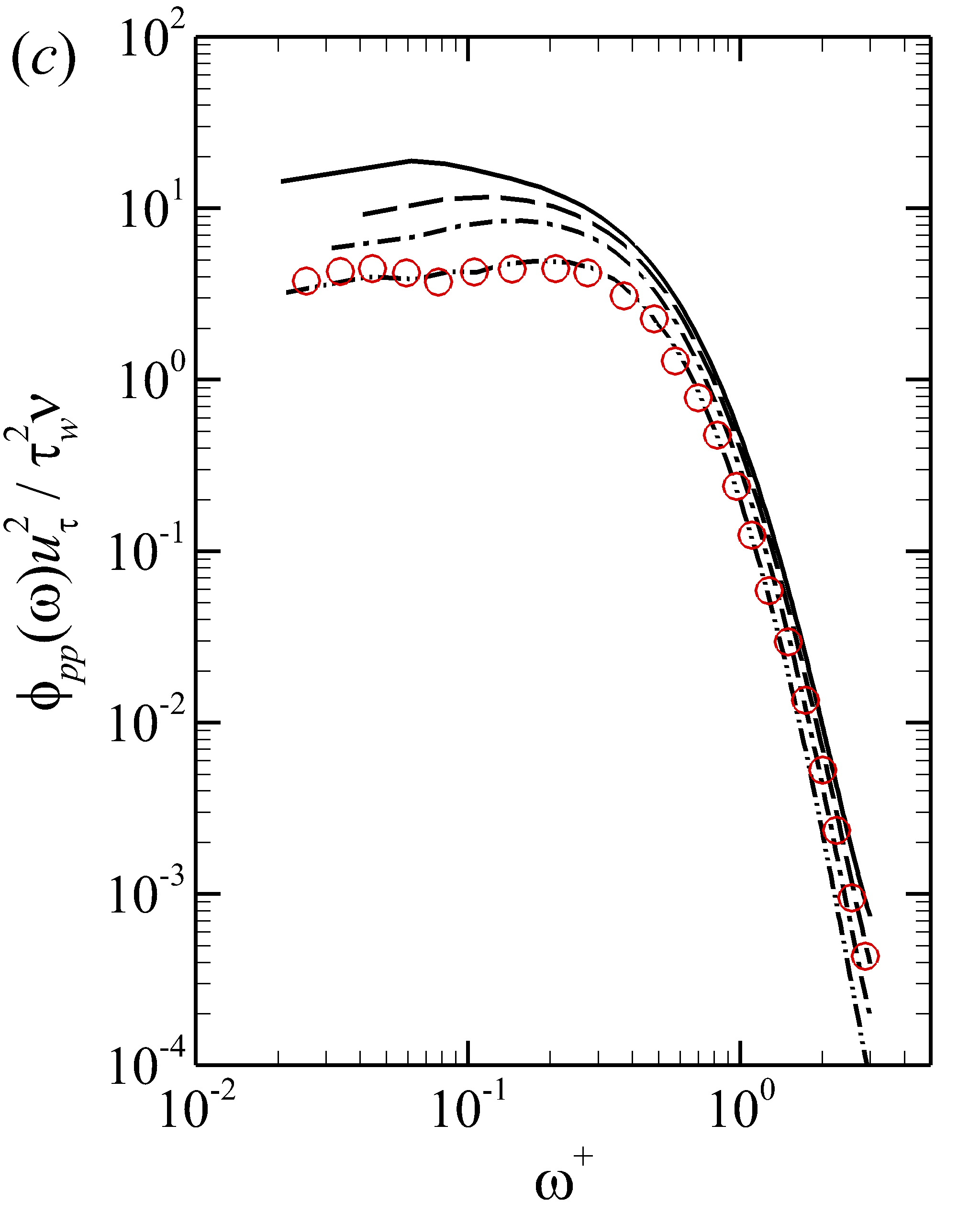}}}
    \caption{One-dimensional spectra of wall pressure fluctuations at $\Rey_\tau=179$, 333, 551 and 998 with respect to (a) streamwise-wavenumber, (b) spanwise-wavenumber and (c) frequency. All spectra are non-dimensionalized using wall units. The results of \citet{Choi90} at $\Rey_\tau=180$ are shown for comparison.}
    \label{fig:fig2}
\end{figure}

Figure~\ref{fig:fig2} compares the one-dimensional streamwise-wavenumber spectrum, spanwise-wavenumber spectrum and frequency spectrum of the wall pressure fluctuation at various Reynolds numbers. The spectra are non-dimensionalized using the wall units. To validate our results, the one-dimensional spectra of  wall pressure fluctuations at $\Rey_\tau=180$ calculated by~\citet{Choi90} are superposed. The present results for case CH180 are in good agreement with the results of~\citet{Choi90}. All of the one-dimensional spectra at different Reynolds numbers are close to each other at high wavenumbers and high frequencies, indicating that the spectra are scaled by the wall units~\citep{Hwang09,Hu06,Farabee91}. At small wavenumbers or frequencies, the magnitudes of all one-dimensional spectra increase monotonically with the Reynolds number, an observation that agrees with the DNS results of~\citet{Abe05}. Figures~\ref{fig:fig2}(a) and~\ref{fig:fig2}(c) show that as the streamwise wavenumber $k_x$ or frequency $\omega$ increases, the magnitude of the corresponding one-dimensional spectrum first increases, and then decreases. However, as depicted in figure~\ref{fig:fig2}(b), the magnitude of the one-dimensional spanwise-wavenumber spectrum $\phi_{pp}(k_z)$ decreases monotonically as $k_z$ increases. 

\subsection{Two-dimensional spectra}\label{subsec:total_two}

\begin{figure}
	\centering{\includegraphics[width=0.40\textwidth]{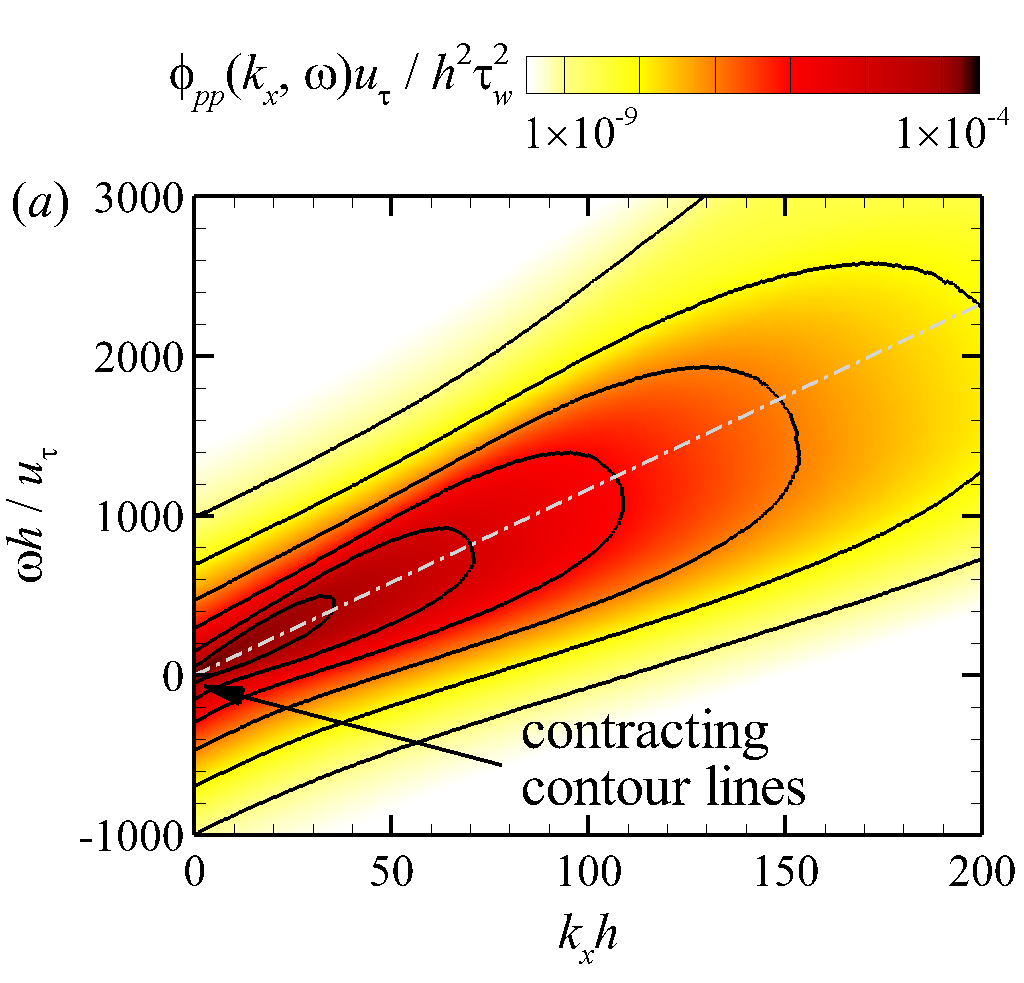}$\quad$
          	  {\includegraphics[width=0.48\textwidth]{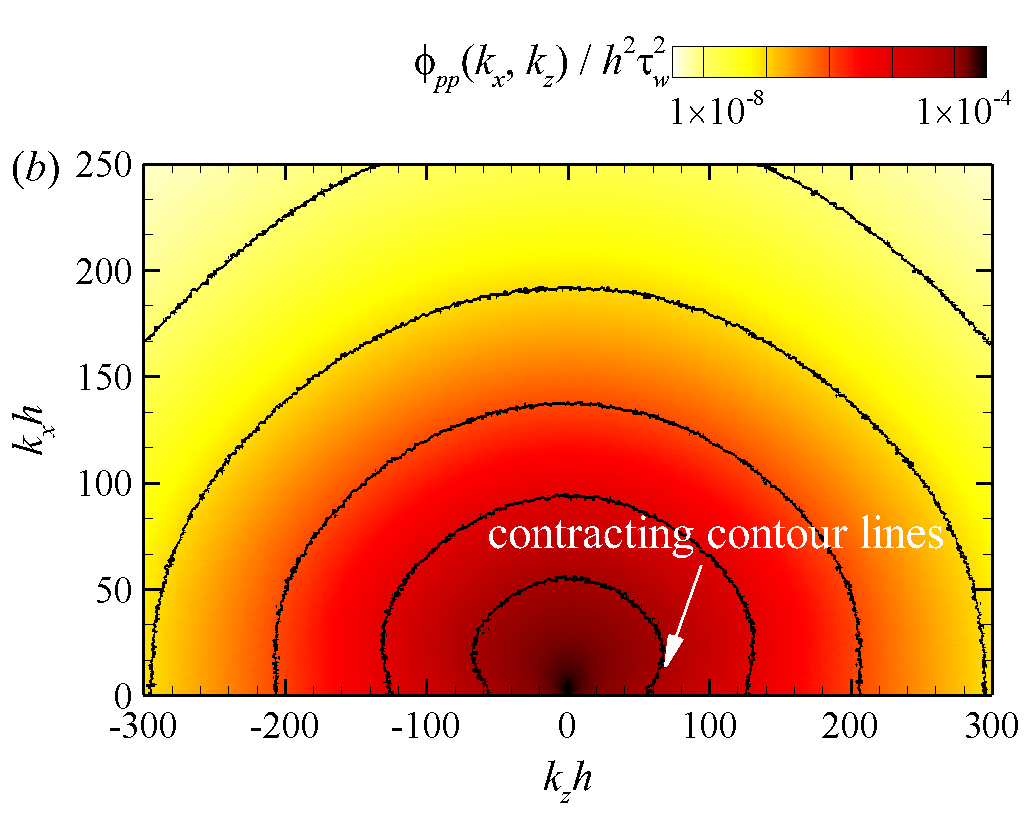}}
              {\includegraphics[width=0.40\textwidth]{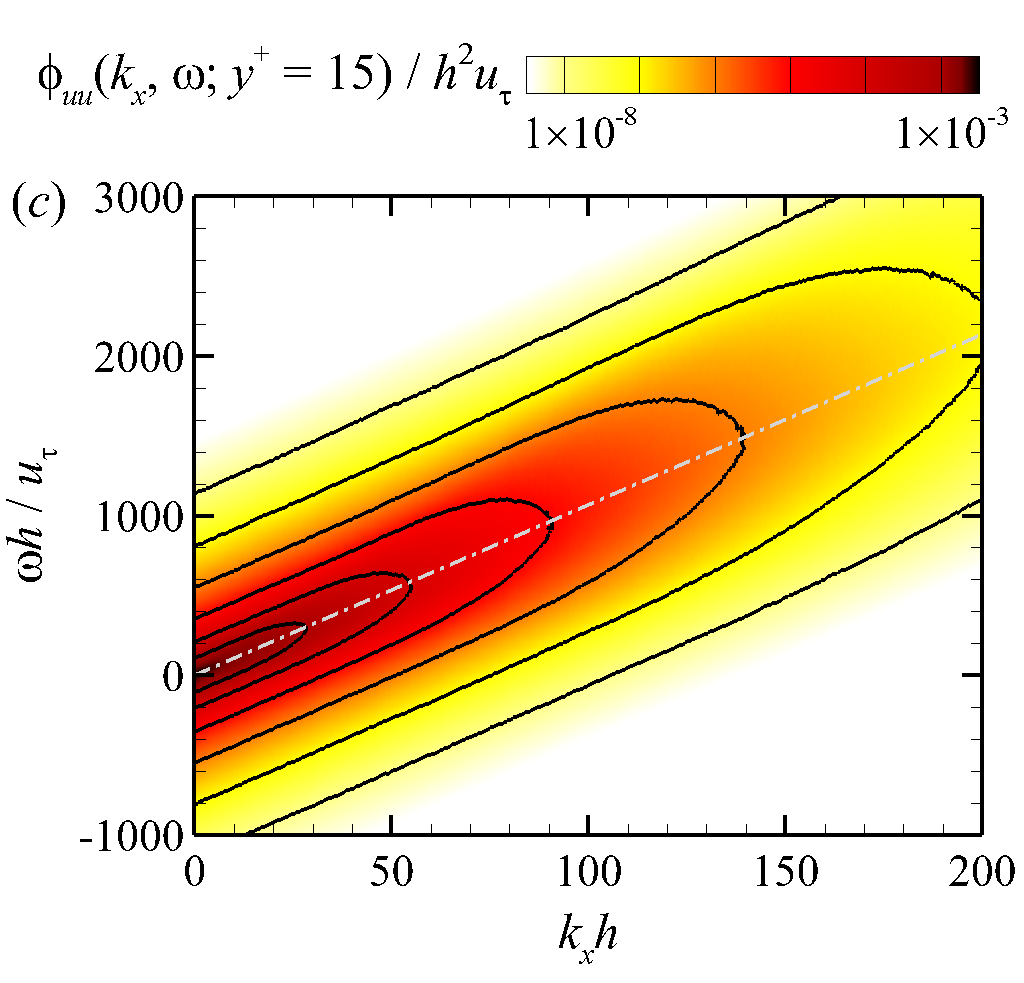}}$\quad$
              {\includegraphics[width=0.48\textwidth]{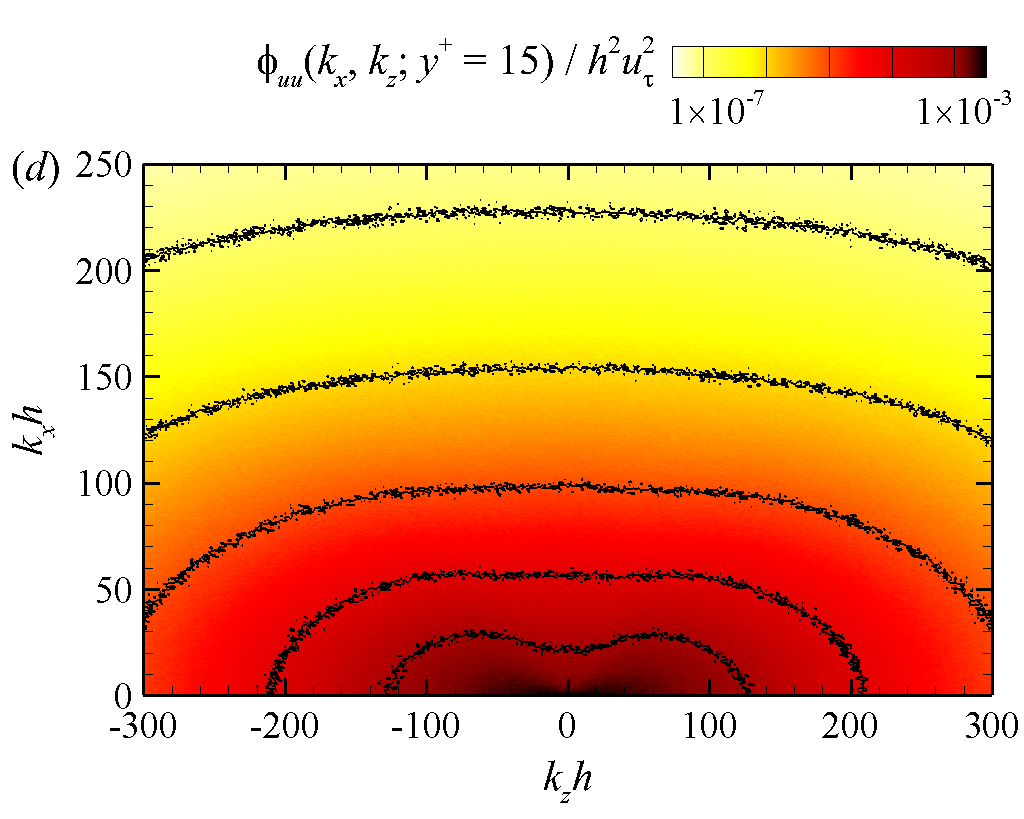}}}
    \caption{Isopleths of two-dimensional spectra of wall pressure fluctuations at $y^+ = 0$ and streamwise velocity fluctuations at $y^+=15$. (a) $k_x$-$\omega$ spectrum of wall pressure fluctuations $\phi_{pp}(k_x,\omega)$; (b) $k_x$-$k_z$ spectrum of wall pressure fluctuations $\phi_{pp}(k_x,k_z)$; (c) $k_x$-$\omega$ spectrum of streamwise velocity fluctuations $\phi_{uu}(k_x,\omega)$; and (d) $k_x$-$k_z$ spectrum of streamwise velocity fluctuations $\phi_{uu}(k_x,k_z)$. The Reynolds number is $\Rey_\tau=998$. }
    \label{fig:fig3}
\end{figure}

In this subsection, we analyze the two-dimensional spectra. To keep the paper concise, we focus on the results at $\Rey_\tau=998$. Figures~\ref{fig:fig3}(a) and~\ref{fig:fig3}(b) show the contours of the two-dimensional $k_x$--$\omega$ and $k_x$--$k_z$ spectra of the wall pressure fluctuation $\phi_{pp}$, respectively. For comparison, the contours of the $k_x$--$\omega$ and $k_x$--$k_z$ spectra of the streamwise velocity fluctuation $\phi_{uu}$ at $y^+=15$ are plotted in figures~\ref{fig:fig3}(c) and~\ref{fig:fig3}(d), respectively. We note that the $k_x$--$\omega$ spectra are symmetric about $(k_x,\omega) = (0,0)$ and the $k_x$--$k_z$ spectra are symmetric about both $k_x=0$ and $k_z=0$. Therefore, only spectra at $k_x\ge 0$ are plotted in figure~\ref{fig:fig3}. As shown in figures~\ref{fig:fig3}(a) and~\ref{fig:fig3}(c), large magnitudes of $\phi_{pp}(k_x,\omega)$ and $\phi_{uu}(k_x,\omega)$ occur around their convection lines $\omega/k_x=U_{bc}$, demarcated using the dash-dotted lines. Here, the bulk convection velocity $U_{bc}$ is defined as~\citep{Alamo09}
\begin{equation}
	{U_{bc}} = \frac{{\int {\int {{k_x}\omega \phi ({k_x},\omega )d\omega d{k_x}} } }}{{\int {\int {k_x^2\phi ({k_x},\omega )d\omega d{k_x}} } }},
	\label{eq:eq_bc}
\end{equation}
where $\phi(k_x,\omega)$ represents the $k_x$--$\omega$ spectrum of either wall pressure fluctuations or velocity fluctuations. The values of the bulk convection velocity for wall pressure fluctuations ($u_{bc}^{+}=11.6$) and streamwise velocity fluctuations at $y^+=15$ ($u_{bc}^{+}=10.7$) are close to each other, indicating that the convection property of  wall pressure fluctuations is similar to the velocity fluctuations in the buffer layer. This is consistent with the conclusion of~\citet{Chang99} that the velocity gradients in the buffer layer are the dominant sources of  wall pressure fluctuations.

Another important observation from figure~\ref{fig:fig3}(a) is that as $k_x$ approaches zero, the isopleths gradually contract towards $(k_x,\omega)=(0,0)$, especially at the energy-containing scales (see the two innermost isopleths of $\phi_{pp}(k_x,\omega)=10^{-4}$ and $10^{-5}$). This contracting feature is not observed from the isopleths of $\phi_{uu}(k_x,\omega)$, which are approximately parallel to the convection line at low streamwise wavenumbers. Similar contracting isopleths also appear in the $k_x$--$k_z$ spectrum of wall pressure fluctuations $\phi_{pp}(k_x,k_z)$ as shown in figure~\ref{fig:fig3}(b), but is absent for $\phi_{uu}(k_x,k_z)$ in figure~\ref{fig:fig3}(d). This contracting behavior suggests that as $k_x$ increases from zero, the spectrum of wall pressure fluctuations first increase in a small range before they decrease. Such a non-monotonic behavior is consistent with the one-dimensional spectrum $\phi_{pp}(k_x)$ plotted in figure~\ref{fig:fig2}(a). Later in \S\,\ref{subsec:two}, we will further show that this contracting behavior is correlated to the rapid pressure fluctuations.

\begin{figure}
	\centering{\includegraphics[width=0.48\textwidth]{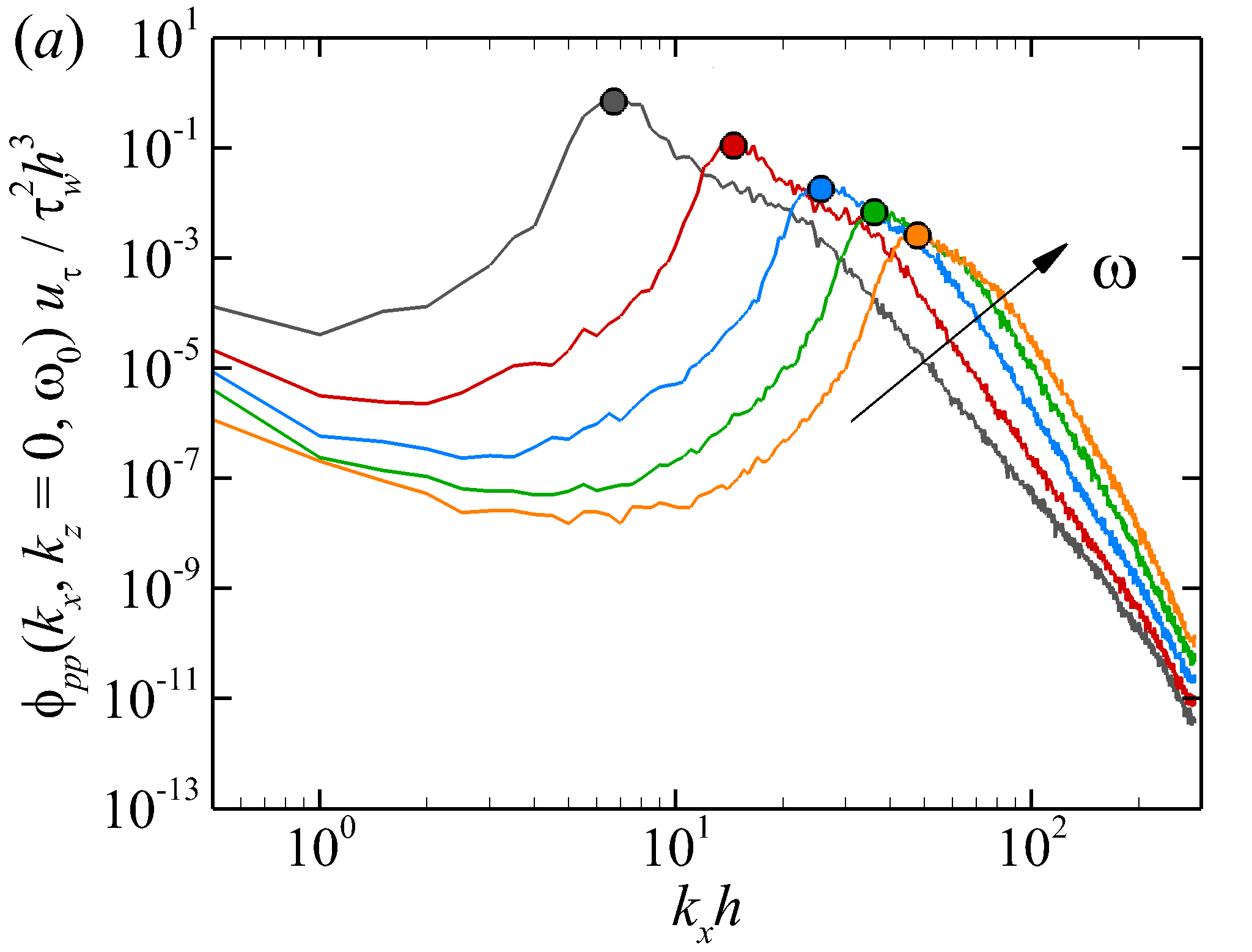}$\quad$
	          {\includegraphics[width=0.48\textwidth]{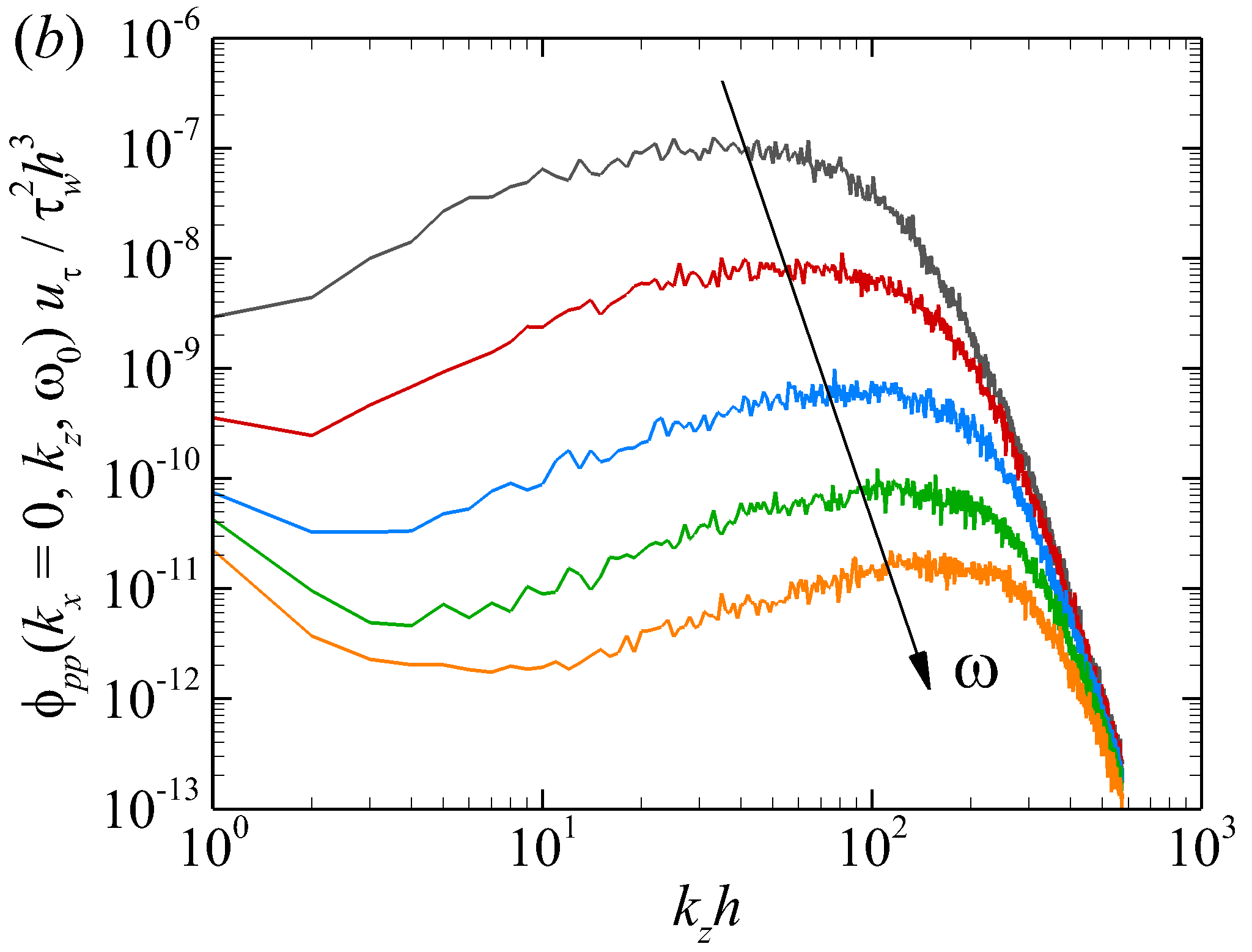}}}
    \caption{Three-dimensional wavenumber--frequency spectrum of wall pressure fluctuations at specific wavenumbers and frequencies: (a) $k_z=0$ and (b) $k_x=0$. The five lines in both panels (a) and (b) correspond to five frequencies $\omega h/u_\tau=123$, 245, 409, 573 and 716. The arrow points to the increasing direction of the frequency. The convective peaks are marked using filled circles in panel (a). The Reynolds number is $Re_\tau = 998$. }
    \label{fig:fig4}
\end{figure}

\subsection{The three-dimensional spectrum}\label{subsec:total_three}
In this section, we continue to analyze the three-dimensional spectrum $\phi_{pp}(k_x,k_z,\omega)$ of wall pressure fluctuations  at $Re_\tau=998$. The variation of $\phi_{pp}(k_x,k_z,\omega)$ with respect to $k_x$ corresponding to $k_z=0$ and five frequencies $\omega h/u_\tau=123$, 245, 409, 573 and 716 are plotted in figure~\ref{fig:fig4}(a), while figure~\ref{fig:fig4}(b) shows the variation of $\phi_{pp}(k_x,k_z,\omega)$ as a function of $k_z$ for $k_x=0$ and the same five frequencies. Convective peaks (denoted by the filled circles) can be identified in all of the five lines in figure~\ref{fig:fig4}(a). As the frequency $\omega$ increases, the location of the convective peak moves to a higher streamwise wavenumber. Focusing on each profile, as the streamwise wavenumber $k_x$ decreases from the convective wavenumber to zero, the value of the $\phi_{pp}(k_x,k_z,\omega)$ first decreases, and then increases slightly at low streamwise wavenumbers, forming a weak local maximum at the lowest resolved streamwise wavenumber. This local maximum is identified as an artificial acoustic mode by~\citet{Choi90}, which is a numerical artifact induced by the periodic boundary condition imposed in the streamwise direction. It is seen from figure~\ref{fig:fig4}(a) that for $\omega h/u_\tau=123$, the artificial acoustic mode only appears at the lowest resolved streamwise wavenumber $k_xh=0.5$. The range of the streamwise wavenumber influenced by the artificial acoustic mode becomes larger as the frequency increases, but is still confined in a small neighboring region around the lowest resolved streamwise wavenumber. Similarly, artificial acoustic modes also occur in the spanwise direction but are confined to small spanwise wavenumbers as shown in figure~\ref{fig:fig4}(b).

\begin{figure}
	\centering{\includegraphics[width=0.3\textwidth]{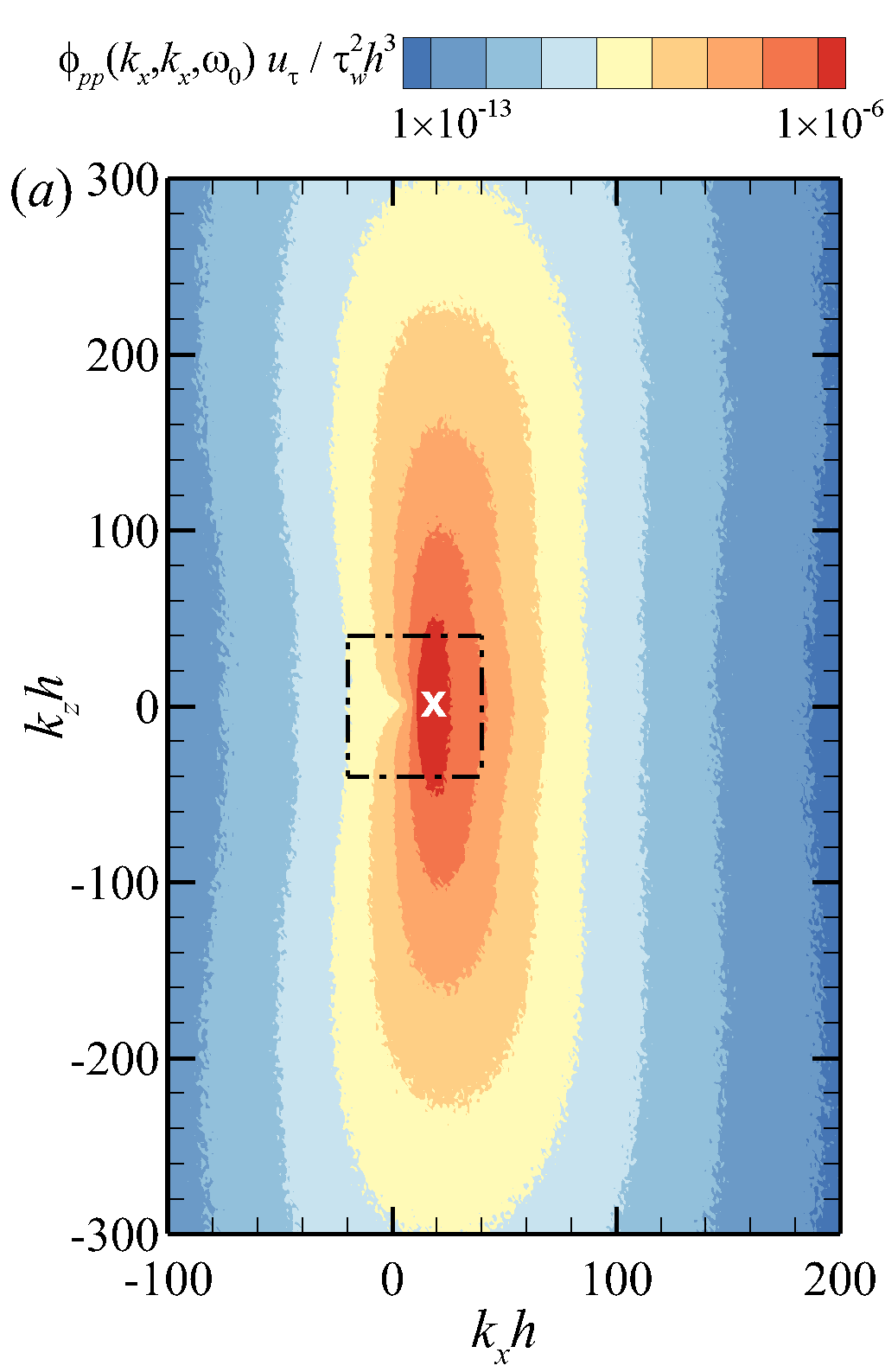}$\quad$
	          {\includegraphics[width=0.3\textwidth]{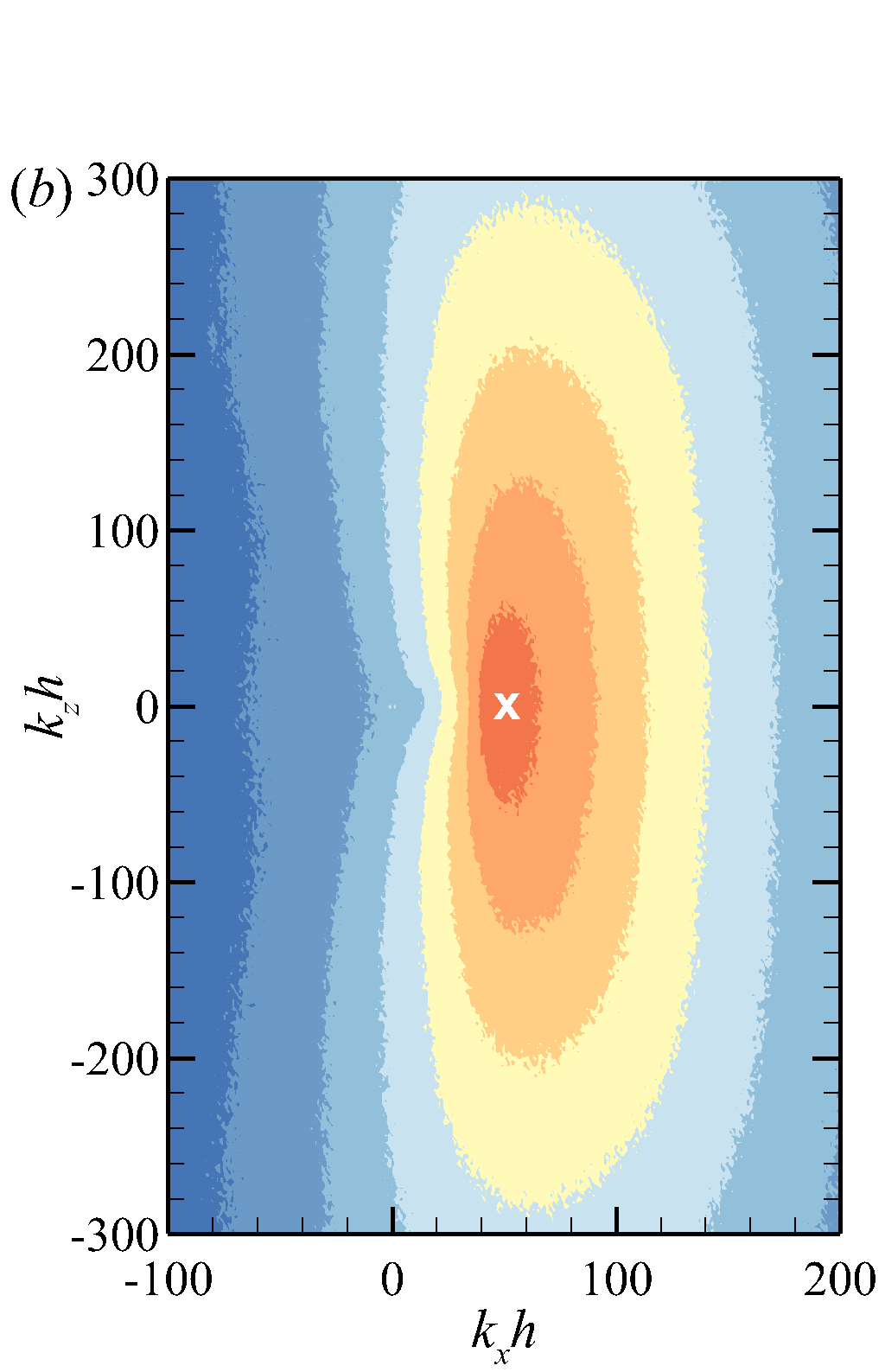}}$\quad$
	          {\includegraphics[width=0.3\textwidth]{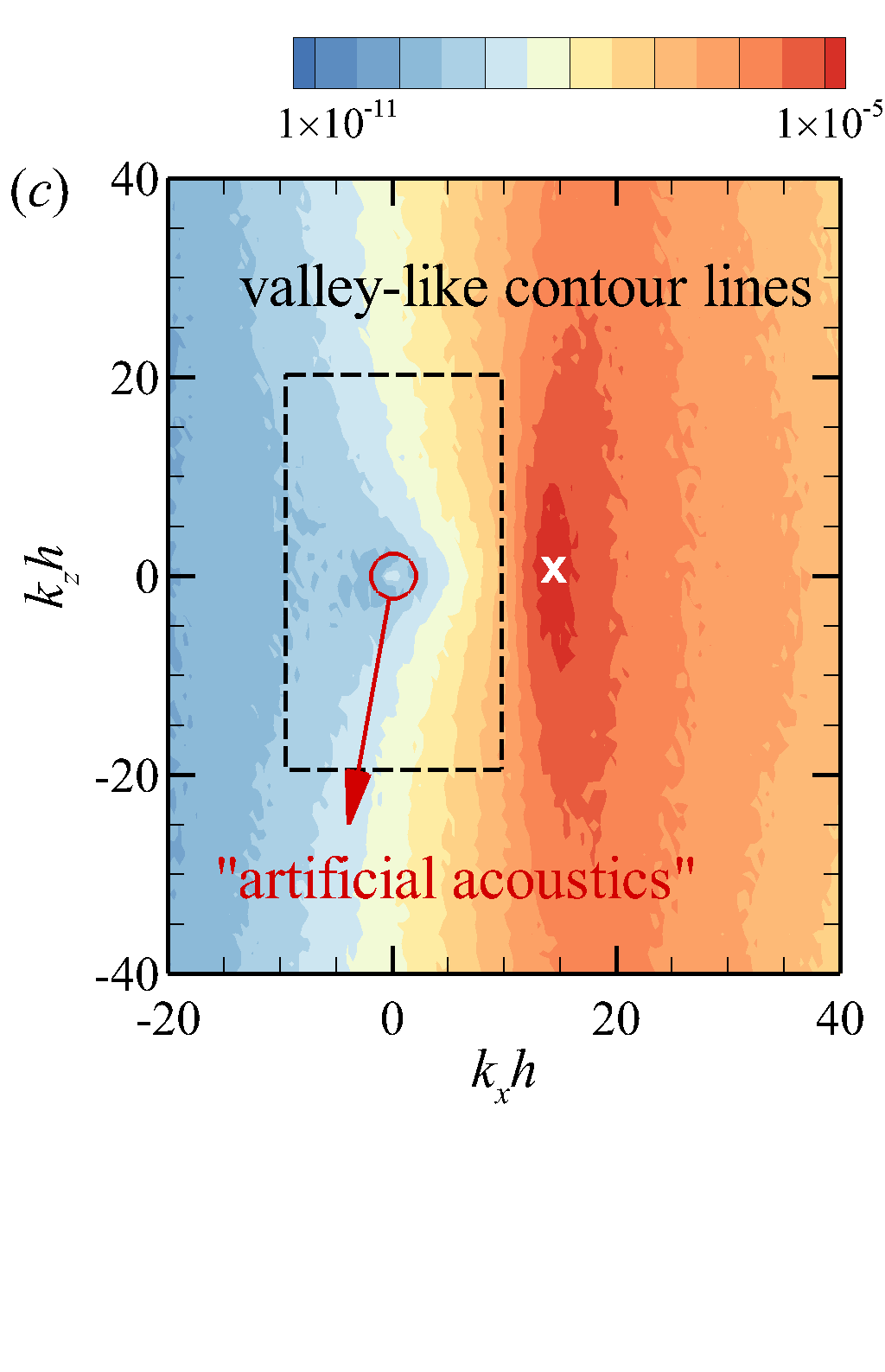}}}
    \caption{Contours of the three-dimensional wavenumber--frequency spectrum of wall pressure fluctuations $\phi_{pp}(k_x,k_z,\omega)$ in the $k_x$-$k_z$ plane for specific frequencies (a) $\omega h/u_\tau=245$ and (b) $\omega h/u_\tau=716$. Panel (c) shows a zoom-in view of the dash-dotted box region around $(k_x,k_z)=(0,0)$ in panel (a). The convective peaks are marked by the cross symbols. In panel(c), the black dashed box denotes the area in which the isopleths form a valley, and the acoustic peak locates in the red circle. Note that in both panels (a) and (b), the isopleth level ranges from $10^{-13}$ to $10^{-6}$, while in panel (c), it ranges from $10^{-11}$ to $10^{-5}$. The Reynolds number is $Re_\tau = 998$.}
    \label{fig:fig5}
\end{figure}

Figures~\ref{fig:fig5}(a) and~\ref{fig:fig5}(b) show the coutours of $\phi_{pp}(k_x, k_z, \omega)$ corresponding to $\omega h/u_\tau=245$ and 716, respectively. Convective peaks can be identified for both frequencies (denoted by the cross symbols in figure~\ref{fig:fig5}), and the spectral value decays as the wavenumber moves away from the convective peaks. The isopleths are closed, ellipse-like curves, except for a small region near $(k_x,k_z)=(0,0)$. Owing to the artificial acoustic modes, the value of $\phi_{pp}(k_x, k_z, \omega)$ obtained from DNS does not decay to zero at $(k_x,k_z)=(0,0)$. To facilitate a clearer observation near $(k_x,k_z)=(0,0)$, figure~\ref{fig:fig5}(c) displays a zoom-in view of the dash-dotted box in figure~\ref{fig:fig5}(a).   It is seen that the artificial acoustic modes only influence a small region around $(k_x,k_z)=(0,0)$ (marked by the red circle in figure~\ref{fig:fig5}c).  Outside this circle, 
the isopleths form a “valley” between $(k_x,k_z)=(0,0)$ and the convective peak.  This observation shows the tendency that the magnitude of $\phi_{pp}(\boldsymbol{k},\omega)$ decays as the spanwise wavenumber approaches $k_z=0$ in the $k_x$-$k_z$ plane, which is in agreement with figure~\ref{fig:fig4}(b). 


\section{Time decorrelation mechanisms of wall pressure fluctuations}\label{sec:decorrelation}
In previous studies of velocity fluctuations, the investigation of the time decorrelation mechanisms is found to be useful for developing predictive models of wavenumber--frequency ($k_x$-$\omega$)  spectra~\citep{Wilczek12,Wilczek15,He17}.  However, the time decorrelation mechanisms of wall pressure fluctuations are rarely discussed in literature. In this section, we examine if the frequency variation of the $k_x$-$\omega$ spectrum of wall pressure fluctuations follows the same functional form as that of  velocity fluctuations.

As shown in figure~\ref{fig:fig3}(a), similar to the velocity fluctuations, the energy-containing region in the $k_x$-$\omega$ spectrum of wall pressure fluctuations is located around the convection line. This means that the convection by the mean flow is one of the main time decorrelation mechanisms of wall pressure fluctuations. On the other hand, the wavenumber--frequency spectrum is not concentrated exactly upon the convection line as predicted by the Taylor's hypothesis~\citep{Taylor38}. Instead, the energy-containing part distributes around the convection line, forming a finite bandwidth due to the Doppler broadening effect~\citep{Wu17,Wu20}. For velocity fluctuations, the Doppler broadening is caused by the large-scale sweeping effect~\citep{Kraichnan64,Tennekes75}. To facilitate the investigation of the time decorrelation mechanisms of wall pressure fluctuations, we briefly review the linear random sweeping model of the wavenumber--frequency spectrum of streamwise velocity fluctuations proposed by~\citet{Wilczek15}.

Considering an arbitrary variable $\psi$ \citep[which is the streamwise velocity fluctuations $u$ in][]{Wilczek15}, in a turbulent channel flow, it is advected in the wall-parallel directions by both mean velocity $\boldsymbol{U}=(U,0)$ and large-scale sweeping velocity $\boldsymbol{v} = (v_x, v_z)$. The sweeping velocity is assumed to be a constant in space and time with a Gaussian ensemble~\citep{He17,Kraichnan64,Yao08}. The spatial Fourier mode of $\psi$ follows a random advection equation
\begin{equation}
\frac{\partial }{{\partial t}}\hat \psi (\boldsymbol{k},t) + \rm{i}(\boldsymbol{U} + \boldsymbol{v}) \cdot \boldsymbol{k}\hat \psi (\boldsymbol{k},t) = 0.
\label{eq:advection}
\end{equation}
The wavenumber--frequency spectrum of $\psi$ can be subsequently derived from the solution of equation~(\ref{eq:advection}) as~\citep{Wilczek15}
\begin{equation}
\phi_{\psi\psi} (\boldsymbol{k},\omega ) = \phi_{\psi\psi} (\boldsymbol{k}) \cdot \frac{1}{{\sqrt {2\pi (\left\langle {v_x^2} \right\rangle k_x^2 + \left\langle {v_z^2} \right\rangle k_z^2)} }}\exp \left[ { - \frac{{{{(\omega  - {k_x}U)}^2}}}{{2(\left\langle {v_x^2} \right\rangle k_x^2 + \left\langle {v_z^2} \right\rangle k_z^2)}}} \right].
\label{eq:spectra}
\end{equation}
In equation~(\ref{eq:spectra}), the wavenumber--frequency spectrum is estimated using the multiplication between the wavenumber spectrum $\phi_{\psi\psi} (\boldsymbol{k})$ and a Gaussian function of the frequency. The numerical results in \citet{Wilczek15} showed that equation~(\ref{eq:spectra}) accurately predicts the wavenumber--frequency spectrum of velocity fluctuations $\phi_{uu}$ in turbulent channel flows, especially at the energy-containing $k_x$--$\omega$ combinations along the convection line. This indicates that sweeping is also a main physical mechanism for the time decorrelation of velocity fluctuations in wall-bounded turbulence. In the following content of this section, we examine if the wavenumber--frequency spectrum of wall pressure fluctuations $\phi_{pp}(k_x,k_z,\omega)$ follows the same functional form as $\phi_{uu}(k_x,k_z,\omega)$.

\begin{figure}
	\centering{\includegraphics[width=0.48\textwidth]{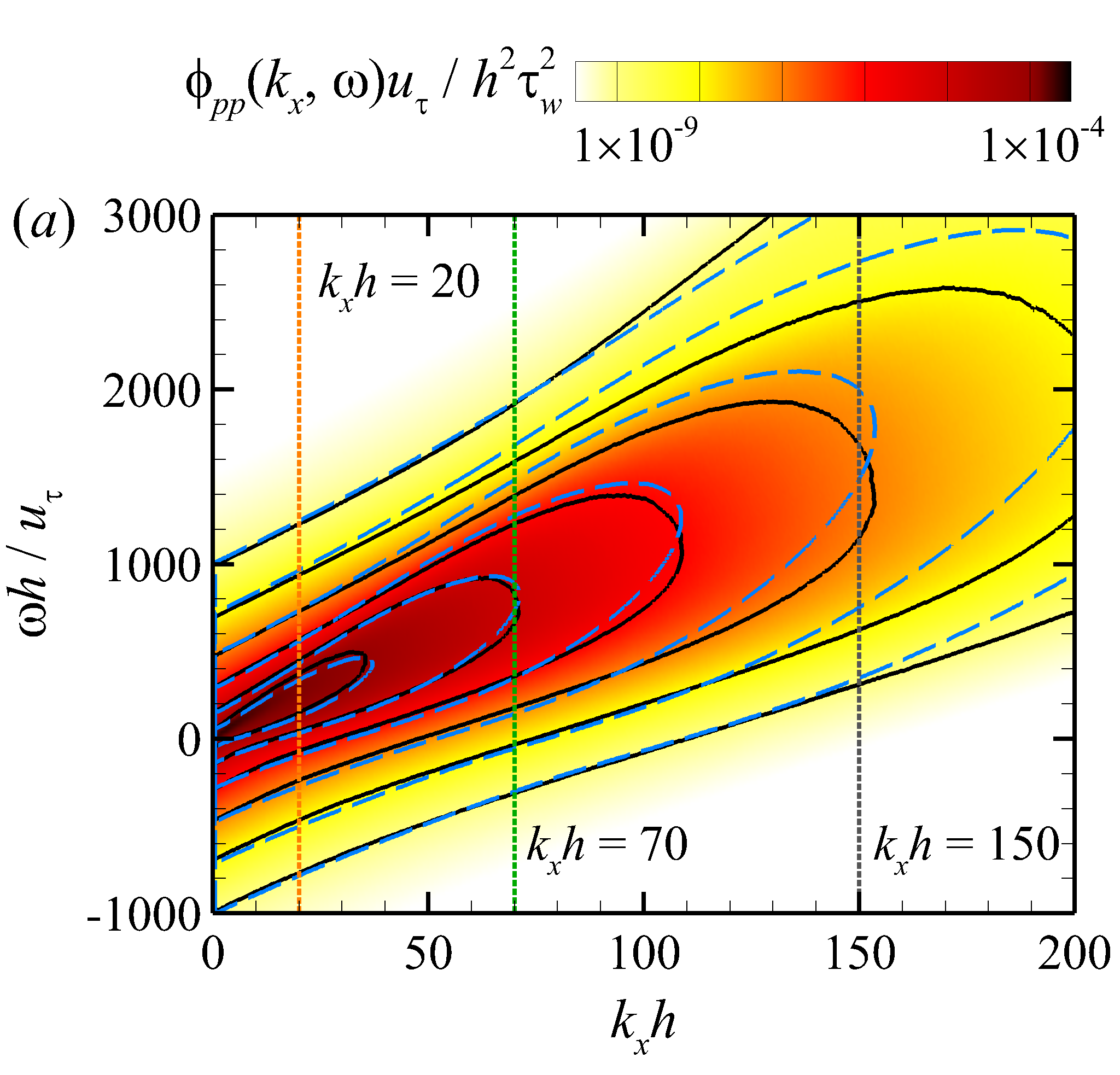}$\quad$
	          {\includegraphics[width=0.48\textwidth]{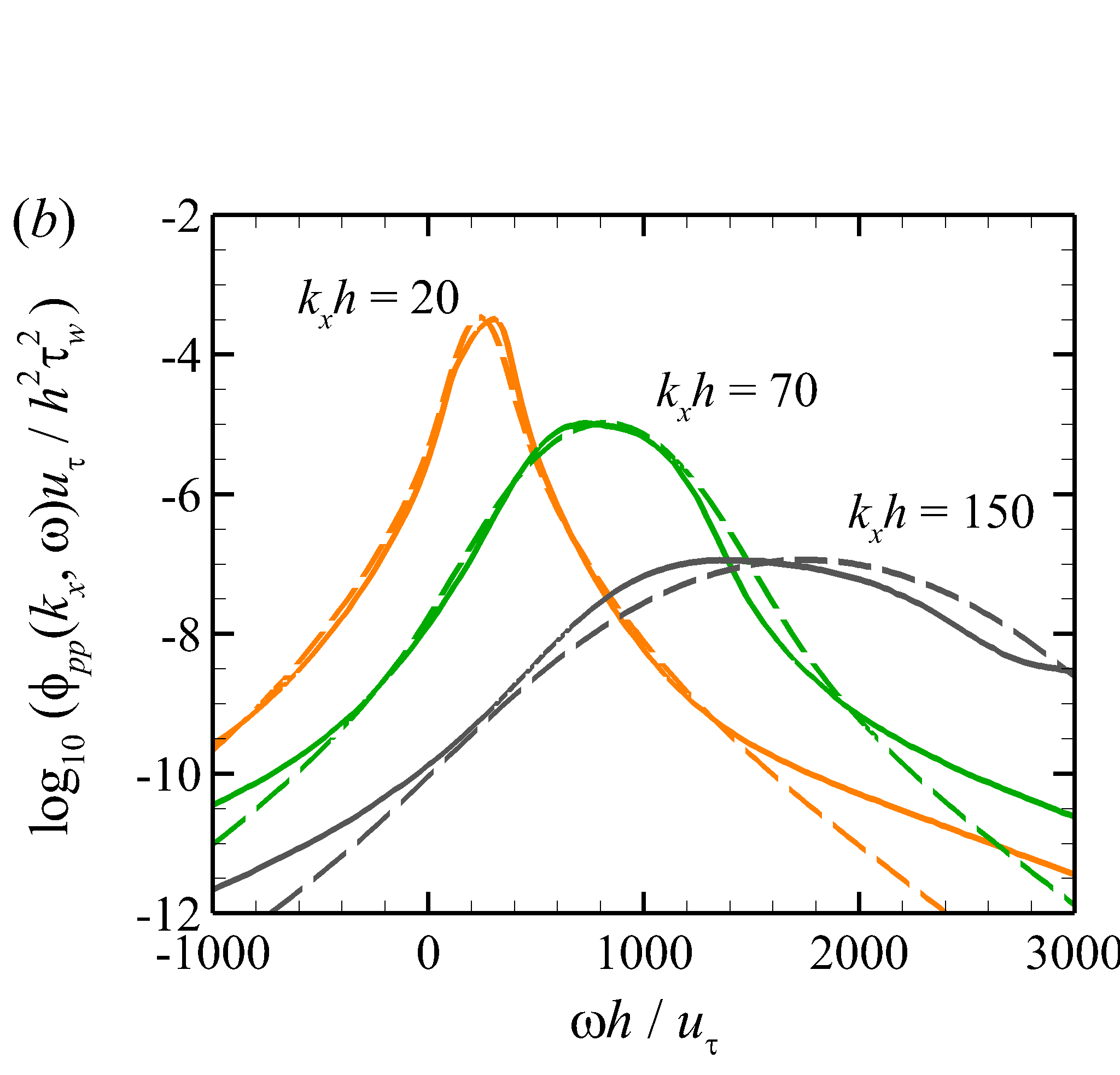}}}
    \caption{ The wavenumber--frequency spectrum of  wall pressure fluctuations $\phi_{pp}(k_x,\omega)$ obtained from DNS and from model~(\ref{eq:spectra}). (a) Contours in the $k_x$-$\omega$ plane. (b) Variation of $\phi_{pp}(k_x,\omega)$ with respect to the dimensionless frequency $\omega h / u_\tau$ for specific streamwise wavenumbers $k_x h = 20$, 70, and 150.  In both panels (a) and (b), the solid and dashed lines represents DNS and model results, respectively. The Reynolds number is $Re_\tau = 998$.}
    \label{fig:fig6}
\end{figure}

Figure~\ref{fig:fig6}(a) compares the isopleths of $\phi_{pp}(k_x,\omega)$ obtained from DNS and  equation~(\ref{eq:spectra}). Here the convection velocity $U$ in equation~(\ref{eq:spectra}) is chosen as the bulk convection velocity $U^+ = 11.6$. ~\citet{Wilczek15} found that the RMS sweeping velocities of the spectrum of streamwise velocity fluctuations can be well approximated by the local RMS velocities. Because the velocity sources in the buffer layer contribute most to wall pressure fluctuations~\citep{Chang99}, we choose the RMS velocities at $y^+=15$ as the RMS sweeping velocities in model~(\ref{eq:spectra}). As such, all the three velocities in this model are chosen as quantities with physical meanings related to wall pressure fluctuations, and no parameter needs to be adjusted in the following comparison. The model~(\ref{eq:spectra}) is in good agreement with the DNS result in the energy-containing spatial--temporal scales. Specifically, the isopleths of $10^{-4}$ and $10^{-5}$ obtained from the model are almost coincident with the DNS result. The contracting feature at low streamwise wavenumbers is also well captured by the model. As is shown in \S\,\ref{subsec:two}, the contracting feature is induced by the scaling of the rapid pressure at low streamwise wavenumbers. Because this scaling behavior is a spatial property embedded in the wavenumber spectrum $\phi_{pp}(\boldsymbol{k})$ as a model input, it is also satisfied by the model of the wavenumber--frequency spectrum $\phi_{pp}(\boldsymbol{k}, \omega)$ given by equation~(\ref{eq:spectra}). Figure~\ref{fig:fig6}(b) further shows three slices of $\phi_{pp}(k_x,\omega)$ at $k_x h = 20$, 70 and 150. For the two lower wavenumbers $k_xh = 20$ and 70, the frequency distributions of $\phi_{pp}(k_x,\omega)$ obtained from the model are close to the DNS results around the convective peaks. For the higher wavenumber $k_xh=150$, the convection peak predicted by the model shifts to a higher frequency in comparison with the DNS result. This indicates that the convection velocity is scale dependent, which is further investigated in \S\,\ref{subsec:two}. We have tested that, if a scale-dependent convection velocity $U(k_x)$ is used in model~(\ref{eq:spectra}), then the peak locations predicted by the model also match the DNS results at high wavenumbers. Since the present model mainly focuses on the energy-containing scales, we choose a constant convection velocity in the model to keep it simple. Furthermore, the model spectrum decreases faster than the DNS result at low and high wavenumbers away from the convective peak. This is because the model assumes a Gaussian decay at all frequencies, while the DNS spectrum indeed shows an approximately exponential decay at high frequencies.

\begin{figure}
	\centering{\includegraphics[width=0.48\textwidth]{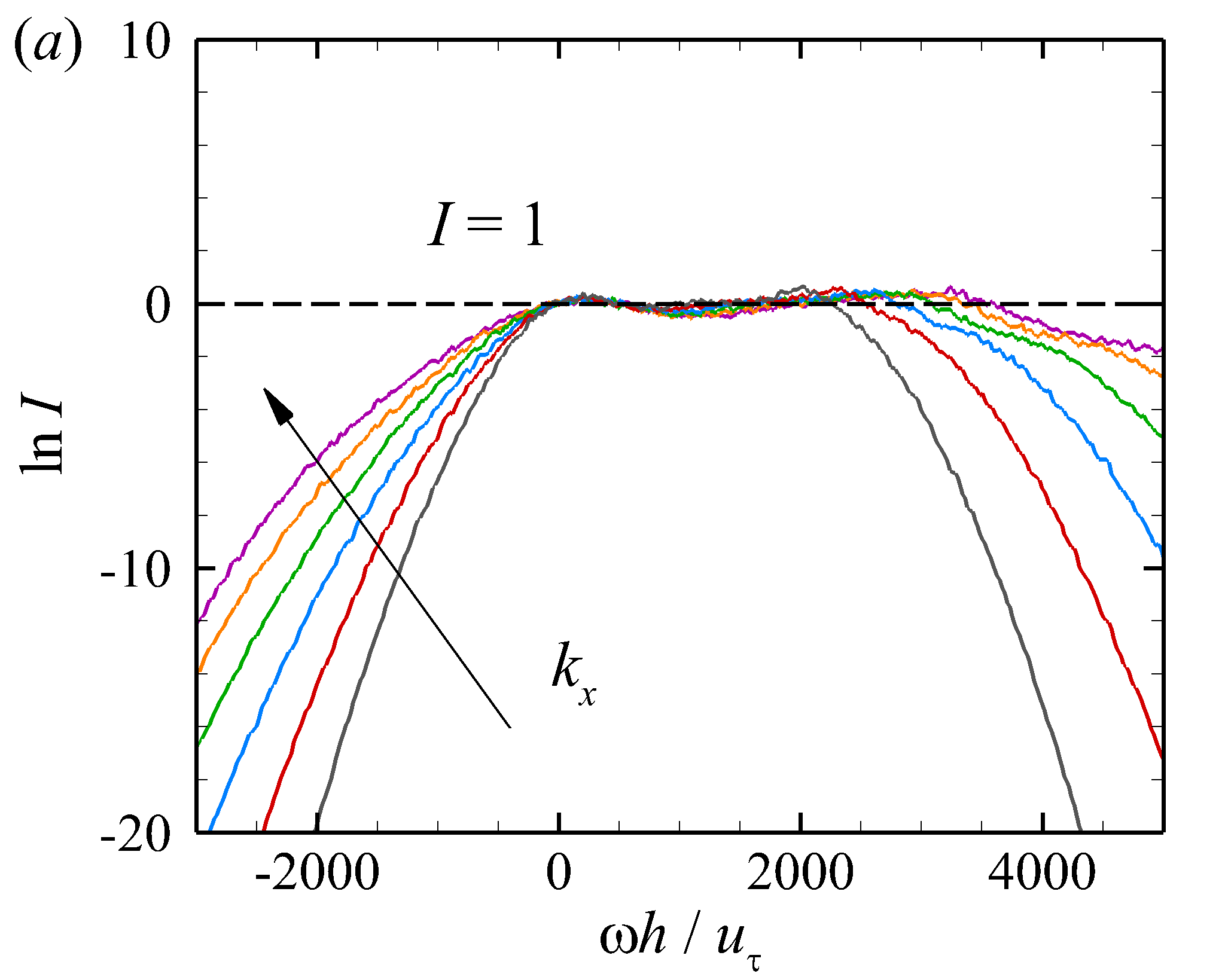}$\quad$
	          {\includegraphics[width=0.48\textwidth]{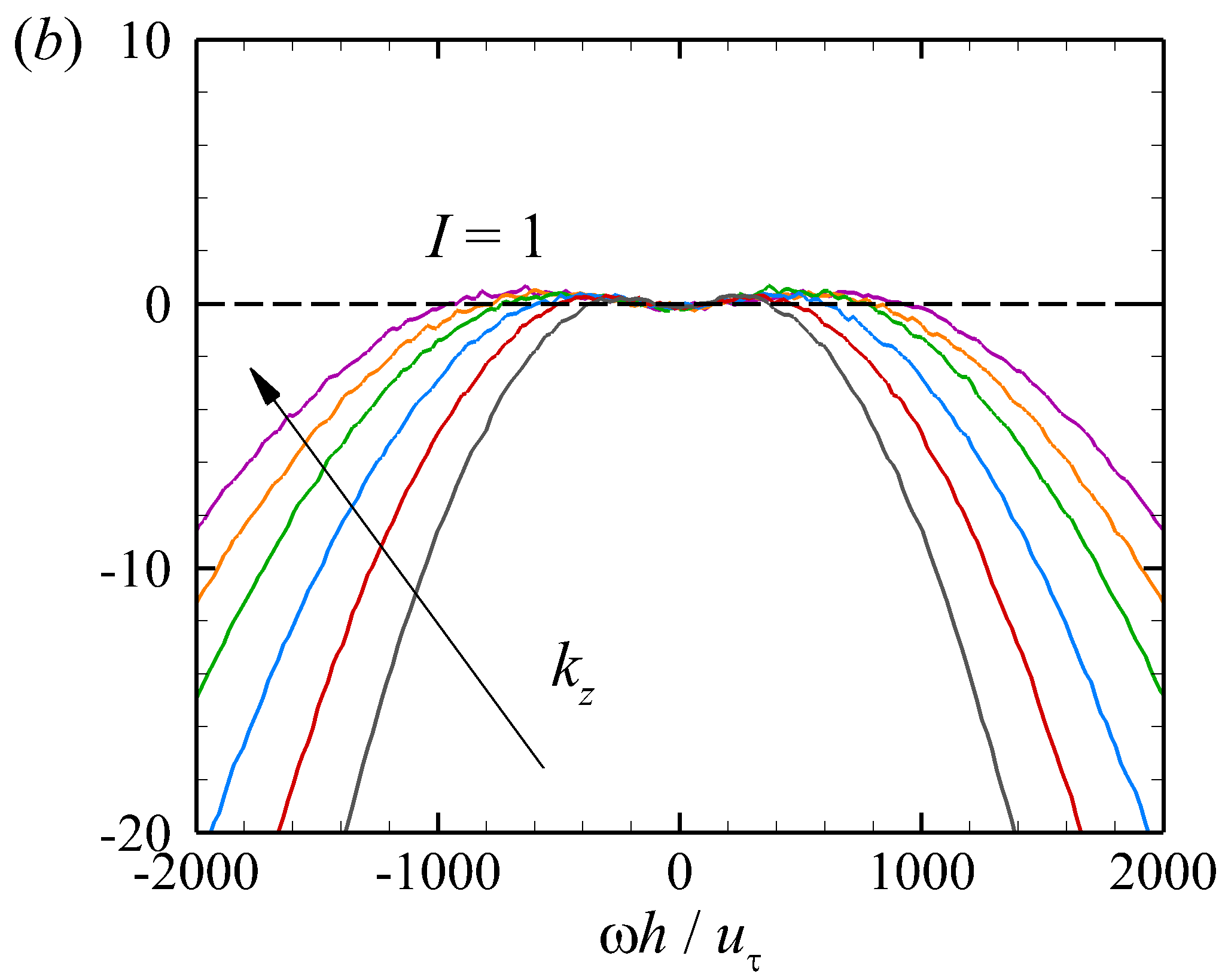}}}
    \caption{Variation of the indicator function $I$ for the spectrum of wall pressure fluctuations obtained from the random sweeping model with respect to the dimensionless frequency $\omega h / u_\tau$ for specific wavenumbers: (a) $k_z=0$ and $k_x = 100$, 120, 140, 160, 180 and 200; (b) $k_x=0$ and $k_z=100$, 120, 140, 160, 180 and 200. The arrow points to the increasing direction of the streamwise or spanwise wavenumber. The horizontal dashed lines denote $I=1$. The Reynolds number is $Re_\tau = 998$.}
    \label{fig:fig7}
\end{figure}

To further show the capability of equation~(\ref{eq:spectra}) in the prediction of the energy-containing part of wavenumber-frequency spectrum of wall pressure fluctuations, we define the following indicator function $I$ as the ratio between the spectra obtained from the model and DNS
\begin{equation}
{I} = \frac{\phi_{pp,model}(\boldsymbol{k},\omega)}{\phi_{pp,DNS}(\boldsymbol{k},\omega)}.
\label{eq:indicator}
\end{equation}
If the wavenumber--frequency spectrum of wall pressure fluctuations follows the functional form of equation~(\ref{eq:spectra}), then the value of $I$ should equal to one. Figures~\ref{fig:fig7}(a) and~\ref{fig:fig7}(b) depict the variation of $I$ with respect to the frequency $\omega$ for $k_z=0$ and $k_x=0$, respectively. It is observed that plateaux with $I = 1$ exist in the energy-containing scales in both figures~\ref{fig:fig7}(a) and~\ref{fig:fig7}(b). This indicates that the wavenumber--frequency spectrum of wall pressure fluctuations follows the random sweeping model~(\ref{eq:spectra}) at energy-containing scales. Similar to the velocity fluctuations, the Doppler broadening effect on the wavenumber--frequency spectrum of wall pressure fluctuations is also induced by the large-scale random sweeping. Noticing that the wall pressure fluctuations can be expressed as an integration of the source terms weighted by the Green's function~\citep{Kim89}, the time decorrelation properties of  wall pressure fluctuations can be regarded as the weighted average properties of the velocity source terms. Thus, it is reasonable that the wavenumber--frequency spectrum of wall pressure fluctuations has the similar functional form as velocity fluctuations.

\section{Decomposition of the wavenumber--frequency spectrum of wall pressure fluctuations}\label{sec:decomposition}

\subsection{Mathematical description of the decomposition}\label{subsec:math}
According to equation~(\ref{eq:eq1.1}), the pressure fluctuations can be decomposed into two components as
\begin{equation}
p = p_r + p_s.
\label{eq:eq4.1}
\end{equation}
The rapid pressure fluctuations $p_r$ and slow pressure fluctuations $p_s$ are governed by the following Poisson equations
\begin{equation}
\left. \begin{array}{l}
\displaystyle
{{\nabla ^2}{p_r} = {f_r} =  - 2\rho\frac{{\partial {u_i}}}{{\partial {x_j}}}\frac{{\partial {U_j}}}{{\partial {x_i}}}}, \\[16pt]
\displaystyle
{{\nabla ^2}{p_s} = {f_s} =  - \rho\frac{{{\partial ^2}}}{{\partial {x_i}\partial {x_j}}}({u_i}{u_j} - \overline {{u_i}{u_j}} )},
\end{array} \right\}
\label{eq:eq4.2}
\end{equation}
respectively, where $f_r$ and $f_s$ denote the corresponding source terms. Because the magnitude of the Stokes pressure induced by the inhomogeneous boundary condition becomes increasingly small as the Reynolds number increases~\citep{Gerolymos13}, we adopt the homogeneous Neumann boundary conditions ${\left. {\partial {p_r}/\partial y} \right|_{y =  \pm h}} = {\left. {\partial {p_s}/\partial y} \right|_{y =  \pm h}} = 0$ for both rapid and slow pressure fluctuations following~\citet{Anantharamu20}. Performing Fourier transform over equation~(\ref{eq:eq4.2}) results in the following Helmholtz equations
\begin{equation}
\left. \begin{array}{l}
\displaystyle
{\frac{{\partial {{\hat p}_r}(\boldsymbol{k},\omega ;y)}}{{\partial {y^2}}} - {k^2}{{\hat p}_r}(\boldsymbol{k},\omega ;y) = {{\hat f}_r}(\boldsymbol{k},\omega ;y)},\quad{\partial {{\hat p}_r}/\partial y{{\rm{|}}_{y =  \pm h}} = 0}, \\[16pt]
\displaystyle
{\frac{{\partial {{\hat p}_s}(\boldsymbol{k},\omega ;y)}}{{\partial {y^2}}} - {k^2}{{\hat p}_s}(\boldsymbol{k},\omega ;y) = {{\hat f}_s}(\boldsymbol{k},\omega ;y)},\quad{\partial {{\hat p}_s}/\partial y{{\rm{|}}_{y =  \pm h}} = 0},
\end{array} \right\}
\label{eq:eq4.3}
\end{equation}
where ${\hat p_r}$, ${\hat p_s}$, ${\hat f_r}$ and ${\hat f_s}$ denote the Fourier modes of ${p_r}$, ${p_s}$, ${f_r}$ and ${f_s}$, respectively. The solution of equation~(\ref{eq:eq4.3}) can be expressed as an integration of the source term $\hat f_r$ weighted by a Green's function as~\citep{Kim89}
\begin{equation}
\left. \begin{array}{l}
\displaystyle
	{\hat p_r}(\boldsymbol{k},\omega ;y) = \int_{ - h}^h {G(y,y',k){{\hat f}_r}(\boldsymbol{k},\omega ;y')dy'},\\[16pt]
\displaystyle
{\hat p_s}(\boldsymbol{k},\omega ;y) = \int_{ - h}^h {G(y,y',k){{\hat f}_s}(\boldsymbol{k},\omega ;y')dy'},
\end{array} \right\}
	\label{eq:eq4.8}
\end{equation}
where the Green's function $G(y,y',k)$ is given as
\begin{equation}
G(y,y',k) = \left\{ {\begin{array}{*{20}{c}}
{\frac{{\cosh (k(y' - h))\cosh (k(y + h))}}{{2k\sinh (kh)\cosh (kh)}},{\rm{   }}y \le y'},\\
{\frac{{\cosh (k(y' + h))\cosh (k(y - h))}}{{2k\sinh (kh)\cosh (kh)}},{\rm{   }}y > y'}.
\end{array}} \right. 
	\label{eq:eq4.9}
\end{equation}
In this paper, we focus on the wall pressure fluctuations, and as such the wall-normal coordinate is fixed to $y=\pm h$ without specific declarations.

From equations~(\ref{eq:eq4.1}) and~(\ref{eq:eq4.3}), it is known that the Fourier modes of wall pressure fluctuations can be also decomposed into rapid and slow components as
\begin{equation}
	\hat p(\boldsymbol{k},\omega ) = {\hat p_r}(\boldsymbol{k},\omega ) + {\hat p_s}(\boldsymbol{k},\omega ).
	\label{eq:eq4.4}
\end{equation}
Substituting~(\ref{eq:eq4.4}) into~(\ref{eq:eq2.2}) yields the following decomposition of the wavenumber--frequency spectrum of wall pressure fluctuations
\begin{equation}
\begin{aligned}
	{\phi _{pp}}(\boldsymbol{k},\omega ) &= {\phi _{rr}}(\boldsymbol{k},\omega ) + {\phi _{ss}}(\boldsymbol{k},\omega ) + 2{\phi _{rs}}(\boldsymbol{k},\omega )\\
	&=\underbrace {\frac{{\overline {{{\hat p}_r}(\boldsymbol{k},\omega )\hat p_r^*(\boldsymbol{k},\omega )} }}{{\Delta {k_x}\Delta {k_z}\Delta \omega }}}_{{\rm{AS}} - {\rm{Rapid}}} + \underbrace {\frac{{\overline {{{\hat p}_s}(\boldsymbol{k},\omega )\hat p_s^*(\boldsymbol{k},\omega )} }}{{\Delta {k_x}\Delta {k_z}\Delta \omega }}}_{{\rm{AS}} - {\rm{Slow}}} + \underbrace {2\frac{{{\mathop{\rm Re}\nolimits} \left( {\overline {{{\hat p}_r}(\boldsymbol{k},\omega )\hat p_s^*(\boldsymbol{k},\omega )} } \right)}}{{\Delta {k_x}\Delta {k_z}\Delta \omega }}}_{{\rm{CS-RS}}},
	\label{eq:eq4.5}
\end{aligned}
\end{equation}
where $\phi_{rr}(\boldsymbol{k},\omega)$, $\phi_{ss}(\boldsymbol{k},\omega)$ and $\phi_{rs}(\boldsymbol{k},\omega)$ represent AS-Rapid, AS-Slow and CS-RS, respectively, and ${\mathop{\rm Re}\nolimits} \left(  \cdot  \right)$ denotes the real part of a complex number.

In early research of wall pressure fluctuations, \citet{Kraichnan56} proposes a hypothesis that the rapid component of  wall pressure fluctuations dominates over the slow component. This hypothesis is supported by the analytical studies of~\citet{Hodgson62} and~\citet{Chase80}, and the experiments of~\citet{Johansson87}. Various predictive models of the spectrum of wall pressure fluctuations are proposed based on Kraichnan's hypothesis~\citep[see for example][]{Kraichnan56,Panton74,Lysak06,Lee05,Ahn10}. In these models, the wavenumber--frequency spectrum of wall pressure fluctuations $\phi_{pp}(\boldsymbol{k},\omega)$ is approximated as the AS-Rapid $\phi_{rr}(\boldsymbol{k},\omega)$. The DNS results of \citet{Kim89} and~\citet{Chang99} show that the slow component of wall pressure fluctuations has the same order of magnitude as the rapid component. Therefore, it is more accurate to account for both AS-Rapid and AS-Slow to predict the spectrum of wall pressure fluctuations~\citep[see for example][]{Chase80,Chase87,Peltier07,Slama18,Grasso19}. In these models, the spectrum of the total wall pressure fluctuations is assumed to be the summation of AS-Rapid and AS-Slow, while CS-RS is neglected as a consequence of the joint-normal distribution assumption of the velocity fluctuations.

In this section, we use the DNS data to examine the influence of CS-RS on the spectrum of the total wall pressure fluctuations. To conduct quantitative analyses, there are two indicators evaluating the error induced by neglecting CS-RS. The first indicator is the ratio between CS-RS and the spectrum of the total wall pressure fluctuations, viz.
\begin{equation}
	R = \frac{{2{\phi_{rs}}(\boldsymbol{k},\omega )}}{{{\phi _{pp}}(\boldsymbol{k},\omega )}}=\frac{\phi_{pp}(\boldsymbol{k},\omega )-(\phi_{rr}(\boldsymbol{k},\omega )+\phi_{ss}(\boldsymbol{k},\omega ))}{\phi_{pp}(\boldsymbol{k},\omega )}\times 100\%.
	\label{eq:eq4.6}
\end{equation}
When $\phi_{rr}(\boldsymbol{k},\omega )+\phi_{ss}(\boldsymbol{k},\omega )$ is used to approximate the `exact value' of $\phi_{pp}(\boldsymbol{k},\omega )$ as is adopted in many models, the ratio $R$ also represents the relative error induced by neglecting $\phi_{rs}$.  It can be proven using the Cauchy-Schwarz inequality that the value of $R$ ranges from $-\infty$ to $50\%$. The second indicator is the decibel-scaled error in $\phi_{pp}$ caused by neglecting $\phi_{rs}$, defined as
\begin{equation}
	E = {10 \times {{\log }_{10}}\frac{{{\phi _{rr}}(\boldsymbol{k},\omega ) + {\phi _{ss}}(\boldsymbol{k},\omega )}}{{{\phi _{pp}}(\boldsymbol{k},\omega )}}} = {10 \times {{\log }_{10}}(1-R)}.
	\label{eq:eq4.7}
\end{equation}
Since the spectrum of wall pressure fluctuations is conventionally presented in a decibel scale, we mainly use $E$ as the indicator in this paper. It is worth noting that the magnitudes of $R$ and $E$ are asymmetric for positive and negative values of $\phi_{rs}$. To be specific, if $\phi_{rs}\ge 0$, then $0\le R\le 50\%$ holds, and the corresponding decibel-scaled error $E$ is no smaller than $-3.0$dB. However, if $\phi_{rs}< 0$, the relative error $R$ can reach $-\infty$ and the corresponding decibel-scaled error $E$ approaches $+\infty$. This suggests that if the value of $\phi_{rs}$ is negative and its magnitude is comparable to $\phi_{rr}+\phi_{ss}$, the wavenumber--frequency spectrum of wall pressure fluctuations can be significantly overestimated.

In the following content of this section, the characteristics of AS-Rapid, AS-Slow and CS-RS are discussed, and the errors in the spectra of the total wall pressure fluctuations caused by neglecting CS-RS are analyzed. In \S\,\ref{subsec:one}-\ref{subsec:three}, the results of one-dimensional, two-dimensional and three-dimensional spectra at $\Rey_\tau=998$ are presented.  The effects of the Reynolds number are examined in \S\,\ref{subsec:Reynolds}.

\subsection{Decomposition of one-dimensional spectra}\label{subsec:one}
\begin{figure}
	\centering{\includegraphics[width=0.3\textwidth]{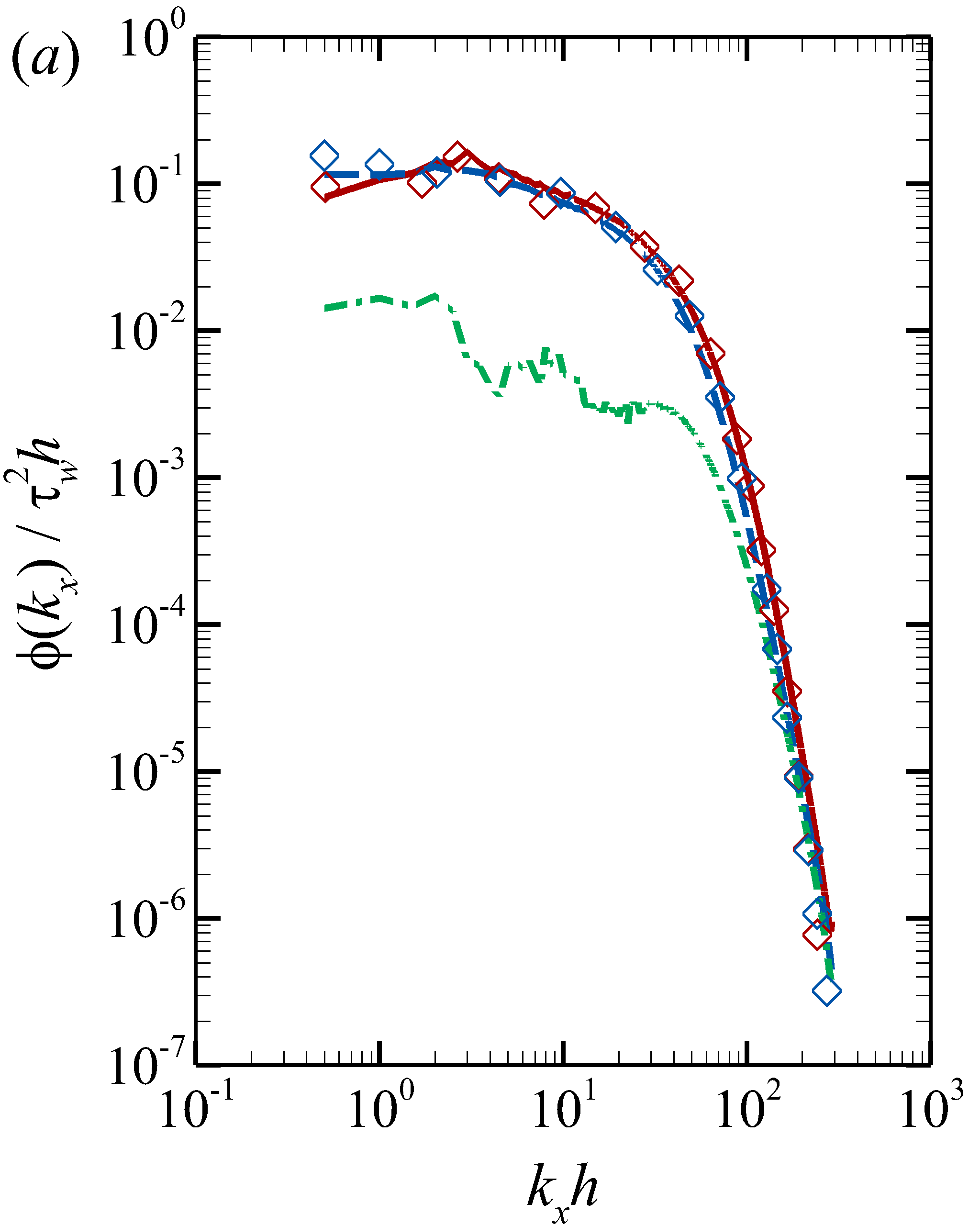}$\quad$
	          {\includegraphics[width=0.3\textwidth]{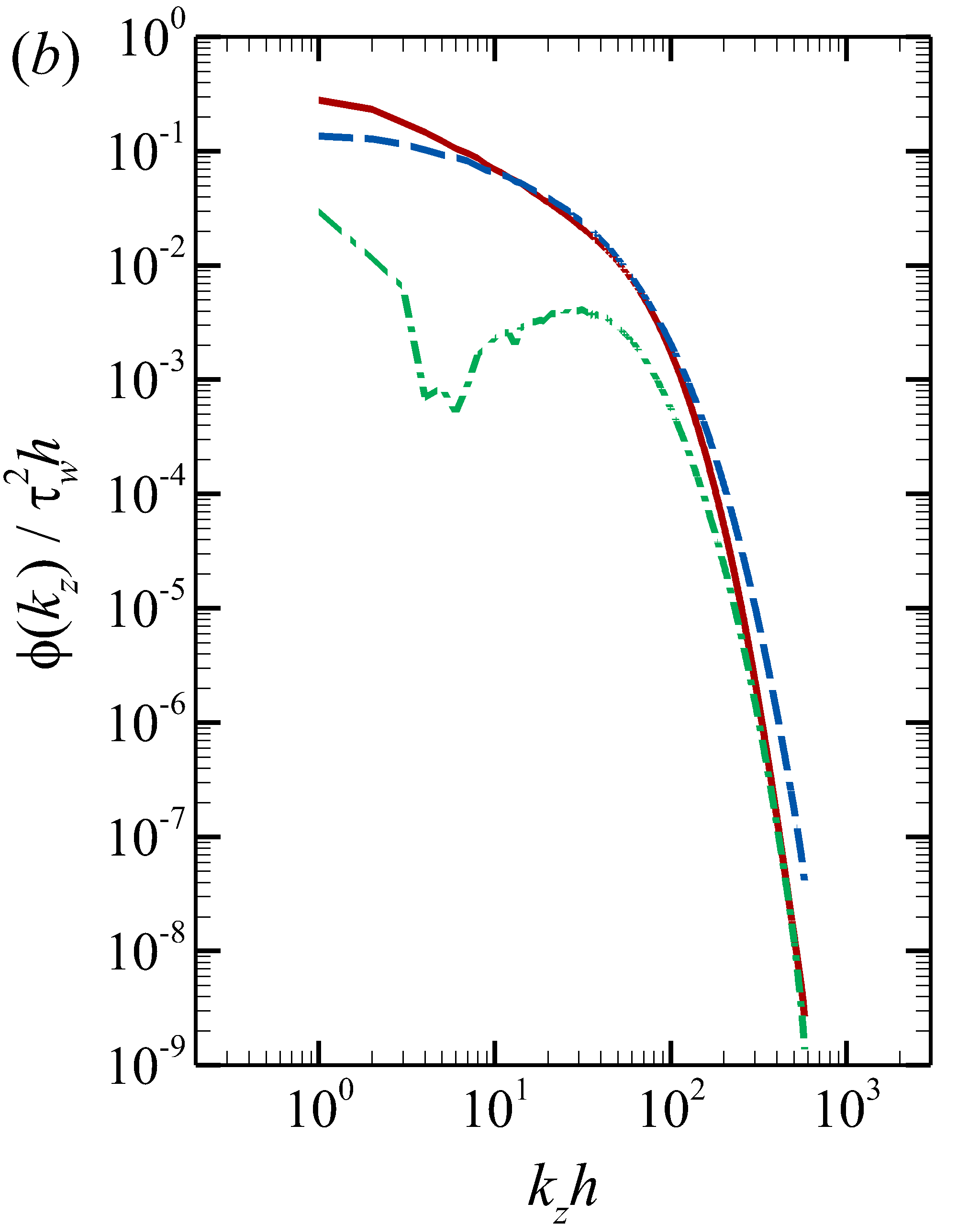}}$\quad$
	          {\includegraphics[width=0.3\textwidth]{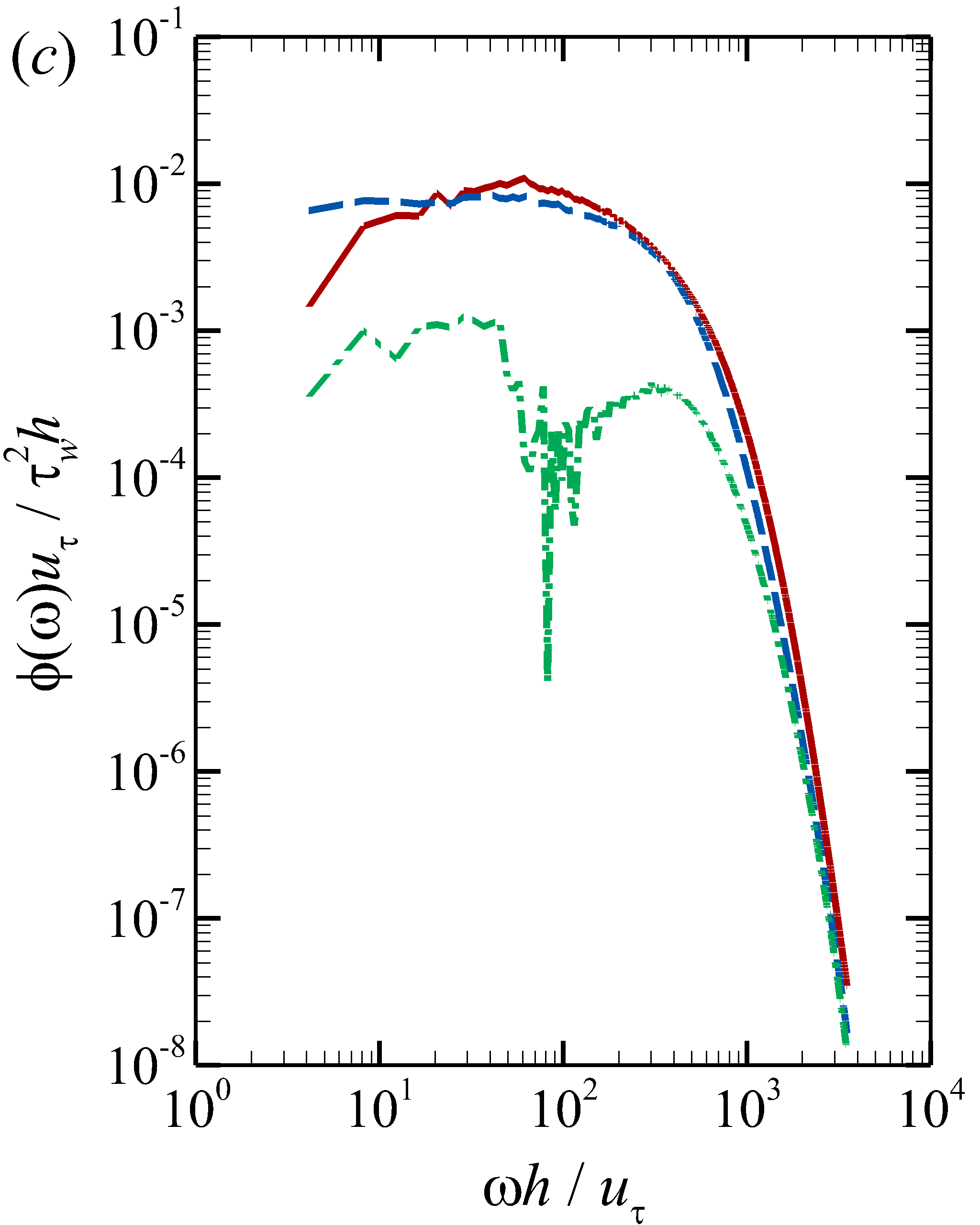}}}
    \caption{One-dimensional AS-Rapid (red solid lines), AS-Slow (blue dashed lines) and the absolute value of CS-RS (green dash-dotted lines) with respect to the (a) streamwise wavenumber $k_x$, (b) spanwise wavenumber $k_z$ and (c) frequency $\omega$. The Reynolds number is $Re_\tau = 998$.  The AS-Rapid (red diamonds) and AS-Slow (blue diamonds) with respect to the streamwise wavenumber $k_x$ of~\citet{Abe05} are superposed in panel (a) for comparison.}
    \label{fig:fig8}
\end{figure}

Figure~\ref{fig:fig8} compares the magnitudes of one-dimensional AS-Rapid, AS-Slow and CS-RS.  Note that $\phi_{rs}$ can be negatively valued,  such that its absolute value $|\phi_{rs}|$ is used to represent its magnitude in figure~\ref{fig:fig8}. To validate our results, the spectra of~\citet{Abe05} for AS-Rapid and AS-Slow with respect to the streamwise wavenumber $k_x$ are superposed in figure~\ref{fig:fig8}(a). It is seen that the present results agree with \citet{Abe05}.

For the $k_x$- and $\omega$-spectra (see figures~\ref{fig:fig8}a and~\ref{fig:fig8}c, respectively), the magnitudes of $\phi_{rr}$ and $\phi_{ss}$ are comparable throughout the entire resolved range of $k_x$ and $\omega$, except that the magnitude of $\phi_{ss}$ is slightly larger than that of $\phi_{rr}$ at small $k_x$ or $\omega$. For the $k_z$-spectra, $\phi_{rr}$ is larger than $\phi_{ss}$ at small wavenumbers for $k_zh<10$, but becomes smaller than $\phi_{ss}$ at large wavenumbers for $k_zh>100$. These observations are consistent with the results of~\citet{Kim89} and~\citet{Chang99} at $\Rey_\tau=180$. The magnitudes of $\phi_{rr}$ and $\phi_{ss}$ are much larger than that of $|\phi_{rs}|$ at low wavenumbers ($k_xh<50$, $k_zh<50$) and low frequencies ($\omega h/u_\tau<500$), indicating that neglecting CS-RS causes little error in the one-dimensional spectra of the wall pressure fluctuation at these wavenumbers and frequencies. As the wavenumber or frequency increases, the magnitudes of $|\phi_{rs}|$, $\phi_{rr}$ and $\phi_{ss}$ become comparable at high wavenumbers ($k_xh>100$, $k_zh>200$) or high frequencies ($\omega h/u_\tau>1000$).

\begin{figure}
	\centering{\includegraphics[width=0.3\textwidth]{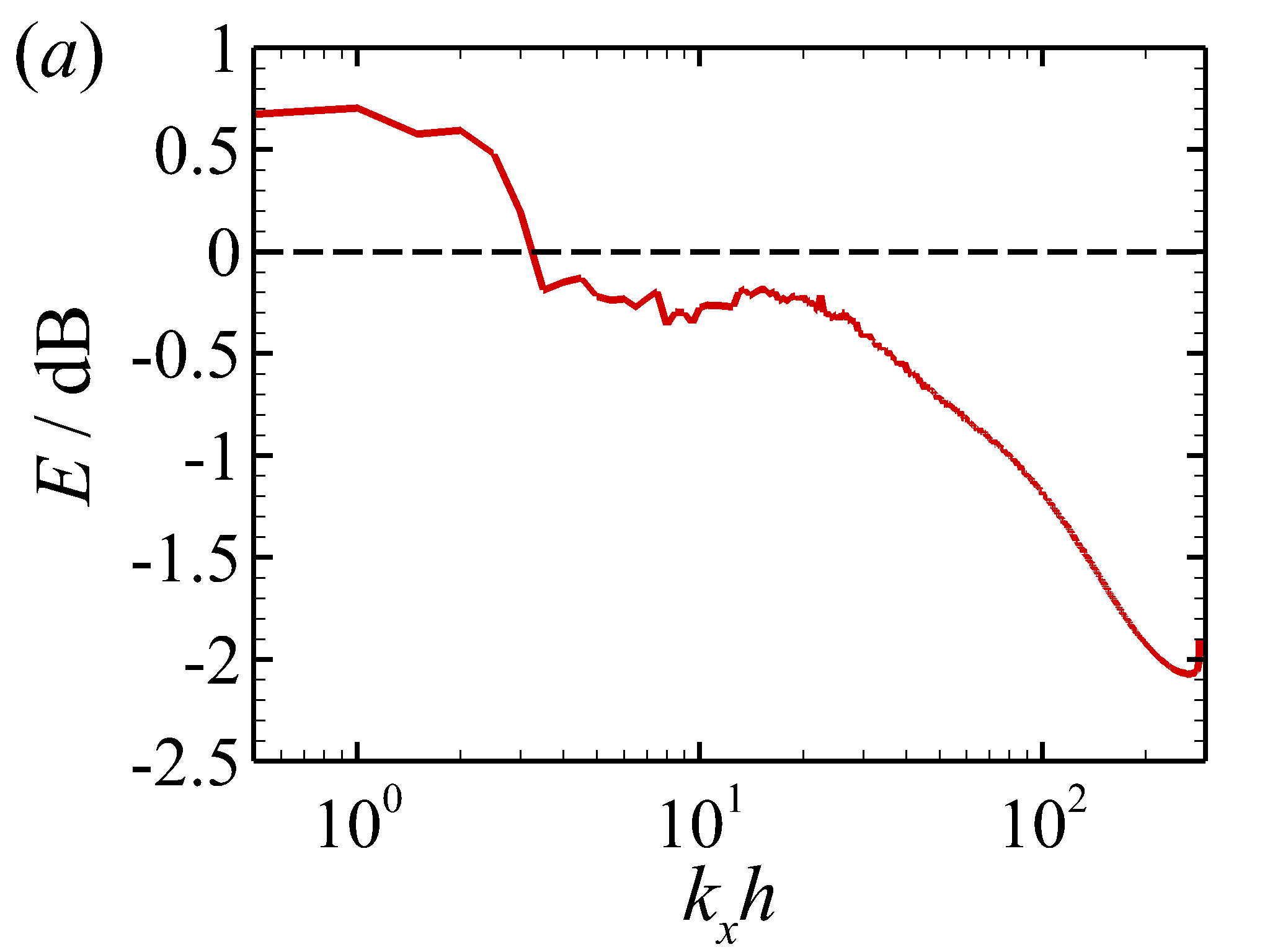}$\quad$
	          {\includegraphics[width=0.3\textwidth]{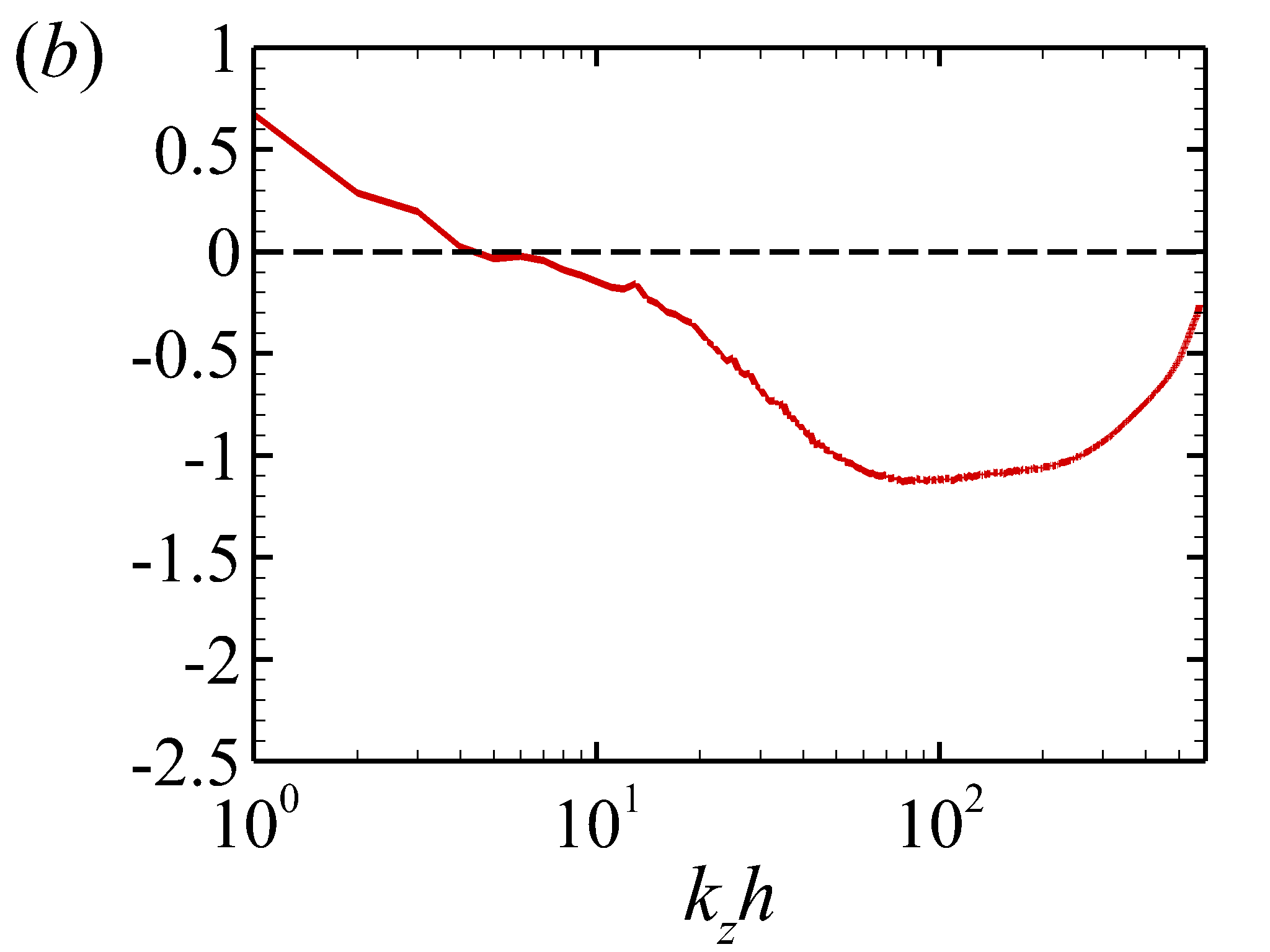}}$\quad$
	          {\includegraphics[width=0.3\textwidth]{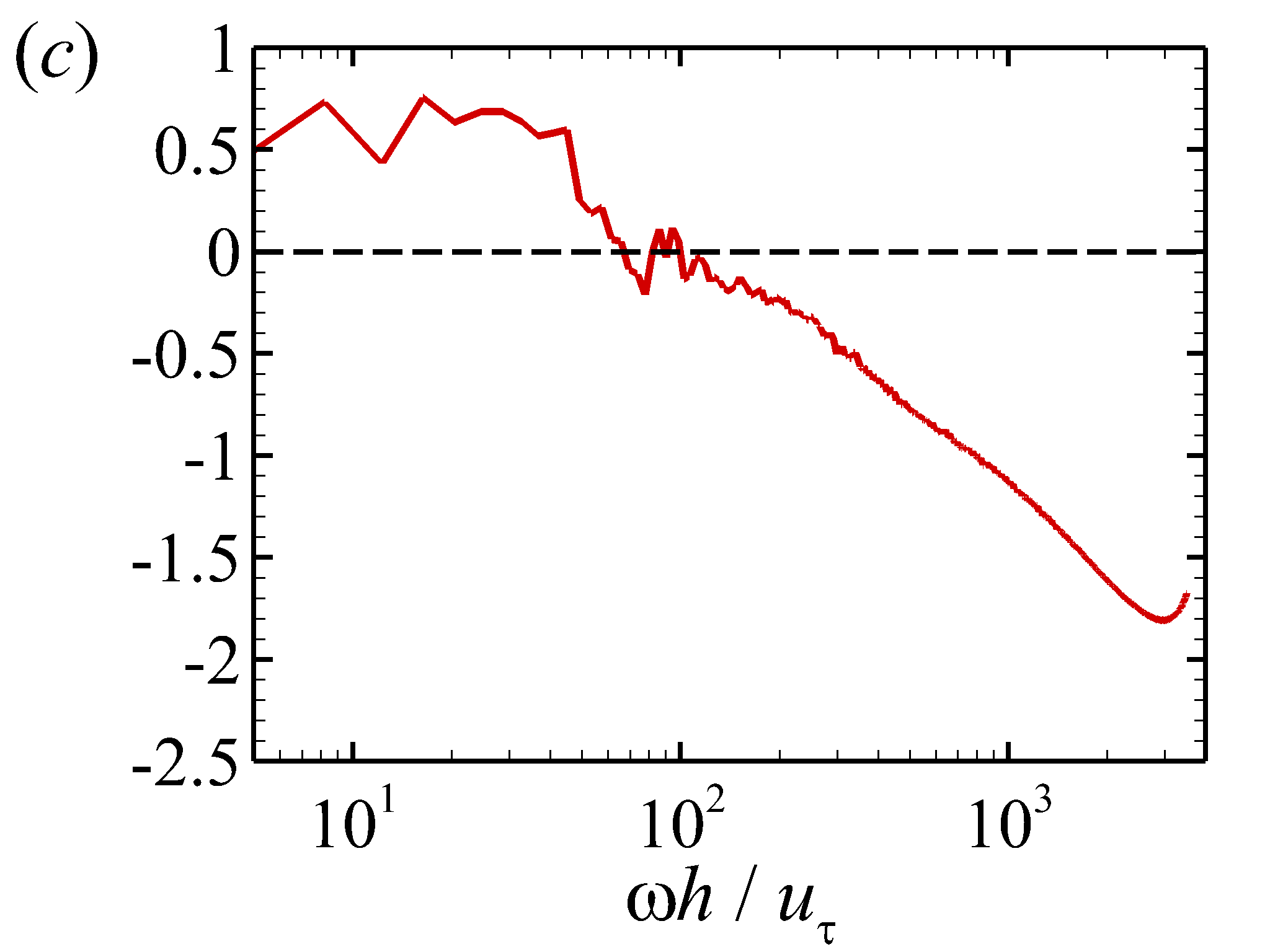}}}
    \caption{Decibel-scaled errors in the one-dimensional (a) $k_x$-, (b) $k_z$- and (c) $\omega$-spectra of the wall pressure fluctuation caused by neglecting CS-RS.}
    \label{fig:fig9}
\end{figure}

Figure~\ref{fig:fig9} shows the decibel-scaled errors in the one-dimensional spectra of the total wall pressure fluctuations caused by neglecting CS-RS. At small wavenumbers or frequencies, $E$ is positively valued, and smaller than 1.0dB. As the wavenumber or frequency increases, the value of $E$ decreases and crosses the dashed line denoting $E=0$ at ${k_x}h \approx 3.5$, ${k_z}h \approx 5.0$ and $\omega h/{u_\tau } \approx 80$. As the wavenumber or frequency continues to increase, the value of $E$ further decreases to its minimum and then increases slightly at very high wavenumbers or frequencies. The minimum values of $E$ for the $k_x$-, $k_z$- and $\omega$-spectra are approximately -2.1dB, -1.1dB and -1.8dB, respectively, suggesting that the assumption of neglecting CS-RS is acceptable for the one-dimensional spectra of wall pressure fluctuations.

\subsection{Decomposition of the two-dimensional spectrum}\label{subsec:two}

In this subsection, we continue to investigate the decomposition of the two-dimensional spectrum of wall pressure fluctuations. To keep the paper concise, we follow~\citet{Slama18} to focus on the $k_x$--$\omega$ spectrum, since the energy-containing convective peak, which is important for the cabin noise generation of a high-speed subsonic civil aircraft~\citep{Graham97}, can be identified from the $k_x$--$\omega$ spectrum.

\begin{figure}
	\centering{\includegraphics[width=0.48\textwidth]{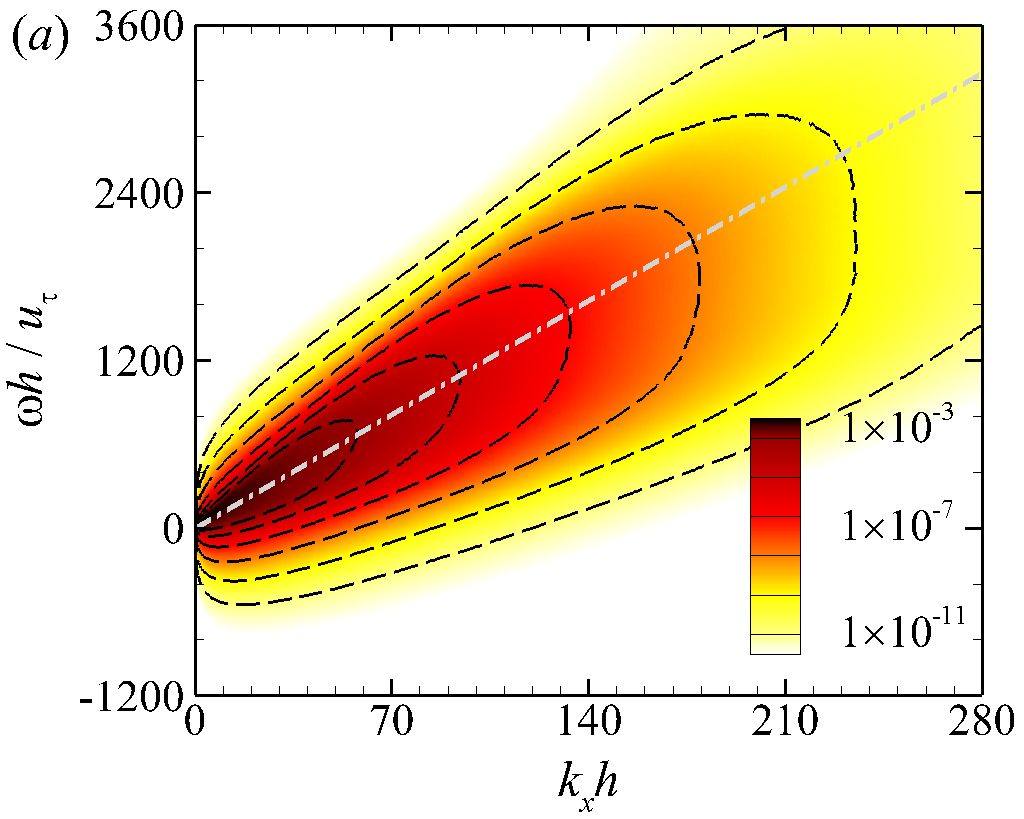}$\quad$
          	  {\includegraphics[width=0.48\textwidth]{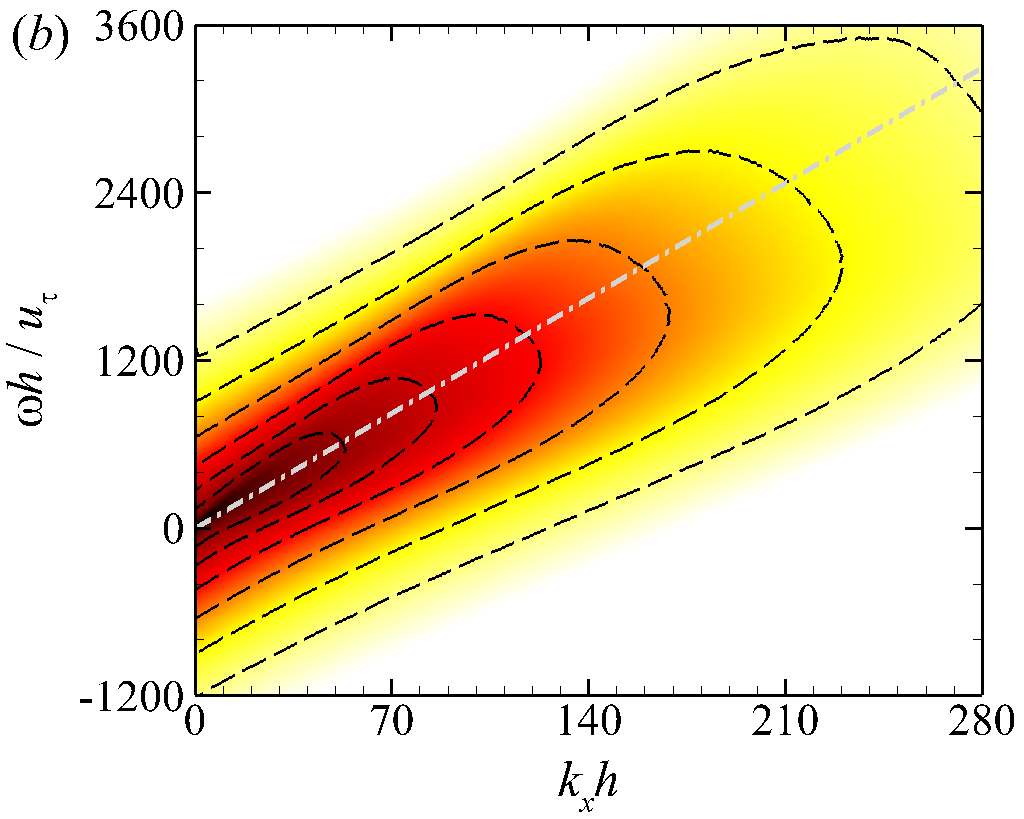}}}
    \caption{Contours of two-dimensional $k_x$--$\omega$ spectra of the (a) rapid and (b) slow components of the wall pressure fluctuation. Six equal-ratio isopleths from $1\times10^{-5}$ to $1\times10^{-10}$ are shown by the dash lines. The convection lines $k_x=\omega/U_c$ are demarcated using the dash-dotted lines. The Reynolds number is $Re_\tau = 998$.}
    \label{fig:fig10}
\end{figure}

The isopleths of two-dimensional AS-Rapid and AS-Slow in the $k_x$--$\omega$ plane are shown in figures~\ref{fig:fig10}(a) and~\ref{fig:fig10}(b), respectively. The difference between $\phi_{rr}$ and $\phi_{ss}$ mainly lies in the low-wavenumber region. As shown in figure~\ref{fig:fig10}(a), the isopleths of $\phi_{rr}$ contract to $({k_x},\omega ) = (0,0)$ as $k_x$ approaches zero. This property of $\phi_{rr}$ can be explained using the Helmholtz equation~(\ref{eq:eq4.3}) of the Fourier modes of the rapid pressure fluctuations. 

The source term of the rapid pressure fluctuations in the wavenumber--frequency space, i.e., ${\hat f}_r(\boldsymbol{k},\omega ;y')$ in equation~(\ref{eq:eq4.8}), can be written as
\begin{equation}
\begin{aligned}
	{\hat f_r}(\boldsymbol{k},\omega ;y) &= \frac{1}{{{{(2\pi )}^3}}}\int {dx} \int {dz} \int {dt }  \cdot \left( { - 2\frac{{dU}}{{dy}}\frac{{\partial v}}{{\partial x}}} \right)\exp [ - i({k_x}x + {k_z}z - \omega t)]\\
	&=  - \frac{{2{\rm{i}}{k_x}}}{{{{(2\pi )}^3}}}\frac{{dU(y)}}{{dy}}\hat v(\boldsymbol{k},\omega ;y),
	\label{eq:eq4.80}
\end{aligned}
\end{equation}
from which it is understood that ${\hat f}_r(k_x=0,k_z,\omega;y)=0$. If $k_z\neq 0$, the denominator of the Green's function approaches a non-zero finite value as $k_x\to0$. As a result, ${\hat p_r}(k_x=0,k_z\neq0,\omega ;y)=0$ also holds. If $k_z=0$, the Green's function is of order $\mathcal{O}(1/k_x^2)$ as $k_x\to0$. The above analyses need further considerations about $\hat v(k_x,k_z=0,\omega ;y)$. Integrating the continuity equation from the wall to $y$, $\hat v$ can be expressed as
\begin{equation}
	\hat v(k_x,k_z=0,\omega ;y) = \int_0^y {\rm{i}}k_x \hat u(k_x,k_z=0,\omega ;y')dy'.
	\label{eq:continuity}
\end{equation}
Note that $\hat u(k_x,k_z=0,\omega ;y')$ is a fluctuating velocity, which approaches zero as $k_x\to0$~\citep{Pope00}. Combining equations~(\ref{eq:eq4.8}) to~(\ref{eq:continuity}), it can be derived that ${\hat p_r}(k_x,k_z=0, \omega ;y)\sim\mathcal{O}(\hat u(k_x,k_z=0,\omega ;y')) \to 0$ as $k_x\to 0$. Since ${\phi _{rr}}(\boldsymbol{k},\omega )$ is a quadratic function of ${\hat p_r}(\boldsymbol{k},\omega ;y)$, it can be inferred that ${\phi _{rr}}(\boldsymbol{k},\omega )$ satisfies the constraint ${\phi _{rr}}({k_x} = 0,{k_z\neq0},\omega ) = 0$ at zero streamwise wavenumber. Integrating $\phi_{rr}(k_x,k_z,\omega)$ over the spanwise wavenumber $k_z$ yields the same constraint for the two-dimensional spectrum $\phi_{rr}(k_x,\omega)$. This constraint also leads to the contracting behavior of the isopleths of $\phi_{pp}$ around $k_x=0$ as shown in figure~\ref{fig:fig3}. In contrast, the source term of the slow pressure fluctuations does not have such a constraint at $k_x=0$. Consequently, the isopleths of AS-Slow do not show a contracting feature around $k_x=0$ (figure~\ref{fig:fig10}b).

\begin{figure}
	\centering{\includegraphics[width=0.7\textwidth]{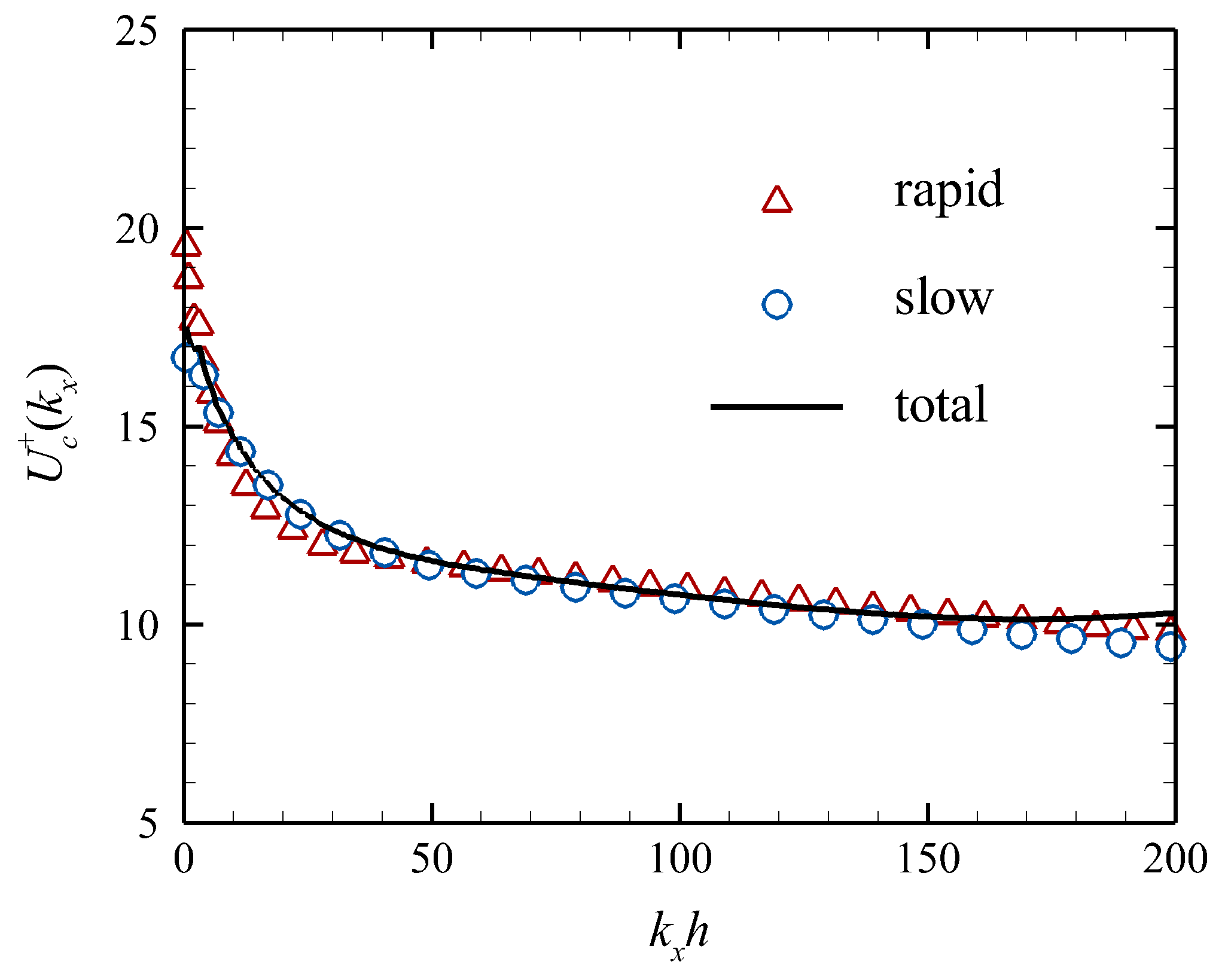}}
    \caption{Wavenumber-dependent convection velocities of the rapid, slow and total wall pressure fluctuations. The Reynolds number is $Re_\tau = 998$.}
    \label{fig:fig11}
\end{figure}

Both $\phi_{rr}$ and $\phi_{ss}$ show a convective property. Specifically, the large spectral values are concentrated around the convection lines demarcated by the dash-dotted lines. The slopes of the convection lines are identical to the bulk convection velocity, which can be determined using equation~(\ref{eq:eq_bc}). The values of the bulk convection velocity for $\phi_{rr}$ and $\phi_{ss}$ are $11.5{u_\tau }$ and $11.7{u_\tau }$, respectively, close to the bulk convection velocity of the total wall pressure fluctuations, i.e.,  $11.6{u_\tau}$ as shown in figure \ref{fig:fig3}(a). Figure~\ref{fig:fig11} further compares the wavenumber-dependent convection velocities of the rapid, slow and total wall pressure fluctuations. Here, the wavenumber-dependent convection velocity of an arbitrary variable is defined using its wavenumber--frequency spectrum $\phi(k_x,\omega)$ as~\citep{Alamo09}
\begin{equation}
	{U_{c}}({k_x}) = \frac{{\int {\omega \phi ({k_x},\omega )d\omega } }}{{{k_x}\int {\phi ({k_x},\omega )d\omega } }}.
	\label{eq:eq4.81}
\end{equation}
From figure~\ref{fig:fig11}, it is seen that the values of ${U_{c}}({k_x})$ for the rapid, slow and total wall pressure fluctuations all decrease monotonically as $k_x$ increases. At low wavenumbers for $k_xh<10$, the convection velocity of the rapid wall pressure fluctuations is larger than that of the slow wall pressure fluctuations. For $k_xh>10$, the convection velocities of the rapid, slow and total wall pressure fluctuations are close to each other. They all approach $10.0u_\tau$ as $k_x$ increases. This value is close to the convection velocity of the total wall pressure fluctuations at $\Rey_{\tau}=180$~\citep{Choi90}. These results indicate that the convective properties of the rapid and slow pressure are similar, and the same convection velocity can be used to model both AS-Rapid and AS-Slow~\citep{Chase80,Chase87}.

\begin{figure}
	\centering{\includegraphics[width=0.7\textwidth]{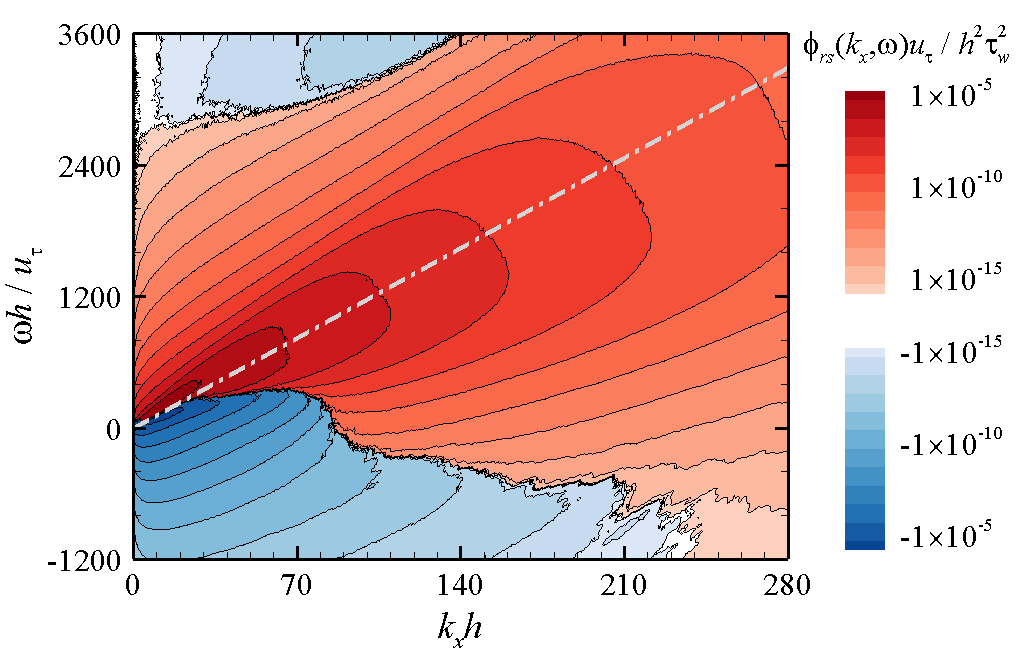}}
    \caption{Contours of two-dimensional CS-RS in the $k_x$--$\omega$ plane. The positive and negative values are colored by red and blue, respectively. The convection line $k_x=\omega/U_c$ is demarcated using the dash-dotted line. The Reynolds number is $Re_\tau = 998$.}
    \label{fig:fig12}
\end{figure}

Figure~\ref{fig:fig12} shows the contours of CS-RS in the $k_x$--$\omega$ plane with the positive and negative values colored by red and blue, respectively. It is seen that CS-RS is positive in most region around the convection line (denoted by the dash-dotted line), except for a small region below the convection line at relatively low $k_x$. Besides, since CS-RS also satisfies the constraint $\phi_{rs}(k_x,\omega)\to 0$ as $k_x$ approaches zero, the isopleths show similar contracting behavior as $\phi_{rr}(k_x,\omega)$ (see figure~\ref{fig:fig10}a) at low streamwise wavenumbers.

\begin{figure}
	\centering{\includegraphics[width=0.7\textwidth]{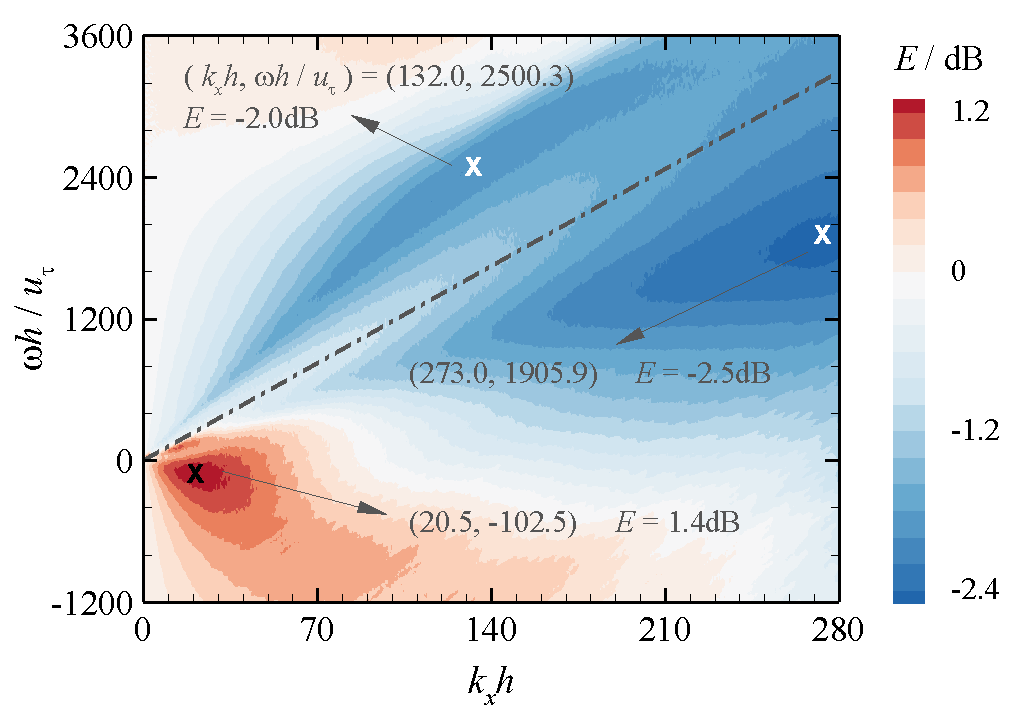}}
    \caption{Contours of the decibel-scaled error $E$ in the $k_x$--$\omega$ spectrum of  wall pressure fluctuations caused by neglecting CS-RS. The local peaks in the regions with negative and positive values of $E$ are marked using the white and black cross symbols, respectively. The Reynolds number is $Re_\tau = 998$.}
    \label{fig:fig13}
\end{figure}


Figure~\ref{fig:fig13} shows the contours of the decibel-scaled error $E$ in the $k_x$--$\omega$ spectrum of  wall pressure fluctuations caused by neglecting CS-RS. The positively and negatively valued regions are colored by red and blue, respectively. It is seen that there are three local peaks, two in the blue region with negative value of $E$ (marked by the white cross symbols), and one in the red region with positive value of $E$ (marked by the black cross symbol). The details of these peaks, including their locations and values, are given in the figure. Specifically, there are two negative peaks, with one located in the sub-convective region and the other in the viscous region. The values of $E$ for these two peaks are -2.0dB and -2.5dB, respectively. The positive peak is located below the convection line, and its decibel-scaled value is 1.4dB.

In summary, neglecting CS-RS leads to a maximum decibel-scaled error of approximately -2.5dB at viscous scales for the $k_x$--$\omega$ spectrum of wall pressure fluctuations. This is still an acceptable error magnitude, though it is slightly larger than that in the one-dimensional spectra. In \S\,\ref{subsec:three}, we continue to show that for the three-dimensional spectrum of wall pressure fluctuations, a much larger decibel-scaled error of approximately 4.7dB in the sub-convective region is induced by neglecting CS-RS.

\subsection{Decomposition of the three-dimensional spectrum}\label{subsec:three}

\begin{figure}
	\centering{\includegraphics[width=0.48\textwidth]{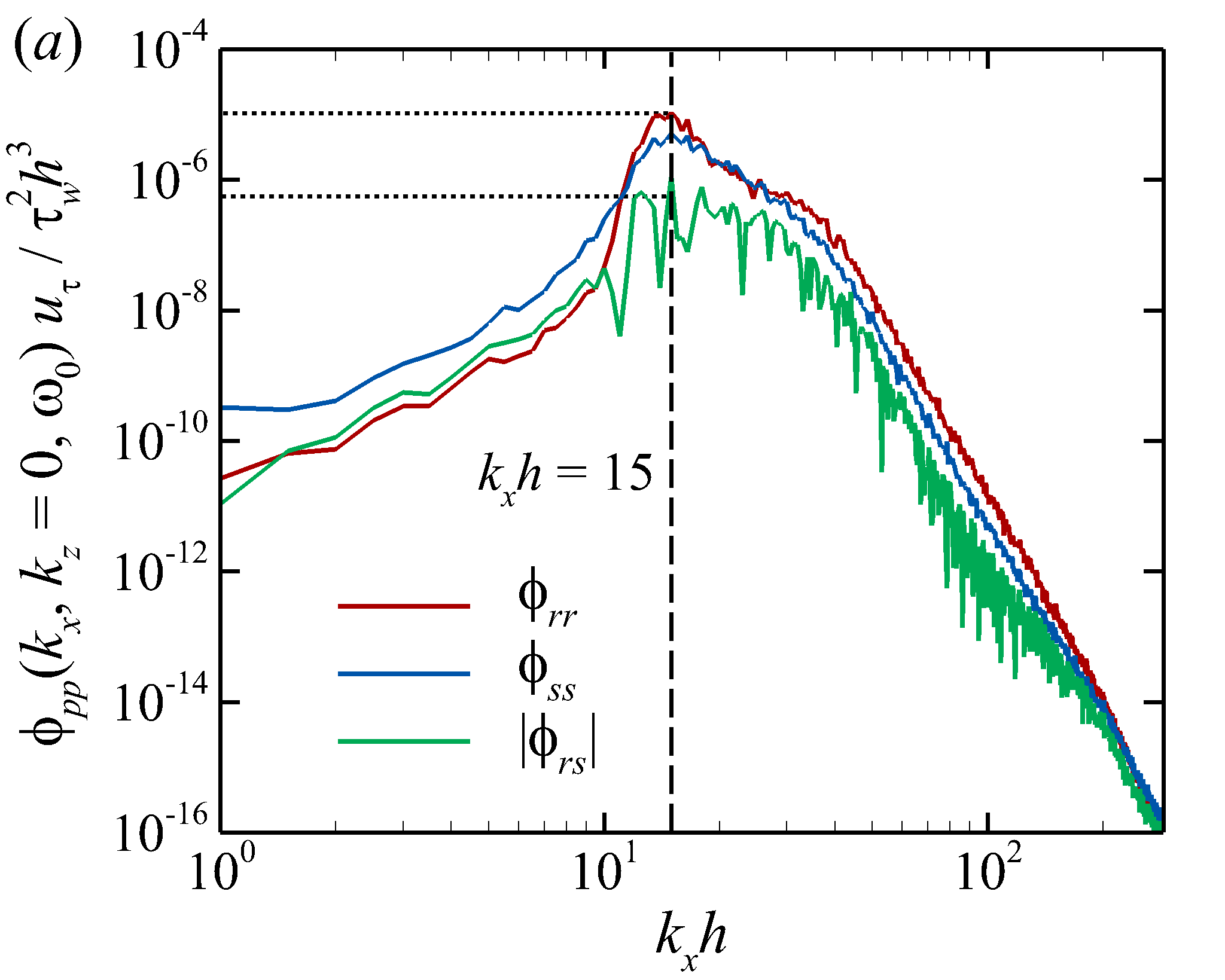}$\quad$
	          {\includegraphics[width=0.48\textwidth]{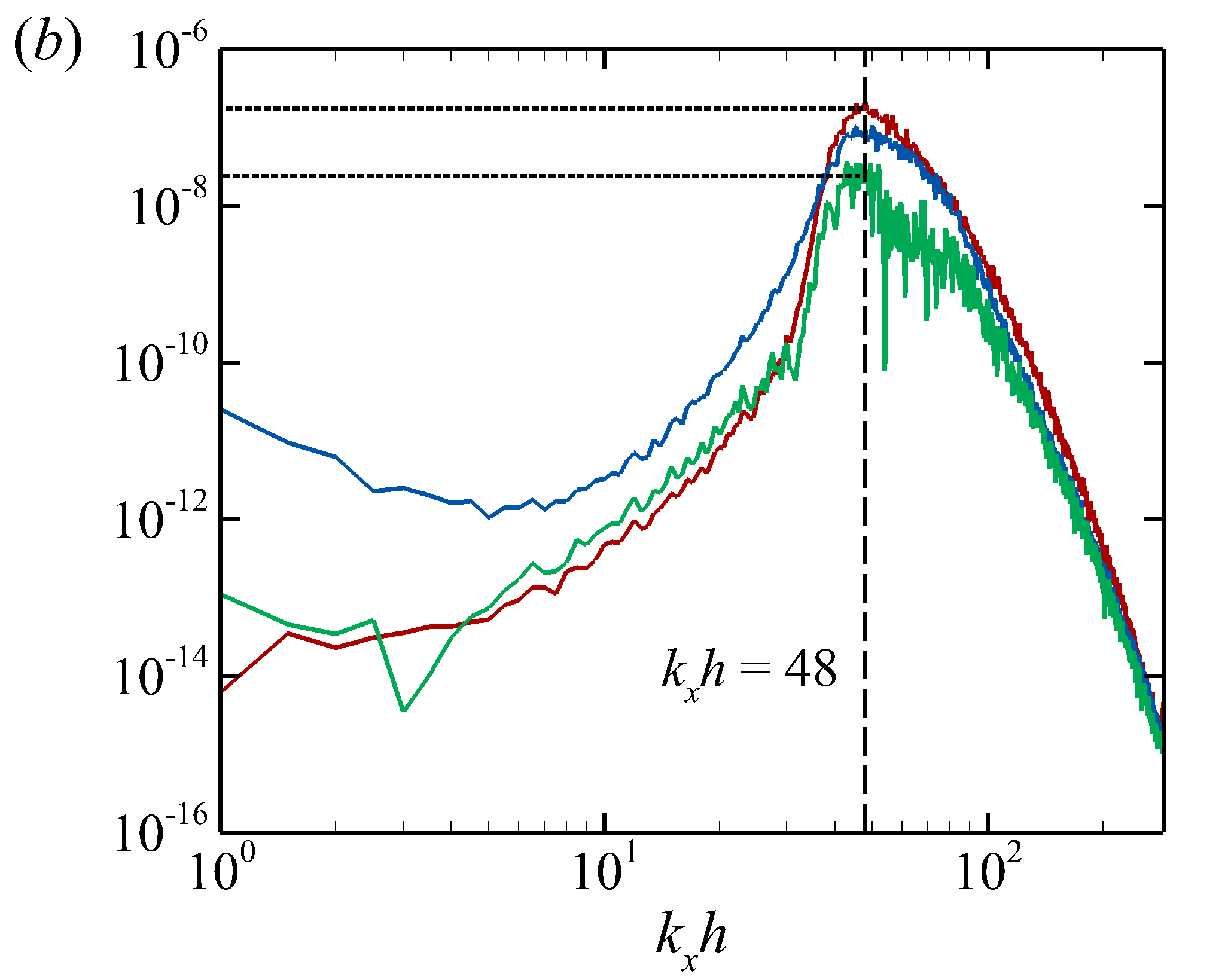}}}
    \caption{Absolute values of AS-Rapid $\phi_{rr}$, AS-Slow $\phi_{ss}$, and CS-RS $|\phi_{rs}|$ for the three-dimensional wavenumber--frequency spectra of wall pressure fluctuations for $k_z=0$ and two given frequencies (a) $\omega h/{u_\tau } = 245$ and (b) $\omega h/{u_\tau } = 716$. The convective wavenumbers are demarcated using vertical dashed lines. The Reynolds number is $Re_\tau = 998$.}
    \label{fig:fig14}
\end{figure}

Figure~\ref{fig:fig14} compares the variations of $\phi_{rr}$, $\phi_{ss}$ and $|\phi_{rs}|$ with respect to $k_x$ at $k_z=0$ and two given frequencies $\omega h/{u_\tau } = 245$ and $\omega h/{u_\tau } = 716$. The corresponding convective wavenumbers demarcated by the vertical dashed lines are $k_xh=15$ and $k_xh=48$, respectively. From these convective wavenumbers and frequencies, it can be calculated that the convection velocities at these two frequencies are $16.3u_{\tau}$ and $14.9u_{\tau}$, respectively. These values are larger than the convection velocity shown in figure~\ref{fig:fig11}. Such a difference is attributed to the fact that figure~\ref{fig:fig14} shows the three-dimensional spectrum at a specific spanwise wavenumber $k_z=0$, while in figure~\ref{fig:fig11}, the convection velocity of the two-dimensional spectrum averaged over all spanwise wavenumbers is depicted.  The above comparison indicates that the convection velocity also depends on the spanwise wavenumber. This is consistent with the results of \citet{Alamo09}.

The spectra at the two frequencies have some common features. First, in the sub-convective region (left side of the vertical dashed lines), the magnitudes of $|\phi_{rs}|$ and $\phi_{rr}$ are comparable, while that of $\phi_{ss}$ is large. Second, as $k_x$ increases to the convective wavenumber, the magnitudes of $\phi_{rr}$ and $\phi_{ss}$  become close to each other gradually, while $|\phi_{rs}|$ becomes one order of magnitude smaller than $\phi_{rr}$ and $\phi_{ss}$ (shown by the horizontal dotted lines in figure~\ref{fig:fig14}). Finally, the magnitudes of $|\phi_{rs}|$, $\phi_{rr}$ and $\phi_{ss}$ become close to each other for $k_xh>200$.

\begin{figure}
	\centering{\includegraphics[width=0.48\textwidth]{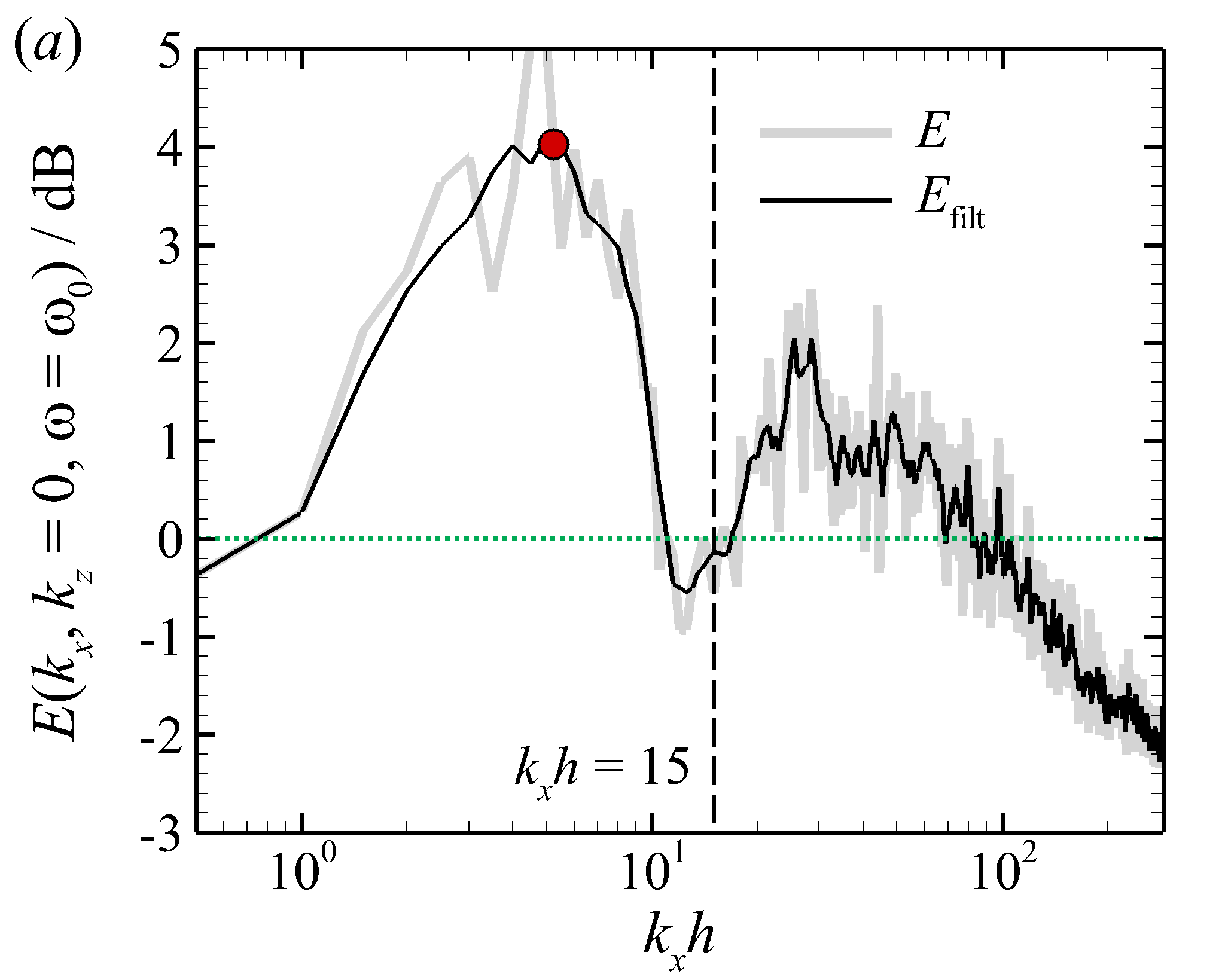}$\quad$
	          {\includegraphics[width=0.48\textwidth]{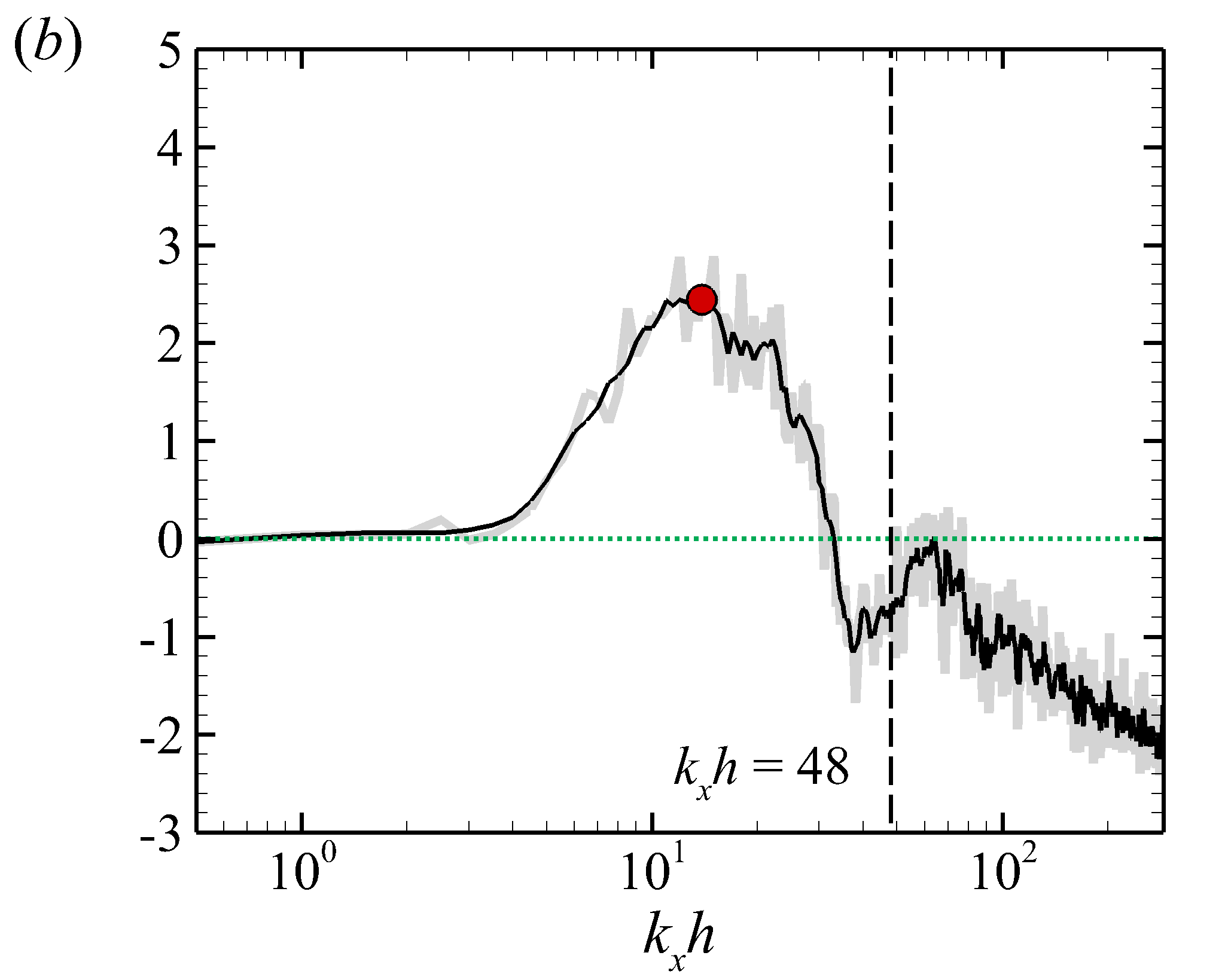}}}
    \caption{Unfiltered (grey lines) and filtered (black lines) error $E$ in the three-dimensional wavenumber--frequency spectrum caused by neglecting CS-RS for $k_z=0$ and two given frequencies (a) $\omega h/{u_\tau } = 245$ and (b) $\omega h/{u_\tau } = 716$. The convective wavenumbers are demarcated using vertical dashed lines. The Reynolds number is $Re_\tau = 998$.}
    \label{fig:fig15}
\end{figure}

Next, we focus on the error in the three-dimensional spectrum of wall pressure fluctuations caused by neglecting CS-RS. By examining the decibel-scaled error in the $k_x$--$k_z$ plane at various frequencies (not shown for brevity), we find that for an arbitrarily given combination of $(k_x, \omega)$, the peak of $E$ always occurs at $k_z = 0$.  Therefore, we focus on $k_z = 0$ in the following error analyses.  Figure~\ref{fig:fig15} shows the $k_x$-variation of the decibel-scaled error $E$ at $k_z=0$ and two frequencies $\omega h/{u_\tau } = 245$ and $\omega h/{u_\tau } = 716$. Since the raw data (denoted by the grey lines in figure~\ref{fig:fig15}) fluctuate violently, we follow \citet{Baars16} to adopt a moving average filter (MAF) to show the qualitative trend. The MAF is expressed as
\begin{equation}
	{E_{{\rm{filt}}}}({k_x}) = \sum\limits_{i =  - l}^l {\frac{1}{{2l + 1}}E({k_x} + l\Delta {k_x})},
	\label{eq:eq4.83}
\end{equation}
where the filter width is chosen to be $l=2$. The filtered results are denoted by the black lines in figure~\ref{fig:fig15}. For both frequencies, the maximum values of $E$ (denoted by red circles in~\ref{fig:fig15}) occur in the sub-convective region. At the lower frequency $\omega h/u_\tau=245$, the maximum values of unfiltered and filtered error reach 5.1dB and 4.1dB, respectively. At the higher frequency, these two values decrease to $2.9$dB and 2.5dB, respectively.  The results shown in figure~\ref{fig:fig15} indicate a significant over-prediction of the spectral magnitude in the sub-convective region.

\begin{figure}
	\centering{\includegraphics[width=0.7\textwidth]{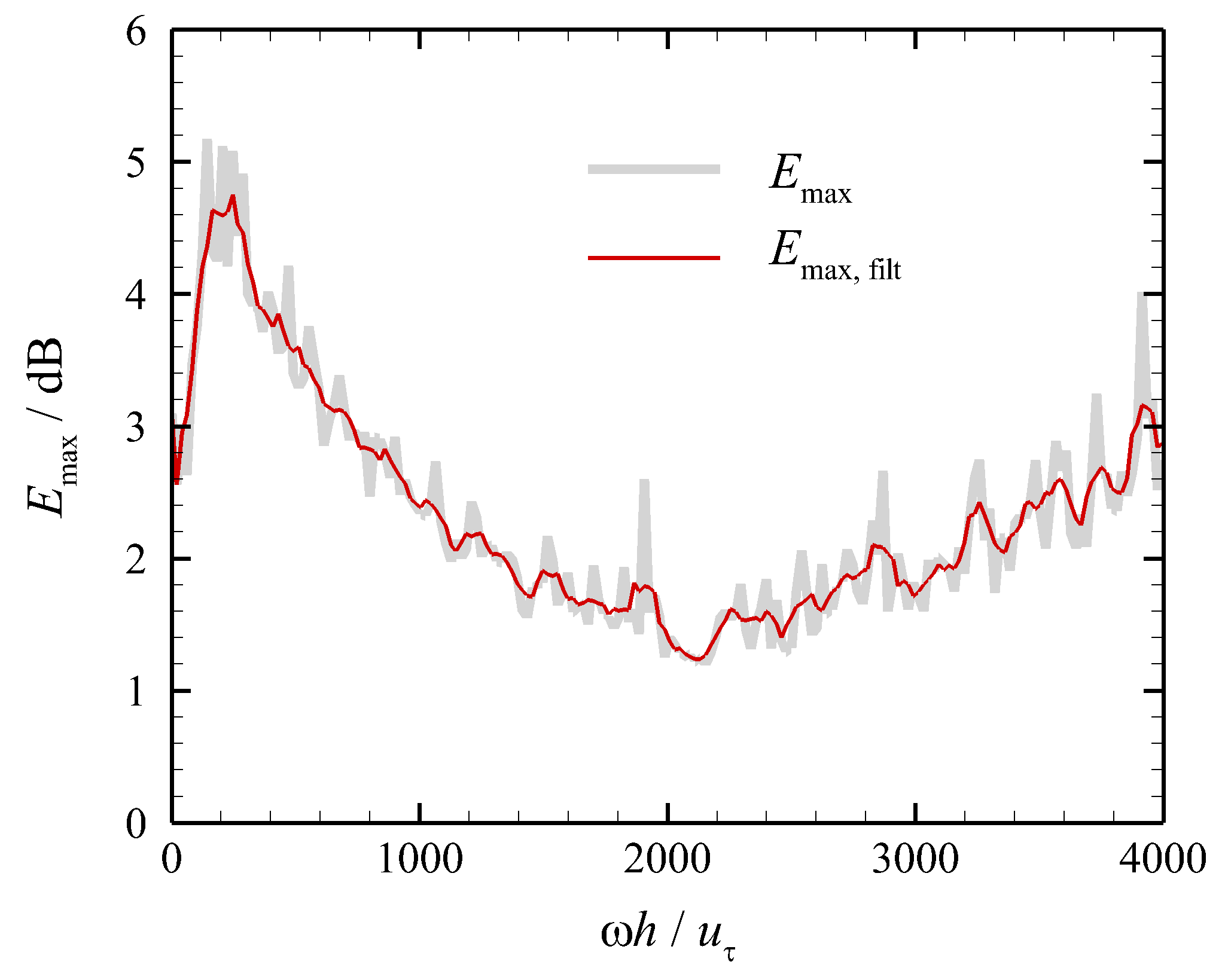}}
    \caption{Unfiltered (grey line) and filtered (red line) frequency-dependent maximum error $E_{\max}$ in the three-dimensional wavenumber--frequency spectrum of wall pressure fluctuations caused by neglecting CS-RS.  The Reynolds number is $Re_\tau = 998$.}
    \label{fig:fig16}
\end{figure}

While figure~\ref{fig:fig15} shows the error at specific frequencies, it is also useful to find the maximum error over all frequencies. For this purpose, we define the frequency-dependent maximum error ${E_{\max }}(\omega )$ as
\begin{equation}
	{E_{\max }}(\omega ) = \mathop {\max }\limits_{{k_x}} E({k_x},{k_z} = 0,\omega ).
	\label{eq:eq4.84}
\end{equation}
Figure~\ref{fig:fig16} shows the variation of $E_{\max}$ with respect to the dimensionless frequency $\omega h / u_\tau$. The raw data and filtered data are denoted by the grey and red lines, respectively. The peak of $E_{\max}$ occurs at a low frequency, of which the value is 4.7dB and 5.2dB with and without the application of the filter, respectively. Such a decibel-scaled error (4.7dB) corresponds to a large relative error of about 195\% caused by neglecting CS-RS in the model of the three-dimensional spectrum of wall pressure fluctuations.

It should be noted here that in practical applications, the total wall pressure fluctuations are weighted integration of their spectrum at all wavenumbers and frequencies.  Therefore, the error in the sub-convective region can be attenuated.  This is supported by the fact that the errors in one- and two-dimensional spectra are smaller than that in the three-dimensional spectrum.  However, the findings of this section cast a doubt on the conventional assumption that CS-RS is negligible compared to AS-Rapid and AS-Slow in some specific applications.  In particular, neglecting CS-RS induces significant errors in the sub-convective region at low frequencies. As pointed out by \citet{Graham97}, the properties of the wavenumber--frequency spectrum model in the sub-convective region are crucial for the prediction of noise generation in low-Mach-number flows over stiff structures.  The omission of CS-RS can be, therefore, inaccurate in corresponding applications, such as automobiles and submarine vehicles, and a modification is expected.

\subsection{Reynolds number effects}\label{subsec:Reynolds}

\begin{figure}
	\centering{\includegraphics[width=0.7\textwidth]{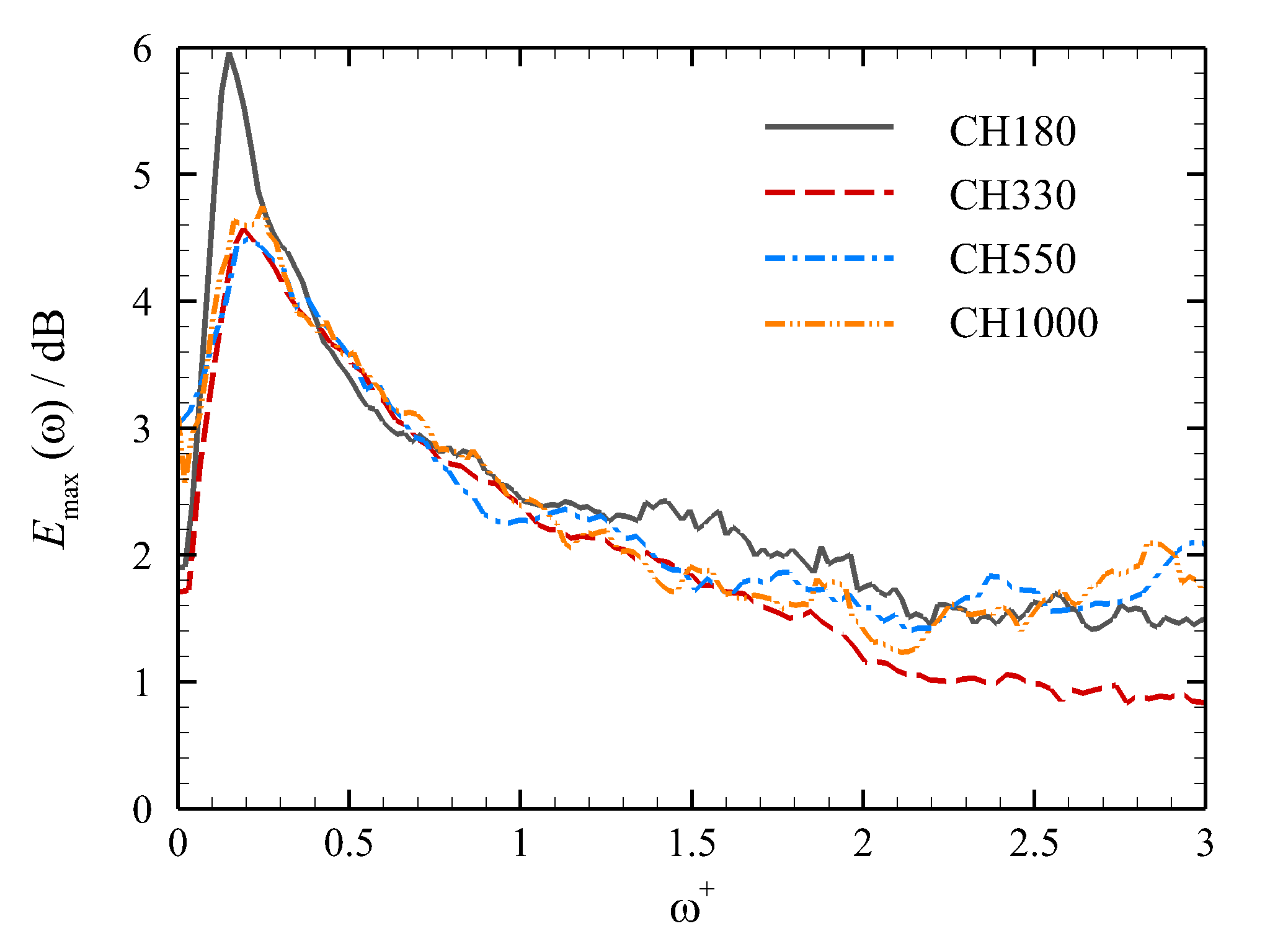}}
    \caption{Variation of filtered $E_{\max}$ with respect to the viscous-scaled frequency $\omega^+$ for different Reynolds numbers.}
    \label{fig:fig17}
\end{figure}

The above analyses of the error in the spectrum of wall pressure fluctuations caused by neglecting CS-RS are conducted at $\Rey=998$. In this subsection, we further examine the effect of the Reynolds number. Because the errors in the one- and two-dimensional spectra are acceptable, while that in the three-dimensional spectrum is significantly larger, we focus on the three-dimensional spectrum in this subsection to keep the paper concise.

Figure~\ref{fig:fig17} compares the variations of the filtered value of $E_{\max}$ (see equation~\ref{eq:eq4.84} for definition) with respect to the viscous-scaled frequency for various Reynolds numbers. It is observed that the distributions of $E_{\max}$ are close to each other for $330\lesssim\Rey_\tau\lesssim 1000$ under the viscous scaling. Specifically, the peak occurs at $\omega^+\approx0.25$ and the peak value is approximately 4.7dB. For the lowest Reynolds number $\Rey_\tau=180$, the peak locates at a slightly lower frequency $\omega^+\approx1.5$ and the peak value is about 6.0dB. The difference might be induced by the low-Reynolds-number effect. This observation indicates that the maximum error in the wavenumber--frequency spectrum of wall pressure fluctuations induced by neglecting CS-RS is correlated to the property of the velocity field in the near-wall region. 

\section{Conclusions}\label{sec:conclusion}

In this paper, we conduct DNS of turbulent channel flows at four Reynolds numbers ($\Rey_\tau=179,333,551,998$) to study the wavenumber--frequency spectrum of wall pressure fluctuations. The characteristics of the wavenumber--frequency spectrum of the total wall pressure fluctuations, along with AS-Rapid, AS-Slow and CS-RS, are analyzed with particular focus on the assumption that CS-RS is negligibly small compared to AS-Rapid and AS-slow, which is commonly used in various models of the wavenumber--frequency spectrum of wall pressure fluctuations. We also conduct an investigation of the time decorrelation of wall pressure fluctuations. The main findings of this paper are summarized as follows.

For the two-dimensional and three-dimensional spectra of the total wall pressure fluctuations, the isopleths gradually contract to the origin as the streamwise wavenumber $k_x$ approaches zero. This contracting behavior is determined by the mathematical constraint of AS-Rapid $\mathop {\lim }\limits_{{k_x} \to 0} {\phi _{rr}}({k_x},k_z,\omega ) =0$, which is proven using the solution of the Helmholtz equation based on the Green's function. 

In the investigation of the time decorrelation mechanisms, it is discovered that the frequency distributions of the wavenumber--frequency spectrum of wall pressure fluctuations at energy-containing scales follow a similar Gaussian form as that of  streamwise velocity fluctuations. The sweeping velocities are close to the RMS velocity fluctuations at the center of the buffer layer. This indicates that the time decorrelation of wall pressure fluctuation is also dominated by the random sweeping, and the energy-containing part of the wavenumber--frequency spectrum of  wall pressure fluctuations can be well predicted using a random sweeping model.

In the analyses of the model error caused by neglecting CS-RS, it is found that the assumption that CS-RS can be neglected in comparison with AS-Rapid and AS-Slow is acceptable for one- and two-dimensional spectra. However, for the three-dimensional spectrum, neglecting CS-RS leads to a significant over prediction of the spectrum in the sub-convective region. The maximal decibel-scaled error is approximately 4.7dB for $330\lesssim\Rey_\tau\lesssim 1000$. Furthermore, the frequency corresponding to the maximum error is scaled by the wall units, indicating that the error is correlated to the velocity sources in the near-wall region. These results show that in the predictive models of the wavenumber--frequency spectrum of wall pressure fluctuations, more efforts are desired to account for CS-RS in the sub-convective region. 

As a final remark of this paper, we note that due to the influence of the artificial acoustic, the wavenumber--frequency spectrum of wall pressure fluctuations at small wavenumbers is not analyzed thoroughly in this paper.  However, the flow-induced low-wavenumber wall pressure fluctuation plays an important role in structure vibration~\citep{Graham97}.  In future studies, it is desired to conduct simulations using sufficiently large computational domain to examine the corresponding theories~\citep{Kraichnan56,Phillips56} and physical models~\citep[e.g.][]{Chase80,Chase87} at small wavenumbers. 
\appendix

\section{Data convergence analysis with respect to the number of averaging samples}\label{appA}

\noindent In this appendix, we examine the data convergence of the present results. As described in \S\,\ref{sec:numerical}, $M=38$ samples are used to conduct time averaging in the calculation of the three-dimensional spectrum of wall pressure fluctuations. Since one- and two-dimensional spectra are further averaged in spatial or temporal direction, they are less sensitive to the number of averaging samples. Therefore, we mainly discuss the data convergence performance of the three-dimensional spectrum, with special focus on whether the frequency-dependent maximum error $E_{\max}$ shown in figure~\ref{fig:fig16} is robust as the number of averaging samples $M$ changes.

\begin{figure}
	\centering{\includegraphics[width=0.7\textwidth]{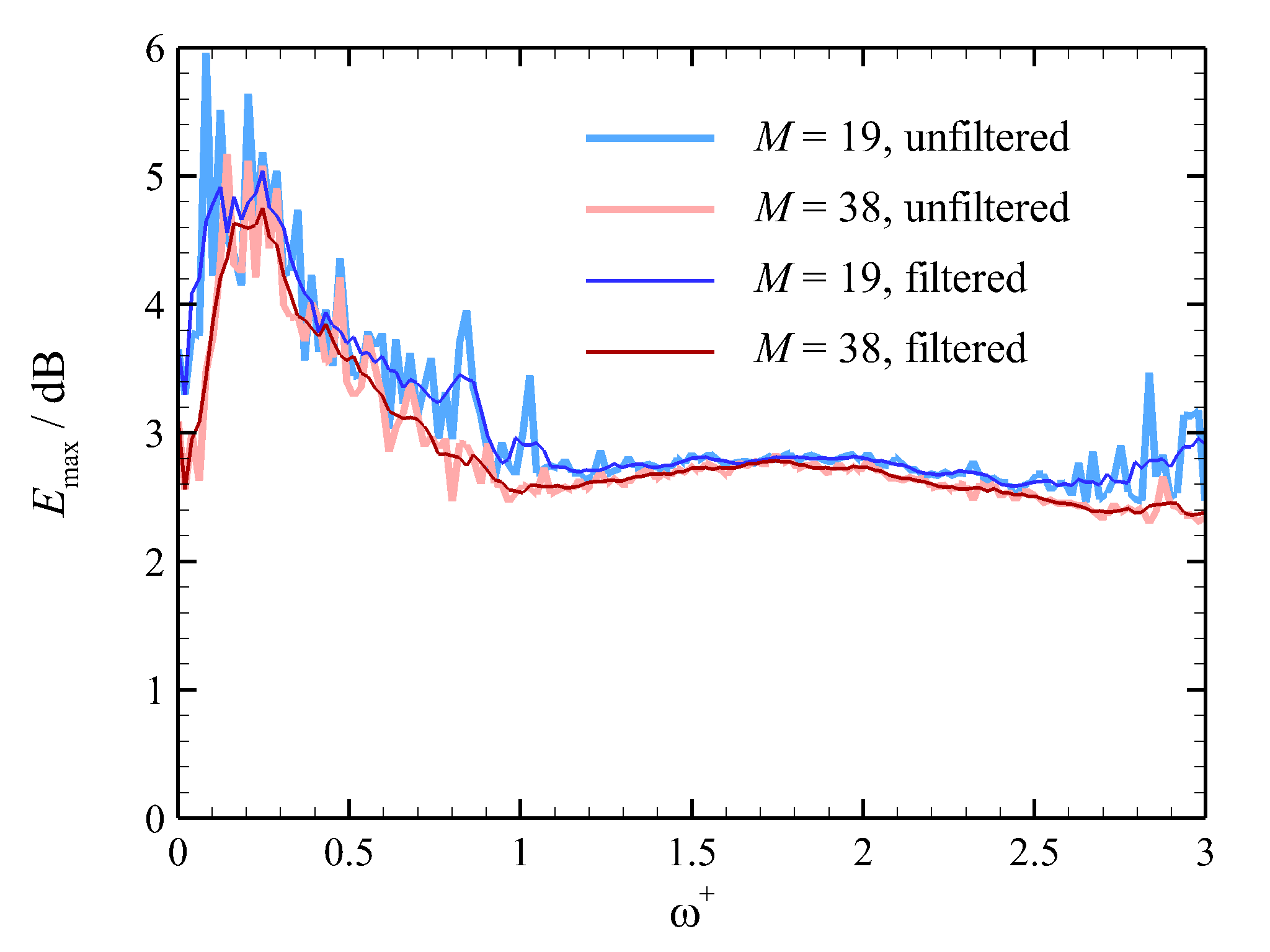}}
    \caption{Unfiltered (thick lines) and filtered (thin lines) frequency-dependent maximum error $E_{\max}$ in the three-dimensional wavenumber--frequency spectrum of wall pressure fluctuations caused by neglecting CS-RS with different number of time-averaging samples. The Reynolds number is $Re_\tau = 998$.}
    \label{fig:fig18}
\end{figure}

Figure~\ref{fig:fig18} compares the frequency-dependent maximum error $E_{\max}$ for $M=19$ (blue lines) and 38 (red lines). Both filtered and unfiltered results are presented. Although the results for $M=19$ are more oscillatory without applying the filter, the results for $M=19$ and 38 are in general consistent. Particularly, both $M=19$ and 38 show a peak at $\omega^+\approx0.25$, and the filtered peak values are 5.0dB and 4.7dB, respectively.  Figure~\ref{fig:fig18} indicates that $E_{\max}$ is robust to the number of averaging samples, and the uncertainty is approximately $6.4\%$.


\backsection[Acknowledgements]{The authors would like to thank Prof.~Guowei He and Dr.~Ting Wu for fruitful discussion.  }

\backsection[Funding]{This research is supported by the NSFC Basic Science Center Program for ``Multiscale Problems in Nonlinear Mechanics'' (No.~11988102), National Key Project (GJXM92579) and the Strategic Priority Research Program (XDB22040104).}

\backsection[Declaration of interests]{The authors report no conflict of interest.}


\backsection[Author ORCID]{ B. Yang, https://orcid.org/0000-0002-4843-6853;
Z. Yang, https://orcid.org/0000-0002-7764-3595}

\backsection[Author contributions]{B.Y. conducts the direct numerical simulation and data processing.  B.Y. and Z.Y. contribute to analysing data, reaching conclusions, and writing the paper.}





\bibliographystyle{jfm}
\bibliography{wall_pressure_spectra}


\end{document}